\documentclass[12pt]{article}%
\pdfoutput=1
\usepackage[nosort]{cite}
\usepackage{graphicx}
\usepackage{multicol}
\usepackage{amsfonts}
\usepackage{amssymb}
\usepackage{amsmath}
\usepackage{heck}
\usepackage{afterpage}
\usepackage{setspace}
\usepackage{verbatim}
\usepackage{color}
\usepackage{longtable}
\usepackage{float}
\usepackage{subcaption}
\usepackage{epsfig}
\usepackage{enumerate}
\usepackage{epstopdf}
\usepackage[enableskew, vcentermath]{youngtab}
\usepackage{adjustbox}
\usepackage{multirow}
\newsavebox{\mysavebox}

\usepackage{tikz}
\usepackage[margin=1in]{geometry}
\usepackage{titletoc}%
\usepackage{hyperref}
\hypersetup{colorlinks,%
citecolor=black,%
filecolor=black,%
linkcolor=black,%
urlcolor=black,%
pdftex}
\providecommand{\U}[1]{\protect\rule{.1in}{.1in}}
\usetikzlibrary{decorations.markings}

\numberwithin{equation}{section}
\def\bZ{\mathbb{Z}}
\def\bR{\mathbb{R}}
\def\bC{\mathbb{C}}
\usepackage{mathtools}
\usetikzlibrary{chains}

\tikzset{node distance=2em, ch/.style={circle,draw,on chain,inner sep=2pt},chj/.style={ch,join},every path/.style={shorten >=4pt,shorten <=4pt},line width=1pt,baseline=-1ex}

\let\dlabel=\alabel

\newcommand{\dnode}[2][chj]{%
\node[#1,label={below:\dlabel{#2}}] {};
}

\newcommand{\dnodebr}[1]{%
\node[chj,label={below right:\dlabel{#1}}] {};
}

\newcommand{\ba}{\begin{eqnarray}}
\newcommand{\ea}{\end{eqnarray}}

\newcommand{\cD}{\mathcal{D}}

\newcommand{\mf}{\mathfrak}

\newcommand{\ov}{\overset }

\newcommand{\D}{\displaystyle}

\newcommand{\be}{\begin{equation}}
\newcommand{\ee}{\end{equation}}

\tikzstyle{startstop} = [rectangle, rounded corners, minimum width=3cm, minimum height=1cm,text centered, draw=black, fill=blue!10]
\tikzstyle{startstop} = [rectangle, rounded corners, minimum width=3cm, minimum height=1cm,text centered, draw=black, fill=blue!10]

\tikzstyle{io} = [trapezium, trapezium left angle=70, trapezium right angle=110, minimum width=3cm, minimum height=1cm, text centered, draw=black, fill=blue!30]

\tikzstyle{process} = [rectangle, minimum width=3cm, minimum height=1cm, text centered, draw=black, fill=orange!30]
\tikzstyle{decision} = [diamond, minimum width=3cm, minimum height=1cm, text centered, draw=black, fill=green!30]

\tikzstyle{arrow} = [thick,->,>=stealth]

\tikzset{->-/.style={decoration={
  markings,
  mark=at position #1 with {\arrow[scale=2.4]{>}}},postaction={decorate}}}

\makeatletter \@addtoreset{equation}{section} \makeatother

\begin{document}

\date{\today}

\title{6D SCFTs and the Classification of Homomorphisms $\Gamma_{ADE} \rightarrow E_8$}

\institution{CEDARVILLE}{\centerline{${}^{1}$Department of Science and Mathematics, Cedarville University, Cedarville, OH 45314, USA}}

\institution{IAS}{\centerline{${}^{2}$School of Natural Sciences, Institute for Advanced Study, Princeton, NJ 08540, USA}}

\authors{Darrin D. Frey\worksat{\CEDARVILLE}\footnote{e-mail: {\tt freyd@cedarville.edu}}
and Tom Rudelius\worksat{\IAS}\footnote{e-mail: {\tt rudelius@ias.edu}}}

\abstract{ We elucidate the correspondence between a particular class of superconformal field theories in six dimensions and homomorphisms from discrete subgroups of $SU(2)$ into $E_8$, as predicted from string dualities.  We show how this match works for homomorphisms from the binary icosahedral group $SL(2,5)$ into $E_8$, correcting previous errors in both the mathematics and physics literature.  We use this correspondence to list the homomorphisms from binary dihedral groups, the binary tetrahedral group, and the binary octahedral group into $E_8$--a novel mathematical result.  The partial ordering specified by renormalization group flows suggests an ordering on these homomorphisms similar to the known ordering of nilpotent orbits of a simple Lie algebra dictated by the Hasse diagram.}

\maketitle

\tableofcontents

\enlargethispage{\baselineskip}

\setcounter{tocdepth}{2}

\newpage

\section{Introduction \label{sec:INTRO}}

In the last century, mathematics has led to incredible progress in theoretical physics.  Differential geometry, linear algebra, and group theory have proven instrumental in modern studies of gravity, quantum mechanics, and particle physics, respectively.  In the last 30 years, however, the pendulum has swung the other way, and theoretical physics has paved the way for new developments in algebraic topology and geometry.

In this work, we will elaborate on one example of this pheomenon: the connection between homomorphisms from discrete subgroups of $SU(2)$ into $E_8$ and a certain class of elliptically-fibered Calabi-Yau three-folds, which are identified in physics terms with six-dimensional superconformal field theories (6D SCFTs).

6D SCFTs are among the most mysterious quantum field theories.  Although they feature tensionless strings as one of their key ingredients, they nonetheless obey the rules of local quantum field theory.  Thus far, the best evidence for the existence of these theories has come from string theory \cite{Witten:1995ex,
Witten:1995zh, Strominger:1995ac, WittenSmall,
Ganor:1996mu,MorrisonVafaII,Seiberg:1996vs, Seiberg:1996qx, Bershadsky:1996nu,
Brunner:1997gf, Blum:1997fw, Aspinwall:1997ye, Intriligator:1997dh,
Hanany:1997gh}, with further evidence provided by the conformal bootstrap \cite{Beem:2015aoa,Chang:2017xmr}.  Nonetheless, the absence of a Lagrangian description makes these theories particularly resistant to investigation.

A seminal work in the study of 6D SCFTs was \cite{Heckman:2013pva}, which showed how these theories could be systematically classified in terms of noncompact, elliptically-fibered Calabi-Yau three-folds using F-theory \cite{Vafa:1996xn}.  Shortly thereafter, \cite{DelZotto:2014hpa} used this F-theory description to study 6D SCFTs arising as worldvolume theories of M5-branes probing a $\mathbb{C}^2/\Gamma$ orbifold singularity and an ``end-of-the-world" $E_8$ wall.  Such theories are uniquely specified by three pieces of data: the number of M5-branes $N$, the choice of orbifold group $\Gamma\subset SU(2)$, and a boundary condition labeled by a homomorphism $\Gamma \rightarrow E_8$.  Thus, for fixed $N$ and $\Gamma$, the resulting class of 6D SCFTs are in 1-1 correspondence with the homomorphisms of interest.

Discrete subgroups $\Gamma \subset SU(2)$ are in 1-1 correspondence with simply-laced Lie groups according to the famed ``McKay correspondence" of \cite{MR604577}.  In particular, one has:
\begin{align}
\Gamma_{A_{k}} &\cong \mathbb{Z}_{k+1}:\mbox{the cyclic group of order $k+1$} \nonumber \\
\Gamma_{D_{k}} &\cong \mbox{Dic}_{k-2}:\mbox{the dicyclic group of order $4k-8$} \nonumber \\
\Gamma_{E_6}   &\cong SL(2,3):\mbox{the binary tetrahedral group}  \nonumber \\
\Gamma_{E_7}   &\cong \mbox{the binary octahedral group}  \nonumber \\
\Gamma_{E_8}   &\cong SL(2,5) :\mbox{the binary icosahedral group}
\end{align}
Homomorphisms $\mathbb{Z}_k \rightarrow E_8$ were classified by Kac in \cite{MR739850}.  Homomorphisms  $\Gamma_{E_8} \cong SL(2,5) \rightarrow E_8$ were classified in \cite{FREY}, as were homomorphisms Dih$_3 \rightarrow E_8$, Dih$_5 \rightarrow E_8$.  In our notation, Dic$_k$ is the double cover $\langle A,B\mid A^{2k}=1, A^k=B^2, B A B^{-1}=A^{-1}\rangle$ of Dih$_k$, which is the dihedral group of order $2k$.  Since Dic$_k$ admits a projection homomorphism to Dih$_k$, a homomorphism Dih$_k \rightarrow E_8$ extends to a homomorphism Dic$_k \rightarrow E_8$ under composition with projection, but not all homomorphisms Dic$_k \rightarrow E_8$ factorize through Dih$_k$ in this way.  Thus, homomorphisms $\Gamma \rightarrow E_8$ have been fully classified for $\Gamma =\mathbb{Z}_k$, $\Gamma = SL(2,5)$ and partially classified for $\Gamma = $Dic$_3$, $\Gamma = $Dic$_5$.

For these particular choices of $\Gamma$ in which the homomorphisms have been classified, \cite{Heckman:2015bfa} checked the conjectured correspondence between 6D SCFTs and homomorphisms by explicitly carrying out the F-theory classification of 6D SCFTs conceptualized in \cite{Heckman:2013pva} and writing down the theories in question.  For $\Gamma = \mathbb{Z}_k$, Dic$_3$, and Dic$_5$, the conjectured correspondence was confirmed spectacularly.  For the more complicated case of $\Gamma = SL(2,5)$, however, there were some slight discrepancies between the list of 6D SCFTs and the list of homomorphisms derived in \cite{FREY}.  Such discrepancies were due in large part to difficulties in 6D SCFTs regarding global symmetries and ``unpaired tensors" (tensor multiplets without a gauge group).  Fortunately, these difficulties have since been largely addressed by the analyses of \cite{Heckman:2015ola, Ohmori:2015pia, global-symmetries, Heckman:2016ssk, Morrison:2016djb, Merkx:2017jey}.  With our improved understanding of 6D SCFTs, we are now in a position to resolve almost all of the discrepancies for the case of $\Gamma = SL(2,5)$, finding a mismatch of only two homomorphisms out of 137 total between the mathematics computation and the physics computation. We extend our conjectured classification of homomorphisms to the dicyclic groups, $SL(2,3) \cong \Gamma_{E_6}$, and $\Gamma_{E_7}$.\footnote{Possibly up to outer automorphism. See subsection \ref{ssec:CAVEAT} for further discussion on this issue.}  We will see that homomorphisms Dic$_{k-2} \cong \Gamma_{D_k} \rightarrow E_8$ can be written in a very simple manner in terms of 6D SCFTs: any such homomorphism is labeled by a nilpotent orbit of $D_k$ together with an appropriate choice of simple Lie algebra.

The rest of this paper is organized as follows: in section \ref{sec:CLASSIFICATION}, we review the classification and global symmetries of 6D SCFTs.  In section \ref{sec:HOMS}, we elaborate on the correspondence between 6D SCFTs and homomorphisms $\Gamma\rightarrow E_8$, reviewing the previously-understood case of $\Gamma =\mathbb{Z}_k$, revising the partially-understood case of $\Gamma=SL(2,5)$, and covering the remaining, novel cases.  In section \ref{sec:RGFLOWS}, we discuss the RG flows between the 6D SCFTs representing these homomorphisms.  In section \ref{sec:CONC}, we end with some conclusions and speculations on future research.  In a set of appendices, we elaborate on the mathematical classification of these homomorphisms, and we display the full list of 6D SCFTs for the specific cases of $\Gamma_{D_4}$, $\Gamma_{D_5}$, $\Gamma_{E_6}$, $\Gamma_{E_7}$, and $\Gamma_{E_8}$.

\section{Classification of 6D SCFTs}\label{sec:CLASSIFICATION}

The classification of 6D SCFTs using F-theory was initiated in \cite{Heckman:2013pva}, carried out in \cite{Heckman:2015bfa}, and reviewed more recently in \cite{Heckman:2018jxk}.  Here, we summarize the aspects of this classification relevant for our purposes.

6D SCFTs are constructed by compactifying 12-dimensional F-theory on a non-compact, elliptically-fibered Calabi-Yau three-fold.  This three-fold consists of a complex 2-dimensional base $B_2$ together with an elliptic fibration.  Given the set of irreducible effective curves $\Sigma_i$ in $B_2$, we can define an intersection matrix $\Omega_{ij} = \Sigma_i \cap \Sigma_j$.  In order to reach the superconformal fixed point, every curve $\Sigma_i$ must be simultaneously contractible, which translates to the statement that $\Omega$ must be negative definite, and every curve $\Sigma_i$ must have genus 0 \cite{MR0146182}.  

The diagonal entries of $\Omega$ represent the self-intersection numbers of the curves $\Sigma_i$.  It turns out that in any 6D SCFT, these self-intersection numbers must satisfy $-1 \geq \Sigma_i \cap \Sigma_i \geq -12$.  Furthermore, the intersection number $\Sigma_i \cap \Sigma_j = \Sigma_j \cap \Sigma_i$ between any two distinct curves can only be 0 or 1.  This allows us to depict $\Omega$ in a simple graphical form: a curve of self-intersection $-n$ is represented by the integer $n$, with neighboring integers indicating curves that intersect at a point.  For instance,
\begin{align}
\Omega& = \left[
\begin{array}
[c]{ccc}%
-2 & 1& 0 \\
1 & -3 & 1 \\
0 & 1& -1 \\
\end{array}
\right] ~~~ \Rightarrow ~~~ 2 \,\, 3 \,\, 1  \label{eq:firstquiver} \\
\Omega& = \left[
\begin{array}
[c]{cccc}%
-1 & 1& 0 & 0 \\
1 & -4 & 1& 1 \\
0 & 1& -1& 0 \\
0 & 1&0&  -1\\
\end{array}
\right] ~~~ \Rightarrow ~~~ 1 \,\, \ov{\text{\normalsize 1}}4 \,\, 1
\end{align}
At generic points in $B_2$, the fiber will be smooth, but over the irreducible curves $\Sigma_i$, it can develop singularities.  The possible types of singular fibers were classified by Kodaira \cite{Kodaira}.  In field theory language, a singular fiber over a curve in $B_2$ indicates a non-Abelian gauge algebra associated with the curve, and the type of singularity dictates the type of simple Lie algebra.  We indicate the gauge algebra $\mf g$ associated with a given curve pictorially by writing the gauge algebra just above the self-intersection number of the curve.  For instance, we might add gauge algebras to the base in (\ref{eq:firstquiver}) as follows:
\begin{equation}
\ov{\mf {su}_2}2 \,\, \ov{\mf {so}_7}3 \,\, \ov{\mf {sp}_1}1
\end{equation}

Every curve has a ``minimal fiber type" that depends on its self-intersection number.  A curve of self-intersection $-1$ or $-2$ may have a smooth fiber, corresponding to the absence of any gauge algebra.  On the other hand, a curve of self-intersection $-3$ or below necessarily has a singular fiber, yielding a non-Abelian gauge algebra.  The minimal gauge algebra for each self-intersection number is as follows:
\begin{equation}
1 \,\, ~~ 2 \,\, ~~ \ov{\mf {su}_3}3\,\, ~~ \ov{\mf {so}_8}4\,\, ~~\ov{\mf {f}_4}5\,\, ~~\ov{\mf {e}_6}6\,\,~~ \ov{\mf {e}_7}7\,\, ~~\ov{\mf {e}_7}8\,\, ~~\ov{\mf {e}_8}{9}\,\, ~~\ov{\mf {e}_8}{10}\,\, ~~\ov{\mf {e}_8}{11}\,\, ~~\ov{\mf {e}_8}{12}.
\label{eq:firstNHC}
\end{equation}
Furthermore, curves of self-intersection $-3$ can intersect curves of self-intersection $-2$ in three distinct patterns.  When this happens, the minimal gauge algebra for each is enhanced to the following:
\begin{equation}
 \ov{\mf {su}_2 }2 \,\,  \ov{\mf {g}_2 }3 \,\, ~~~~~~2 \,\, \ov{\mf {su}_2 }2 \,\,  \ov{\mf {g}_2 }3 \,\, ~~~~~~ \ov{\mf {su}^{(L)}_2 }2 \,\, \ov{\mf {so}_7 }3 \,\,  \ov{\mf {su}^{(R)}_2}2.
\label{eq:secondNHC}
\end{equation}
Whenever gauge algebras appear on intersecting curves, as they do here, there must be hypermultiplets charged under some mixed representation of the two gauge algebras.  In the first two cases above, there is a half-hypermultiplet charged under the bifundamental $(\bf{7}, \bf{2})$ of $\mf{g}_2 \oplus \mf{su}_2$.  In the second case, there is one half-hypermultiplet in the mixed representation $(\bf{8}, \bf{2})$ of $\mf{so}_7 \oplus \mf{su}_2^{(L)}$ and another in the mixed representation $\mf{so}_7 \oplus \mf{su}_{2}^{(R)}$.

The list of curves in (\ref{eq:firstNHC}) of self-intersection $-2$ or below together with the configurations of curves in (\ref{eq:secondNHC}) form the full list of so-called ``non-Higgsable clusters" (NHCs) that arise in 6D SCFTs \cite{Morrison:2012np}.  The base of any 6D SCFT consists of a set of NHCs linked together by curves of self-intersection $-1$.  For instance, the base
$$
 3 \,\, 1 \,\, 5 \,\, 1 \,\, 3 \,\, 2 \,\, 2 \,\, 1 \,\, 8
$$ 
consists of four NHCs ($3$, $5$, $322$, $8$) linked together by three curves of self-intersection $-1$.  The gauge algebras allowed on this base are minimally given by (\ref{eq:firstNHC}) and (\ref{eq:secondNHC}):
\begin{equation}
 \ov{\mf {su}_3}3 \,\, 1 \,\,  \ov{\mf {f}_4}5 \,\, 1 \,\,  \ov{\mf {g}_2}3 \,\,  \ov{\mf {su}_2}2 \,\, 2 \,\, 1 \,\,  \ov{\mf {e}_7}8
\label{eq:firsttheory}
\end{equation}
However, these gauge algebras may be ``enhanced" to larger non-Abelian Lie algebras consistent with certain conditions.  First of all, the anomaly cancellation conditions discussed in section 6 of \cite{Heckman:2015bfa} (see also \cite{Green:1984bx,
Nishino:1985xp, Sagnotti:1992qw, Schwarz:1996, Sadov:1996zm, Kumar:2010ru,Ohmori:2014kda,Ohmori:2014pca}) must be obeyed.  These anomaly cancellation conditions are very constraining, and in all but one instance they uniquely fix the spectrum of massless hypermultiplets charged under a given gauge algebra.  For instance, the gauge algebra on a $-3$ curve can be enhanced from $\mf{su}_3 $ to $ \mf{g}_2$, but there is necessarily a single fundamental charged under this $\mf g_2$.  This $\mf g_2$ can be further enhanced to $\mf {so}_7$ with two spinors.  We indicate this charged matter with a subscript:
\begin{equation}
{ \ov{\mf {su}_3}3} ~~~ \Rightarrow~~~ \underset{[N_f=1]}{ \ov{\mf {g}_2}3} ~~~ \Rightarrow~~~ \underset{[N_s=2]}{ \ov{\mf {so}_7}3} 
\end{equation}
Moving from left to right here corresponds in F-theory to enhancing the singularity, while moving from right to left corresponds in field theory to the process of Higgsing.

There are also conditions on ``unpaired" $-1$ and $-2$ curves, which do not have an associated gauge algebra.  If an unpaired $-1$ curve meets curves carrying gauge algebra $\mf g_L $ and $\mf g_R$, we must have
\begin{equation}
\mf g_L \oplus \mf g_R \subset \mf e_8.
\end{equation}
This is often referred to as the ``$E_8$ gauging condition."  We see that the $\mf {su}_3$ gauge algebra in the theory of (\ref{eq:firsttheory}) can be enhanced to $\mf g_2$ consistently with the $E_8$ gauging condition,
\begin{equation}
\underset{[N_f=1]}{ \ov{\mf {g}_2}3} \,\, 1 \,\,  \ov{\mf {f}_4}5 \,\, 1 \,\,  \ov{\mf {g}_2}3 \,\,  \ov{\mf {su}_2}2 \,\, 2 \,\, 1 \,\,  \ov{\mf {e}_7}8
\label{eq:secondtheory}
\end{equation}
However, a further enhancement to $\mf{so}_7$ is impossible, because $\mf{so}_7 \oplus \mf f_4 \not\subset \mf{e}_8$:
\begin{equation}
\text{NOT POSSIBLE: } ~~~\underset{[N_s=2]}{ \ov{\mf {so}_7}3} \,\, 1 \,\,  \ov{\mf {f}_4}5 \,\, 1 \,\,  \ov{\mf {g}_2}3 \,\,  \ov{\mf {su}_2}2 \,\, 2 \,\, 1 \,\,  \ov{\mf {e}_7}8
\end{equation}

Similarly, an unpaired $-2$ curve that meets curves carrying gauge algebra $\mf g_L $ and $\mf g_R$ must satisfy an $SU(2)$ gauging condition,
\begin{equation}
\mf g_L \oplus \mf g_R \subset \mf {su}_2.
\end{equation}
This $SU(2)$ gauging condition is necessary but not quite sufficient \cite{Morrison:2016djb}.  Further constraints on unpaired $-2$ curves have been understood from F-theory, but their field theory interpretation is not yet clear.

Thus, every known 6D SCFT is uniquely specified by a diagram like the ones in (\ref{eq:firsttheory}) and (\ref{eq:secondtheory}), which encode the intersection pairing $\Omega$ for curves in the F-theory base $B_2$, the spectrum of gauge algebras associated with these curves, and the matter charged under these gauge algebras.  We will sometimes refer to these diagrams as ``quiver diagrams."  Further information on the classification of 6D SCFTs in terms of these quiver diagrams can be found in \cite{Heckman:2015bfa}.

\subsection{Global symmetries of 6D SCFTs}\label{ssec:GLOBAL}

Global symmetries will play a crucial role in the match between 6D SCFTs and homomorphisms.  The global symmetry of a given 6D SCFT is typically easy to compute from its quiver, but there are several cases in which it is quite nontrivial.  Fortunately, recent progress in \cite{Heckman:2015ola, Ohmori:2015pia, global-symmetries, Heckman:2016ssk, Morrison:2016djb, Apruzzi:2020eqi} has resolved almost all of these subtle cases.  In this subsection, we summarize the rules for computing the global symmetry of a given 6D SCFT.  This is in fact the first time such a summary has appeared in the literature.

For a given 6D SCFT, global symmetries can arise in one of three ways:
\begin{enumerate}[(i)]
\item As the unbroken subgroup of $E_8$ associated with an ``unpaired" $-1$ curve.
\item As the unbroken $SU(2)$ associated with one or more consecutive ``unpaired" $-2$ curves.
\item As the flavor symmetry rotating hypermultiplets charged under a given gauge algebra.
\end{enumerate}

We begin by examining case (i).  A given 6D SCFT may have unpaired $-1$ curves intersecting either one or two curves, both of which may carry a gauge algebra:
$$
\overset{\mf{g}_L}L \,\, 1 \,\, \overset{\mf{g}_R}R
$$
As discussed previously, a necessary condition on these gauge algebras is $\mf g_L \oplus \mf g_R \subset \mf e_8$.  This $\mf e_8$ can be interpreted as the global symmetry of the $-1$ curve, which is gauged by the combination $\mf g_L \oplus \mf g_R$.  The leftover global symmetry on the $-1$ curve is then the maximal subalgebra $\mf h \subset \mf e_8$ left ungauged.  More precisely, the global symmetry on the $-1$ curve is given by the maximal $\mf h$ such that $\mf h \oplus \mf g_L \oplus  \mf g_R \subset \mf e_8$.  This holds true even if the unpaired $-1$ curve meets an unpaired $-2$ curve.  Thus, the global symmetry associated with each $-1$ curve in the below quiver is $G_2$, since $\mathfrak{g}_2 \oplus \mf f_4 $ is a maximal subalgebra of $\mf e_8$:
$$
[SU(2)] \,\, 2 \,\, \underset{[G_2]}1 \,\, \overset{\mf f_4}5 \,\, \underset{[G_2]}1 .
$$
Here and throughout this paper, we use upper-case letters to indicate flavor symmetries in order to distinguish them from gauge symmetries.  Unless otherwise stated, however, we are referring to the global symmetry \emph{algebra} rather than the \emph{group}.  Note also that we depict the flavor symmetry associated with a given curve with a subscript if the curve is in the interior of the quiver, but we place it on the side if the curve is on the end of the quiver to make it easier to read.

The above rule suffices to compute the global symmetry of a $-1$ curve in just about every case.  An exception to this is the case where the $-1$ curve meets an $\mf {su}_2$ gauge algebra and an $\mf {so}_8$ gauge algebra:
$$
 ... \,\,{\overset{\mathfrak{su}_2}2} \,\,  1 \,\,  \overset{\mathfrak{so}_8}4 \,\,  ...
$$
Here, the $\mathfrak{so}_8$ gauge algebra leaves an $\mathfrak{so}_8$ unbroken inside of $\mathfrak{e}_8$, but now there is a subtlety: $\mathfrak{so}_8$ has both $\mathfrak{su}_2 \oplus \mathfrak{su}_2 \oplus \mathfrak{su}_2 \oplus \mathfrak{su}_2$ and $\mathfrak{su}_2 \oplus \mathfrak{sp}_2$ as maximal subalgebras.  This means that the flavor symmetry leftover after gauging the $\mathfrak{su}_2$ could be either $SU(2) \times SU(2) \times SU(2)$ or $Sp(2)$.  This ambiguity shows up in one case in the list of theories in appendix \ref{sec:LISTS}.

We now consider case (ii), the global symmetry associated with one or more unpaired $-2$ curves.  In any 6D SCFT, a chain of more than one consecutive unpaired $-2$ curves always has an associated $SU(2)$ flavor symmetry:
$$
[SU(2)] \,\, 2 \,\, 2 \,\, ... \,\, 2 \,\, ...
$$
A single unpaired $-2$ curve can meet at most one curve carrying $\mf {su}_2$ gauge algebra.  If it does meet such a curve, it has no global symmetry.  If it does not, it has $SU(2)$ global symmetry.  For instance, the unpaired $-2$ curve at the left of the following quiver has $SU(2)$ global symmetry, while the one at the right does not:
$$
[SU(2)] \,\, 2 \,\, \underset{[G_2]}1 \,\, \overset{\mf f_4}5 \,\, 1  \,\, \overset{\mf g_2}3 \,\, \overset{\mf {su}_2}2 \,\,  2.
$$

Finally, we consider case (iii), in which the global symmetry comes from hypermultiplets charged under a gauge group.  In a typical quiver, many of the hypermultiplets needed for gauge anomaly cancellation will transform in mixed representations under two different gauge algebras.  These mixed representations never contribute to the global symmetry.  The leftover hypermultiplets, which are charged under a single gauge algebra, will then dictate the global symmetry associated with this curve.  As an example, consider the quiver
$$
[SO(10)]\,\, \overset{\mf{sp}_1}1 \,\, \overset{\mf{so}_{10}}4 \,\, \overset{\mf{sp}_1}1 \,\, [SO(10)]
$$ 
Here, gauge anomaly cancellation implies there are ten hypermultiplets charged under each $\mf {sp}_1$ gauge algebra.  However, mixed anomaly cancellation implies that there is a half-bifundamental $\frac{1}{2}(\textbf{2},\textbf{10})$ charged under the first and second gauge algebra, and another half-bifundamental between the second and third gauge algebras.  After accounting for these bifundamentals, there are five ``leftover" fundamentals charged under the first gauge algebra, none under the second, and five under the third.  These leftover doublets transform as half-bifundamentals under respective $SO(10)$ global symmetries on the left and right of the quiver.

For a given gauge algebra and set number of hypermultiplets, the flavor symmetry (in almost all cases) may be read off from Table 5.1 of \cite{global-symmetries}.  We reproduce the relevant aspects here in Table \ref{tab:global}.

\begin{table}
\begin{center}
\begin{tabular}{ ccc } \hline
 \textbf{Gauge Algebra} &\textbf{ Relevant Representations} & \textbf{Global Symmetry}  \\  \hline
$\mf{su}_2$ & Fundamental $\textbf{2}$ & $SO(2 N_F )^*$ \\
$\mf{su}_3$ & Fundamental $\textbf{3}$ & $U( N_F)$ \\
$\mf{su}_4$ & Fundamental $\textbf{4}$, Antisymmetric $\textbf{6}$ & $U( N_F) \times Sp(N_A)$ \\
$\mf{su}(N), N \geq 5$ & Fundamental $\textbf{N}$, Antisymmetric $\frac{\textbf{N}(\textbf{N}-1)}{2}$ & $U(N_{F}) \times U(N_A)$ \\\hline
$\mf{so}_7$ & Vector $\textbf{7}$, Spinor $\textbf{8}$ & $Sp(N_{V}) \times Sp(N_S)$ \\
$\mf{so}_8$ & Vector $\textbf{8}_v$, Spinor $\textbf{8}_s$, Conjugate $\textbf{8}_c$ & $Sp(N_{V}) \times Sp(N_S) \times Sp(N_C)$ \\ 
$\mf{so}_9$ & Vector $\textbf{9}$, Spinor $\textbf{16}$ & $Sp(N_{V}) \times Sp(N_S) $ \\ 
$\mf{so}_{10}$ & Vector $\textbf{10}$, Spinor $\textbf{16}$ & $Sp(N_{V}) \times U(N_S) $ \\
$\mf{so}_{11}$ & Vector $\textbf{11}$, Spinor $\textbf{32}$ & $Sp(N_{V}) \times SO(N_S) $ \\
$\mf{so}_{12}$ & Vector $\textbf{12}$, Spinor $\textbf{32}$ & $Sp(N_{V}) \times SO(N_S) $ \\
$\mf{so}(M), M \geq 13$ & Vector $\textbf{M}$ & $Sp(N_{V}) $ \\ \hline
$\mf{sp}(P), P \geq 1$ & Fundamental $\textbf{2P}$ & $SO(2 N_F) $ \\ \hline
$\mf{g}_2$ & Fundamental $\textbf{7}$ & $Sp( N_F) $ \\ \hline
$\mf{f}_4$ & Fundamental $\textbf{26}$ & $Sp( N_F) $ \\ \hline
$\mf{e}_6$ & Fundamental $\textbf{27}$ & $U(N_F) $ \\ 
$\mf{e}_7$ & Fundamental $\textbf{56}$ & $SO(N_F) $ \\ 
$\mf{e}_8$ & N/A & N/A  \\ \hline
\end{tabular}
\caption{Flavor Symmetries for hypermultiplets in 6D SCFTs. Note that $\mf{su}_2$ on a $-2$ curve yields only an $SO(7)$ flavor symmetry, rather than the na\"ively expected $SO(8)$.}
\label{tab:global}
\end{center}
\end{table}

Most of the study of 6D SCFTs has ignored abelian flavor symmetries. However, thanks to recent progress in the study of abelian symmetries \cite{Lee:2018ihr, Apruzzi:2020eqi}, we can now classify the $U(1)$s associated with a given quiver. In general, these $U(1)$s arise either as the abelian part of some $U(N)$ or as an $SO(2) \simeq U(1)$, as determined by Table \ref{tab:global}. However, there is one subtlety: in the case of a chain of $\mf{su}(N)$ gauge algebras, an overall $U(1)$ center-of-mass is projected out. For instance, consider the quiver:
$$
[U(3)]\,\, \overset{\mf {su}_3}2 \,\, \overset{\mf {su}_3}2 \,\, \overset{\mf {su}_3}2 \,\, [U(3)]
$$ 
Here, the overall global symmetry is not simply $U(3) \times U(3)$, but rather $S[U(3) \times U(3)] \simeq SU(3) \times SU(3) \times U(1)$. This projection does not occur in other cases. For instance, the following theory has a $U(2) \simeq SU(2) \times U(1)$ global symmetry:
\begin{equation}
\underset{[U(2)]}{\overset{\mf e_6}4}
\end{equation}

The only remaining subtleties with flavor symmetries involve unpaired $-2$ curves or the gauge group $\mf {su}_2$.  First, as conjectured in \cite{Heckman:2015bfa} and proven in \cite{Ohmori:2015pia}, for a $-2$ curve carrying $\mf {su}_2$, the 8 half-hypermultiplets transform as a spinor under an $SO(7)$ flavor symmetry, rather than the na\"ively expected $SO(8)$:
$$
[SO(7)] \,\, \overset{\mf {su}_2}2 \,\, 1 \,\, \overset{\mf e_7}8 \,\, ...
$$
Furthermore, an empty $-2$ curve sitting next to a curve carrying $\mf {su}_2$ gauge algebra sucks away a single half-hypermultiplet.  So, for instance, we have an $SO(19)$ flavor symmetry in the theory
$$
[SU(2)] \,\, 2 \,\, \overset{\mathfrak{sp}_1}1 \,\, [SO(19)]
$$
rather than the $SO(20)$ in the theory without the unpaired $-2$ curve,
$$
 \overset{\mathfrak{sp}_1}1 \,\, [SO(20)]
$$
By the same reasoning, a $-2$ curve with $\mf{su}_2$ gauge algebra has its global symmetry reduced from $SO(7)$ to $G_2$ by an unpaired $-2$ curve, with 7 half-hypermultiplets transforming in the fundamental of $G_2$ \cite{Ohmori:2015pia}:
$$
2 \,\, \overset{\mathfrak{su}_2}2 \,\, [G_2]
$$

There are a few remaining situations in which the flavor symmetry is difficult to read off from the quiver, but may be determined using the connection between 6D SCFTs and nilpotent orbits discovered in \cite{Heckman:2016ssk}.  A $-2$ curve carrying $\mf{su}_2$ gauge algebra that meets \emph{two} unpaired $-2$ curves carries an $SU(3)$ gauge algebra:
$$
2 \,\, \underset{[SU(3)]}{\ov{\mf{su}_2}2} \,\, 2
$$
One might have thought that the six half-hypermultiplets of $\mf {su}_2$ would transform under $\mf {so}_6$, but this turns out not to be the case.  
Next, a chain of multiple $-2$ curves carrying $\mf{su}_2$ gauge algebra has $SU(2) \times SU(2) \times SU(2)$ global symmetry:
$$
[SU(2) \times SU(2)] \,\, \overset{\mf {su}_2}2 \,\, \underset{[SU(2)]}{\overset{\mf {su}_2}2 }\,\, 1 \,\, \overset{\mf e_7}8 \,\, ...
$$
One might have expected an $SO(4) \cong SU(2) \times SU(2)$ on both the left- and right-hand sides of the chain of $-2$ curves, but again this intuition is wrong.  If one side of the chain has an empty $-2$ curve, the flavor symmetry is reduced to $SU(2) \times SU(2)$:
$$
[SU(2) \times SU(2)] \,\, {\overset{\mf {su}_2}2 }\,\, \overset{\mf {su}_2}2 \,\, {\overset{\mf {su}_2}2 }\,\, 2 \,\, 1 \,\, \overset{\mf e_8}{12} \,\, ...
$$
If both sides of a chain of two $-2$ curves have an empty $-2$ curve, the flavor symmetry is reduced to $SU(2)$,
$$
2 \,\, {\overset{\mf {su}_2}2 }\,\, \underset{[SU(2)]}{\overset{\mf {su}_2}2} \,\, 2 \,\, 1 \,\, \overset{\mf e_8}{12} \,\, ...
$$
For three $\mf{su}_2$s, there is an additional $U(1)$:
$$
2 \,\, {\overset{\mf {su}_2}2 }\,\, \underset{[SU(2) \times U(1)]}{\overset{\mf {su}_2}2} \,\, {\overset{\mf {su}_2}2 }\,\, 2 \,\, 1 \,\, \overset{\mf e_8}{12} \,\, ...
$$

These rules appear to be sufficient to determine the global symmetry of any 6D SCFT in the classification of \cite{Heckman:2015bfa}.  Many more examples can be found in the appendix of \cite{Heckman:2016ssk} or appendix \ref{sec:LISTS} of the current paper.

\section{6D SCFTs and Hom($\Gamma,E_8 $)}\label{sec:HOMS}

In the present work, we are concerned with a particular class of 6D SCFTs that admit a construction in heterotic M-theory.  In particular, we take a stack of $N$ M5-branes to probe an $E_8$ wall as well as an orbifold singularity $\mathbb{C}^2/\Gamma$, as shown in Table \ref{tab:braneconfig}.

\begin{table}
\begin{center}
\begin{tabular}{ c|ccccccccccc } 
 & 0 & 1 & 2 & 3 & 4 & 5 & 6 & 7 & 8 & 9 & 10  \\  \hline
$N$ M5s & $\times$ & $\times$ &$\times$ &$\times$ &$\times$ &$\times$ & & & & &  \\ 
$ E_8$ Wall & $\times$ & $\times$ &$\times$ &$\times$ &$\times$ &$\times$ &$ \times$ &$\times$ &$\times$ & &$\times$  \\
$\mathbb{C}^{2}/\Gamma$ & &&&&&& $\times$& $\times$& $\times$& $\times$& 
\end{tabular}
\caption{The brane configuration.}
\label{tab:braneconfig}
\end{center}
\end{table}

The worldvolume theory of this stack of M5-branes is a 6D SCFT with a partial tensor branch with quiver description of the form
\begin{equation}
[E_8] \,\, \underset{N}{\underbrace{\overset{\mathfrak{g}}1 \,\,\overset{\mathfrak{g}}2 \,\, ...\,\, \overset{\mathfrak{g}}2}}\,\, [G]
\label{eq:heterotic}
\end{equation}
Here, $\mathfrak{g}$ and $G$ represent the simply-laced Lie algebra and Lie group associated to the orbifold singularity.
%To get to the full tensor branch of these theories, one must introduce conformal matter, which in F-theory language corresponds to resolving the the base at points of intersection of neighboring curves.  For instance, for the case of a $\Gamma_{D_{4}} \cong$Dic$_2$ singularity, one must resolve the base once at each interior point and five times on the left-hand side of the quiver to get the full tensor branch:
%$$
%[E_8] \,\, {{\overset{\mathfrak{so}_8}1 \,\,\overset{\mathfrak{so}_8}2 \,\, ...\,\, \overset{\mathfrak{so}_8}2}}\,\, [SO(8)] ~~\rightarrow ~~[E_8] \,\, 1 \,\, 2  \,\, {\overset{\mathfrak{su_2}}2} \,\,\overset{\mathfrak{g_2}}3 \,\, 1 \,\,  \overset{\mathfrak{so_8}}4 \,\,    1 \,\, {\overset{\mathfrak{so_8}}4} \,\,  ...  \,\, {\overset{\mathfrak{so_8}}4}\,\, 1 \,\, [SO(8)]
%$$

In addition to the number of M5-branes $N$ and the type of orbifold singularity $\Gamma$, there is a choice of boundary condition.  Namely, we may choose to turn on a flat $E_8$ connection at the ``infinity" of $\mathbb{C}^2/\Gamma \cong S^3/\Gamma$ \cite{DelZotto:2014hpa}.  Flat connections on this space are in 1-1 correspondence with homomorphisms $\pi_1(S^3/\Gamma) \rightarrow E_8$, which are in turn in a 1-1 correspondence with homomorphisms $\Gamma \rightarrow E_8$.  Thus, for each such homomorphism, we expect to find a corresponding 6D SCFT.  These 6D SCFTs arise as deformations of the theory in (\ref{eq:heterotic}).  In field theory terms, this deformation involves moving onto the ``Higgs branch" of the theory by giving vevs to scalars in hypermultiplets.  In F-theory terms, it involves a complex structure deformation.  This means in particular that the resulting configuration of curves in the base will always blow down to the same $1,2,2,...,2$ configuration as in the original undeformed theory of (\ref{eq:heterotic}).

As in the case of homomorphisms $\mathfrak{su}_2 \rightarrow \mathfrak{g}$ discussed in \cite{Heckman:2016ssk}, global symmetries are the key to matching 6D SCFTs with homomorphisms $\Gamma \rightarrow E_8$.  A given homomorphism $\rho: \Gamma \rightarrow E_8$ will generically have an image $\rho(\Gamma) \subset E_8$ that commutes with some subgroup $H(\rho) \subset E_8$, and the associated 6D SCFT will have global symmetry $H(\rho)$.  This allows us to identify the corresponding 6D SCFTs by a computation of their global symmetries.

In \cite{Heckman:2015bfa}, this conjectured correspondence was verified explicitly for homomorphisms $\mathbb{Z}_k \cong \Gamma_{A_{k-1}} \rightarrow E_8$ and for the known homomorphisms Dic$_3 \rightarrow E_8$, Dic$_5 \rightarrow E_8$.  Further, many 6D SCFTs representing homomorphisms $SL(2,5) \cong \Gamma_{E_8} \rightarrow E_8$ were identified, but subtleties involving global symmetries prevented a perfect match.  In subsequent subsections, we will review the $\mathbb{Z}_k$ case, revisit the $SL(2,5)$ case, and extend the match to the remaining cases.

\subsection{Review of Hom($\mathbb{Z}_k, E_8$)}

We consider first the $A_{k-1}$-type case, in which we consider M5-branes probing an $E_8$ wall and a $\mathbb{Z}_k$ orbifold singularity.  This case was considered in section 7 of \cite{Heckman:2015bfa}, and as a first example, we review it here.  As shown in \cite{MR739850}, homomorphisms from $\mathbb{Z}_k$ into $E_8$ can be classified by deleting nodes from the affine $E_8$ Dynkin diagram according to a simple rule.\footnote{Technically, the heterotic M-theory setup involves the real form of $E_8$, whereas the classification of \cite{MR739850} deals with the complex form of the Lie group $E_8$. The match between 6D SCFTs and homomorphisms nonetheless works perfectly.}
  In particular, given the Dynkin diagram, one numbers the nodes as follows:

$$
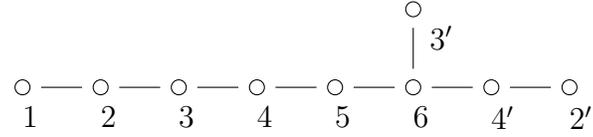
\begin{tikzpicture}
\begin{scope}[start chain]
{
\dnode{1}
\dnode{2}
\dnode{3}
\dnode{4}
\dnode{5}
\dnode{6}
\dnode{4'}
\dnode{2'}
}
\end{scope}
\begin{scope}[start chain=br going above]
\chainin (chain-6);
\dnodebr{3'}
\end{scope}
\end{tikzpicture}
$$

Now, to classify the homomorphisms from $\mathbb{Z}_k$ into $E_8$, one considers all lists of nodes such that the sum of the numbers of these nodes equals $k$, where any given node may be used multiple times.  For instance, for $k=4$, we have the following choices of nodes:
\begin{equation}
1+1+1+1,~~1+1+2,~~1+1+2',~~1+3,~~1+3',~~2+2,~~2+2',~~2'+2',~~4,~~4'.
\end{equation}
The maximal subgroup of $E_8$ that commutes with the image of the homomorphism is then given simply by the diagram remaining after deleting the corresponding nodes from the affine $E_8$ Dynkin diagram.  This subgroup is isomorphic to the flavor symmetry of the corresponding 6D SCFT.  So, for each of the above homomorphisms, we have the following correspondence between homomorphisms, Dynkin diagrams, and 6D SCFTs:
$$
1+1+1+1 ~~\leftrightarrow~~ \begin{tikzpicture}
\begin{scope}[start chain]
{
\dnode{2}
\dnode{3}
\dnode{4}
\dnode{5}
\dnode{6}
\dnode{4'}
\dnode{2'}
}
\end{scope}
\begin{scope}[start chain=br going above]
\chainin (chain-5);
\dnodebr{3'}
\end{scope}
\end{tikzpicture}~~
\leftrightarrow~~
[E_8] \,\, 1 \,\,2 \,\, \overset{\mathfrak{su}_2}2 \,\, \overset{\mathfrak{su}_3}2  \,\, {\overset{\mathfrak{su}_4}2} \,\,... [SU(4)]
$$
$$
1+1+2 ~~\leftrightarrow ~~\begin{tikzpicture}
\begin{scope}[start chain]
{
\dnode{3}
\dnode{4}
\dnode{5}
\dnode{6}
\dnode{4'}
\dnode{2'}
}
\end{scope}
\begin{scope}[start chain=br going above]
\chainin (chain-4);
\dnodebr{3'}
\end{scope}
\end{tikzpicture}
~~\leftrightarrow~~
[E_7] \,\, 1  \,\,  \underset{[N_f=1]}{\overset{\mathfrak{su}_2}2} \,\, \overset{\mathfrak{su}_3}2  \,\, {\overset{\mathfrak{su}_4}2} \,\,... [SU(4)]
$$
$$
1+1+2' ~~\leftrightarrow ~~\begin{tikzpicture}
\begin{scope}[start chain]
{
\dnode{2}
\dnode{3}
\dnode{4}
\dnode{5}
\dnode{6}
\dnode{4'}
}
\end{scope}
\begin{scope}[start chain=br going above]
\chainin (chain-5);
\dnodebr{3'}
\end{scope}
\end{tikzpicture}
~~\leftrightarrow~~
[SO(14)] \,\,  \overset{\mathfrak{sp}_1}1  \,\,  \overset{\mathfrak{su}_3}2  \,\, {\overset{\mathfrak{su}_4}2} \,\,... [SU(4)]
$$
$$
1+3 ~~\leftrightarrow ~~\begin{tikzpicture}
\begin{scope}[start chain]
{
\dnode{2}
}
\end{scope} 
\end{tikzpicture}~~~~~~~
\begin{tikzpicture}
\begin{scope}[start chain]
{
\dnode{4}
\dnode{5}
\dnode{6}
\dnode{4'}
\dnode{2'}
}
\end{scope}
\begin{scope}[start chain=br going above]
\chainin (chain-3);
\dnodebr{3'}
\end{scope}
\end{tikzpicture}
~~\leftrightarrow~~
[E_6] \,\,  1  \,\,  \underset{[  SU(2) ]}{\overset{\mathfrak{su}_3}2 } \,\, {\overset{\mathfrak{su}_4}2} \,\,... [SU(4)]
$$
$$
1+3' ~~\leftrightarrow ~~
\begin{tikzpicture}
\begin{scope}[start chain]
{
\dnode{2}
\dnode{3}
\dnode{4}
\dnode{5}
\dnode{6}
\dnode{4'}
\dnode{2'}
}
\end{scope}
\end{tikzpicture}
~~\leftrightarrow~~
[SU(8)] \,\,   {\overset{\mathfrak{su}_3}1 }  \,\,  {\overset{\mathfrak{su}_4}2 } \,\, {\overset{\mathfrak{su}_4}2} \,\,... [SU(4)]
$$
$$
2+2 ~~\leftrightarrow ~~
\begin{tikzpicture}
\begin{scope}[start chain]
{
\dnode{1}
}
\end{scope} 
\end{tikzpicture}~~~~~~~
\begin{tikzpicture}
\begin{scope}[start chain]
{
\dnode{3}
\dnode{4}
\dnode{5}
\dnode{6}
\dnode{4'}
\dnode{2'}
}
\end{scope}
\begin{scope}[start chain=br going above]
\chainin (chain-4);
\dnodebr{3'}
\end{scope}
\end{tikzpicture}
~~\leftrightarrow~~
[E_7] \,\,   1 \,\,  {\overset{\mathfrak{su}_2}2 } \,\, \underset{[SU(2)]}{\overset{\mathfrak{su}_4}2 }  \,\,... [SU(4)]
$$
$$
2+2' ~~\leftrightarrow ~~
\begin{tikzpicture}
\begin{scope}[start chain]
{
\dnode{1}
}
\end{scope} 
\end{tikzpicture}~~~~~~~
\begin{tikzpicture}
\begin{scope}[start chain]
{
\dnode{3}
\dnode{4}
\dnode{5}
\dnode{6}
\dnode{4'}
}
\end{scope}
\begin{scope}[start chain=br going above]
\chainin (chain-4);
\dnodebr{3'}
\end{scope}
\end{tikzpicture}
~~\leftrightarrow~~
[SO(12)] \,\,   \overset{\mathfrak{sp}_1}1 \,\,  \underset{[SU(2)]}{\overset{\mathfrak{su}_4}2 } \,\, {\overset{\mathfrak{su}_4}2 }  \,\,... [SU(4)]
$$
$$
2'+2' ~~\leftrightarrow ~~
\begin{tikzpicture}
\begin{scope}[start chain]
{
\dnode{1}
\dnode{2}
\dnode{3}
\dnode{4}
\dnode{5}
\dnode{6}
\dnode{4'}
}
\end{scope}
\begin{scope}[start chain=br going above]
\chainin (chain-6);
\dnodebr{3'}
\end{scope}
\end{tikzpicture}
~~\leftrightarrow~~
[SO(16)] \,\,   \overset{\mathfrak{sp}_2}1 \,\,  {\overset{\mathfrak{su}_4}2 } \,\, {\overset{\mathfrak{su}_4}2 }  \,\,... [SU(4)]
$$
$$
4 ~~\leftrightarrow ~~
\begin{tikzpicture}
\begin{scope}[start chain]
{
\dnode{1}
\dnode{2}
\dnode{3}
}
\end{scope}
\end{tikzpicture}~~~~~
\begin{tikzpicture}
\begin{scope}[start chain]
{
\dnode{5}
\dnode{6}
\dnode{4'}
\dnode{2'}
}
\end{scope}
\begin{scope}[start chain=br going above]
\chainin (chain-2);
\dnodebr{3'}
\end{scope}
\end{tikzpicture}
~~\leftrightarrow~~
[SO(10)] \,\,   1 \,\,  \underset{[SU(4)]}{\overset{\mathfrak{su}_4}2 } \,\, {\overset{\mathfrak{su}_4}2 }  \,\,... [SU(4)]
$$
$$
4' ~~\leftrightarrow ~~
\begin{tikzpicture}
\begin{scope}[start chain]
{
\dnode{1}
\dnode{2}
\dnode{3}
\dnode{4}
\dnode{5}
\dnode{6}
}
\end{scope}
\begin{scope}[start chain=br going above]
\chainin (chain-6);
\dnodebr{3'}
\end{scope}
\end{tikzpicture}~~~~~
\begin{tikzpicture}
\begin{scope}[start chain]
{
\dnode{2'}
}
\end{scope}
\end{tikzpicture}
~~\leftrightarrow~~
[SU(8) \times SU(2)] \,\,    {\overset{\mathfrak{su}_4}1 } \,\,  {\overset{\mathfrak{su}_4}2 } \,\, {\overset{\mathfrak{su}_4}2 }  \,\,... [SU(4)]
$$

This correspondence may be easily extended to arbitrary $\mathbb{Z}_k$, though the number of homomorphisms grows rapidly with $k$.

\subsubsection{Hom($\mathbb{Z}_k,E_8$) and the F-theory Swampland}

It is worth mentioning that this match between 6D SCFTs and Hom($\mathbb{Z}_k,E_8$) may be useful for resolving apparent discrepancies between a geometric F-theory analysis and a field theoretic analysis of 6D SCFTs.  In particular, a $-1$ curve carrying $\mf{su}_3$ gauge algebra features 12 fundamental hypermultiplets transforming under an $SU(12)$ flavor symmetry.  But as noted in \cite{global-symmetries}, (see also \cite{Merkx:2017jey}), no F-theory geometry has yet been constructed in which the full $SU(12)$ flavor symmetry is realized.  In fact, if one further attempts to gauge some $\mf{su}(M)$ of this $SU(12)$ symmetry with a $-2$ curve, the known F-theory constructions only account for $M \leq 9$ \cite{Johnson:2016qar}.  Does this point to a limitation in the field theory analysis, which suggests that all 12 can be gauged consistently with anomaly cancellation, or a limitation in the F-theory analysis, which suggests that they cannot?

In this case, the match with homomorphisms decides in favor of field theory.  The $\mathbb{Z}_{11} \rightarrow E_8$ homomorphism labeled by $3' + 3' + 3' + 2'$ yields an $SU(8)$ flavor symmetry with corresponding 6D SCFT
\begin{equation}
    \underset{[N_f=1]}{\overset{\mathfrak{su}_3}1 } \,\,  \underset{[SU(8)]}{\overset{\mathfrak{su}_{11}}2 } \,\, {\overset{\mathfrak{su}_{11}}2 }  \,\,... [SU(12)]
\end{equation}
Similarly, the $\mathbb{Z}_{12} \rightarrow E_8$ homomorphism labeled by $3' + 3' + 3' + 3'$ yields an $SU(9)$ flavor symmetry.  One can check that the corresponding theory must be
\begin{equation}
    {\overset{\mathfrak{su}_3}1 } \,\,  \underset{[SU(9)]}{\overset{\mathfrak{su}_{12}}2 } \,\, {\overset{\mathfrak{su}_{12}}2 }  \,\,... [SU(12)]
\end{equation}
In particular, all 12 of the fundamentals of $\mf{su}_3$ have been gauged by the neighboring $\mf{su}_{12}$.  Constructing these theories in F-theory therefore remains an open challenge.

\subsubsection{Hom($\mathbb{Z}_k,E_8$) and the 6D $\theta$ Angle}

Another important issue regarding this match with homomorphisms is the possibility of distinct theories with a single 6D SCFT quiver, labeled by a choice of discrete $\theta$ angle \cite{Mekareeya:2017jgc}.  Consider a quiver of the form
\begin{equation}
{\overset{\mathfrak{sp}(P)}1 } \,\, {\overset{\mathfrak{su}(2P+8)}2 } \,\,... 
\label{eq:thetaquiv}
\end{equation}
where the remainder of the quiver on the right-hand side is free to vary.  An unpaired $-1$ curve can be thought of as the special case of $P=0$.  As discussed in \cite{Mekareeya:2017jgc}, this quiver actually corresponds to \emph{two} distinct 6D SCFTs, which differ by the embedding of $\mf{su}(2P+8)$ into the flavor symmetry $\mf{so}(4P+16)$ of the $-1$ curve. This comes about because these theories have instanton strings for the $\mf{sp}(P)$ gauge symmetry, which transform under a spinor representation of the $\mf{so}(4P+16)$ flavor symmetry; the two theories in question are distinguished by the chirality of this spinor, as the two spinors of $\mf{so}(4P+16)$ decompose differently under $\mf{su}(2P+8) \subset \mf{so}(4P+16)$. As a result, the flavor symmetries of the theories differ: one has $SU(8)$ flavor symmetry, while the other yields $SU(8) \times SU(2)$.

This subtlety only shows up for the particular case of $\mf{su}(2P+8) \subset \mf{so}(4P+16)$ depicted in (\ref{eq:thetaquiv}). As noted in \cite{Mekareeya:2017jgc}, a chiral spinor of $\mf{so}(4P+16)$ decomposes into nonchiral spinors of some proper subalgebra $\mf{so}(2x)$, so there is no subtlety for embedding $\mf{su}(x) \subset \mf{so}(4P+16)$. And if the flavor symmetry is instead a unitary or symplectic group, there is no such distinction between decompositions of chiral and nonchiral representations, so the issue does not arise at all. As a result, this subtlety involving a discrete angle only shows up for Hom($\mathbb{Z}_k, E_8$), and not in any of the dicyclic or binary polyhedral group homormorphisms we will consider in the rest of the paper.

\subsection{Reconsideration of Hom($SL(2,5), E_8$)}

The case of a $\mathbb{Z}_k$ orbifold above was rather straightforward due to the simplicity of the 6D SCFTs involved and the elegant classification of \cite{MR739850}.  On the other hand, the $D_k$ and $E_k$ cases involve significantly more complicated theories, and the corresponding homomorphisms have not been classified in general.  The one exception to this is the case of $E_8$: homomorphisms from $\Gamma_{E_8} \cong SL(2,5)$ were classified in \cite{FREY}.  In \cite{Heckman:2015bfa}, a large class of 6D SCFTs associated to these homomorphisms were identified.  However, the primitive understanding of global symmetries in 6D SCFTs at the time prevented a complete match.  Now, using our improved understanding of global symmetries in 6D SCFTs, as reviewed in section \ref{ssec:GLOBAL}, we may revisit the correspondence between 6D SCFTs and Hom($SL(2,5), E_8$).  Our analysis reveals minor errors in both the physics and mathematics literature, which upon correction result in a near-perfect match between the relevant homomorphisms and the relevant 6D SCFTs.

\begin{table}
\begin{center}
\begin{tabular}{|c|c|c|c|} \hline
\multirow{2}{*}{Centralizer} & \multicolumn{3}{|c|}{Number of Homomorphisms} \\ \cline{2-4}
 & Math&Physics&Difference \\\hline
$\emptyset$ & 1 & 1 & 0 \\ 
$U(1)^2$ & 4 & 4 & 0 \\
$SU(2)$ & 12 & 14 & -2 \\
$SU(2)  \times U(1)$ & 7 & 7 & 0 \\
$SU(2)^2$ & 21 & 21 & 0 \\
$SU(2)^2  \times U(1)$ & 2 & 2 & 0 \\
 $SU(2)^3$ & 10 & 10 & 0 \\
  $Sp(2)$ & 4 & 4 & 0 \\
  $Sp(2) \times U(1)$ & 4 & 4 & 0 \\
    $Sp(2) \times SU(2)$ & 2 & 2 & 0 \\
$SU(3)$ & 4 & 4 & 0 \\
$SU(3) \times U(1)$ & 4 & 4 & 0 \\
$SU(3) \times SU(2)$ & 2 & 2 & 0 \\
$G_2$ & 4 & 4 & 0 \\
$G_2 \times SU(2)$ & 7 & 7 & 0 \\
$Sp(2)^2$ & 1 & 1 & 0 \\
$Sp(3)$ & 4 & 4 & 0 \\
$Sp(3)\times SU(2)$ & 2 & 2 & 0 \\
$Sp(4)$ & 2 & 2 & 0 \\
$SO(7)$ & 1 & 1 & 0 \\
$SO(7) \times U(1)$ & 2 & 2 & 0 \\
$SO(7) \times SU(2)$ & 4 & 4 & 0 \\
$SU(4)$ & 1 & 1 & 0 \\
$SU(4) \times U(1)$ & 2 & 2 & 0 \\
$SU(4) \times SU(2)$ & 2 & 2 & 0 \\
$SU(5)$ & 1 & 1 & 0 \\
$SU(6)$ & 4 & 4 & 0 \\
$SU(3)^2$ & 1 & 1 & 0 \\
$G_2^2$ & 2 & 2 & 0 \\
$SO(8)$ & 2 & 2 & 0 \\
$F_4$ & 2 & 2 & 0 \\
$F_4 \times SU(2)$ & 2 & 2 & 0 \\
$SO(9) \times SU(2)$ & 2 & 2 & 0 \\
$SO(10)$ & 1 & 1 & 0 \\
$SO(11)$ & 1 & 1 & 0 \\
$SO(12)$ & 1 & 1 & 0 \\
$SO(13)$ & 2 & 2 & 0 \\
$E_6$ & 2 & 2 & 0 \\
$E_7$ & 2 & 2 & 0 \\
$E_8$ & 1 & 1 & 0 \\\hline
\bf{Total} & \bf{135} & \bf{137} & \bf{-2} \\\hline
\end{tabular}
\end{center}
\caption{The number of homomorphisms $SL(2,5) \rightarrow E_8$ for each centralizer type. Math and physics agree except for two homomorphisms with centralizer $SU(2)$.}
\label{tab:numberscheck}
\end{table}

On the mathematical side, the classification of Hom($SL(2,5), E_8$) was first performed in \cite{FREY}, and the results are listed in Tables 7.6 and 8.2.  More precisely, these tables list the conjugacy classes of $SL(2,5)$ subgroups of $E_8$. In most cases, a single conjugacy class corresponds to two distinct homomorphisms, which are related by an outer automorphism.\footnote{This fact was not adequately appreciated in \cite{Heckman:2015bfa}, which led to errors in the claimed correspondence.}  In some cases, however, the $\mathbb{Z}_2$ outer automorphism takes the homomorphism back to itself, and there is only one homomorphism associated with the conjugacy class.  More details on the mathematical classification of conjugacy classes and their associated homomorphisms can be found in appendix \ref{sec:MATHEMATICAL}.

On the physics side, the classification of Hom$(SL(2,5),E_8)$ via 6D SCFTs works similarly to the Hom$(\mathbb{Z}_k,E_8)$ case considered previously.  The theories in question feature a chain of $\mf{e}_8$ gauge algebras and a global symmetry which matches the centralizer of the homomorphism inside $E_8$.  For instance, there is one homomorphism in this set with centralizer $SO(12)$. It corresponds to a 6D SCFT of the form
\begin{equation}
[SO(12)] \,\, {\overset{\mathfrak{sp_1}}1}  \,\, \overset{\mathfrak{so_7}}3 \,\, \overset{\mathfrak{su_2}}2\,\, 1\,\, \overset{\mathfrak{e_7}}8\,\,1 \,\, \overset{\mathfrak{su_2}}2\,\, \overset{\mathfrak{g_2}}3 \,\, 1 \,\, \overset{\mathfrak{f_4}}5 \,\, 1 \,\, \overset{\mathfrak{g_2}}3 \,\,{\overset{\mathfrak{su_2}}2} \,\, 2 \,\, 1 \,\, \overset{\mathfrak{e_8}}{(12)} \,\, ...[E_8]
\end{equation}
As another example, there is a homomorphism with centralizer $SU(2)$ corresponding to a 6D SCFT of the form
\begin{equation}
 2 \,\, \underset{[SU(2)]}{\overset{\mathfrak{su_2}}2}\,\, {\overset{\mathfrak{su_2}}2} \,\, 2 \,\, 1 \,\, \overset{\mathfrak{e_8}}{(11)} \,\, ...[E_8]
\end{equation}
It is nontrivial to read off the $SU(2)$ global symmetry of this theory, and as a result the theory was originally omitted from the claimed correspondence of \cite{Heckman:2015bfa}. However, using the correspondence between 6D SCFTs and nilpotent orbits in \cite{Heckman:2016ssk} as a Rosetta Stone, we can now verify that the global symmetry is indeed $SU(2)$.

The full list of 6D SCFTs in this correspondence can be found in appendix \ref{sec:LISTS}. The number of homomorphisms of each centralizer subgroup of $E_8$, computed both from the physics side and from the mathematics side, is shown in Table \ref{tab:numberscheck}. Clearly, the match between mathematics and physics is remarkable, and far too close to be a coincidence. However, it is not perfect: the correspondence with 6D SCFTs suggests the existence of two homomorphisms with $SU(2)$ centralizer that are not observed from the mathematical perspective. We are not sure what accounts for this discrepancy.\footnote{One possible source of the discrepancy is the difference between the real and complex forms of $E_8$, since our physics classification deals with the former while the mathematical classification deals with the latter. In Table~\ref{evenmult}, we show that all the image groups are conjugate to groups that live in $E_8(\bR)$, so the discrepancy is not due to subgroups of $E_8(\bC)$ that are not conjugate to subgroups of $E_8(\bR)$.  However, two subgroups of $E_8(\bR)$ could be conjugate in $E_8(\bC)$ but not conjugate in $E_8(\bR)$.  So if the $E_8(\bC)$ classification were to differ from the $E_8(\bR)$ classification, we would expect the $E_8(\bR)$ classification to have a larger number of classes (which it currently does).}

It is also worth noting that the match with 6D SCFTs has revealed several minor errors in the mathematical classification of \cite{FREY}. In particular, Fusion pattern 19 of Table 8.2 has centralizer dimension 17 rather than the claimed 11 and corresponds to $G_2 A_1$ rather than $A_2 A_1$ (note that 17 matches the result given in Table 4.9 of \cite{FREY}.)  In addition, cases 1310 and 1328 in Table 8.2 have centralizer dimension 21 and rank 3 but correspond to $C_3$ rather than $B_3$.  Similarly, case 2324 has centralizer dimension 24 and rank 4 but corresponds to $C_3 A_1$ rather than $B_3 A_1$.

\subsection{Classification of Hom(Dic$_{k-2}, E_8$)}\label{ssec:DICYCLIC}

Having understood the connection between 6D SCFTs and Hom$(SL(2,5),E_8)$, we are now in a position to do the same for homomorphisms from dicyclic groups into $E_8$.  Some of these homomorphisms were computed for the particular cases of Dic$_3$ and Dic$_5$ in \cite{FREY}, and the analogous 6D SCFTs were identified in \cite{Heckman:2015bfa}.  However, these homomorphisms have not been classified in full generality.  In this subsection, we will describe the 6D SCFTs corresponding to Hom(Dic$_{k-2} \cong \Gamma_{D_k},E_8)$ for $k \geq 4$.  The special cases of Dic$_2$ and Dic$_3$ are written explicitly in appendix \ref{sec:LISTS}.

Our starting point is the theory of (\ref{eq:heterotic}) for $G=SO(2k)$, which corresponds to the trivial homomorphism:
\begin{equation}
[E_8] \,\, \overset{\mathfrak{so}_{2k}}1 \,\,\overset{\mathfrak{so}_{2k}}2 \,\, ...\,\, \overset{\mathfrak{so}_{2k}}2\,\, [SO(2k)] ~~\rightarrow ~~[E_8] \,\, 1 \,\, 2  \,\, {\overset{\mathfrak{su_2}}2} \,\,\overset{\mathfrak{g_2}}3 \,\, 1 \,\,  \overset{\mathfrak{so_9}}4 \,\,    {\overset{\mathfrak{sp_1}}1} \,\, {\overset{\mathfrak{so_{11}}}4} \,\,  ...\overset{\mathfrak{so}_{2k}}4 \,\, ... \overset{\mathfrak{so}_{2k}}4 \,\, \overset{\mathfrak{sp}_{k-4}}1 \,\,[SO(2k)].
\label{eq:Dicstart}
\end{equation}
As usual, we assume that the quiver is sufficiently long that we can ignore any deformations of the right-hand side of the quiver.  The resolved theory involves a ramp of the type discovered in \cite{DelZotto:2014hpa} that starts with $\mathfrak{so}_9$ gauge algebra and builds up to $\mathfrak{so}_{2k}$ before leveling off.  Clearly, theory has F-theory base,
\begin{equation}
1 \,\, 2 \,\, 2 \,\, 3 \,\, 1 \,\, 4 \,\, 1 \,\, 4 \,\,...4 \,\, 1.
\end{equation}
All other homomorphisms are labeled by deformations of this theory and involve one of the following F-theory bases:
\begin{align}
1 \,\, 2 \,\, 2 \,\, 3 \,\, 1 \,\, 4 \,\, 1 \,\, 4 \,\,...4 \,\, 1  \label{eq:1223base} \\
1 \,\, 2 \,\, 3 \,\, 1 \,\, 4 \,\, 1 \,\, 4 \,\,...4 \,\, 1 \label{eq:123base} \\
1 \,\,  3 \,\, 1 \,\, 4 \,\, 1 \,\, 4 \,\,...4 \,\, 1 \label{eq:131base}\\
2 \,\,  1 \,\, 4 \,\, 1 \,\, 4 \,\,...4 \,\, 1 \label{eq:214base} \\
 1 \,\, \overset{1}4 \,\, 1 \,\, 4 \,\,...4 \,\, 1 \label{eq:treebase}
\end{align}
Our task of classifying these homomorphisms thus amounts to identifying the ways in which these bases may be decorated with gauge groups consistent with the usual rules of 6D SCFTs, ending with a sequence of $\mathfrak{so}_{2k}$ gauge algebras on the right-hand side of the quiver.
Here, we show that the solutions to these constraints, which are in 1-1 correspondence with homomorphisms Dic$_k \rightarrow E_8$, may be given a simple combinatoric interpretation in terms of ``D-partitions" of $2k$ supplemented with an additional gauge algebra.

\subsubsection{Hom(Dic$_{k-2}, E_8$) from Nilpotent Orbits}

Our starting point is to notice the similarity between the 6D SCFTs related to Hom(Dic$_{k-2}, E_8$) and the 6D SCFTs related to nilpotent orbits of $D_k$ studied in \cite{Heckman:2016ssk,Mekareeya:2016yal}.  These nilpotent orbits are classified by ``D-partitions" of $2k$, which are partitions of $2k$ subject to the constraint that any even number must appear an even number of times.  So for instance, the 6D SCFT corresponding to the partition $\mu = [2k-1,1]$ is given by
\begin{equation}
 2  \,\, {\overset{\mathfrak{su_2}}2} \,\,\overset{\mathfrak{g_2}}3 \,\, 1 \,\,  \overset{\mathfrak{so_9}}4 \,\,    {\overset{\mathfrak{sp_1}}1} \,\, {\overset{\mathfrak{so_{11}}}4} \,\,  ...\overset{\mathfrak{so}_{2k}}4 \,\, ... \overset{\mathfrak{so}_{2k}}4 \,\, \overset{\mathfrak{sp}_{k-4}}1 \,\,[SO(2k)].
\end{equation}
This looks very similar to the theory in (\ref{eq:Dicstart})!  The only difference between the two is that the $-1$ curve at the far left of the theory in (\ref{eq:Dicstart}) has now disappeared.  At the other extreme, the nilpotent orbit $\mu = [1^{2k}]$ corresponds to the 6D SCFT
\begin{equation}
[SO(2k)] \,\, \overset{\mathfrak{sp}_{k-4}}1 \,\, \overset{\mathfrak{so}_{2k}}4 \,\, \overset{\mathfrak{sp}_{k-4}}1 ...\overset{\mathfrak{so}_{2k}}4 \,\, ... \overset{\mathfrak{so}_{2k}}4 \,\, \overset{\mathfrak{sp}_{k-4}}1 \,\,[SO(2k)].
\end{equation}
By adding a single curve of self-intersection $-2$ to the left-hand side of this quiver or a $-1$ curve above the left-most $-4$ curve, we can get a base that looks like the ones in (\ref{eq:214base}) or (\ref{eq:treebase}), respectively.  This is the idea behind the full classification of homomorphisms Dic$_{k-2} \rightarrow E_8$: we start with an SCFT quiver corresponding to a nilpotent orbit of $SO(2k)$ and ``affinize" it by adding a single curve.\footnote{All little string theories are related to 6D SCFTs by this ``affinization" process of adding a single node to the quiver \cite{Bhardwaj:2015oru}.  Note that here, our end result of affinization is not a little string theory, but simply another SCFT.}  We are then left with a set of choices for the gauge algebra on this additional node, each of which corresponds to a distinct homomorphism.  Thus, any homomorphism Dic$_{k-2} \rightarrow E_8$ is labeled by a D-partition of $2k$ with an additional choice of gauge algebra.

We now explain how this works in detail.  For reasons that will become clear shortly, we may split our analysis of D-partitions $\mu$ into three cases: (i) $\mu^T_1 \geq 8$, (ii) $\mu^T_1 < 8, \mu^T_1 +\mu^T_2 \geq 6$, and (iii) $\mu^T_1 + \mu^T_2 < 6$.  Here, $\mu^T$ indicates the transpose of the partition $\mu$, and $\mu^T_i$ is the $i$th entry of $\mu^T$.

\vspace{.2cm}
\noindent
\textbf{Case (i): $\mu^T_1 \geq 8$.}

6D SCFTs corresponding to nilpotent orbits with $\mu^T_1 \geq 8$ take the form \cite{Mekareeya:2016yal}
\begin{equation}
 {\overset{\mathfrak{sp}(P_1)}1} \,\,  \overset{\mathfrak{so}(M_1)}4 \,\,    {\overset{\mathfrak{sp}(P_2)}1} \,\,  ...\overset{\mathfrak{so}_{2k}}4 \,\, ... \overset{\mathfrak{so}_{2k}}4 \,\, \overset{\mathfrak{sp}_{k-4}}1 \,\,[SO(2k)].
\end{equation}
We may ``affinize" this quiver in two ways: we can either (a) add a $-2$ curve to the far left, giving us the base in (\ref{eq:214base}), or (b) add a $-1$ curve on top of the left-most $-4$ curve, giving us the base in (\ref{eq:treebase}).  At least one of these is always possible for any D-partition with $\mu_1^T \geq 8$.
In case (a), we have several possibilities for how to decorate this $-2$ curve with a gauge algebra: it may in general hold $\mathfrak{su}(N)$, $\mathfrak{so}(M)$, $\mathfrak{g}_2$, or be empty of any gauge algebra, but the actual set of possibilities is constrained by the partition $\mu$.  In particular, if the $-2$ curve has $\mathfrak{su}(N), N \geq 2$:
$$
{\overset{\mathfrak{su}(N)}2} \,\, {\overset{\mathfrak{sp}(P_1)}1} \,\,  \overset{\mathfrak{so}(M_1)}4 \,\,    {\overset{\mathfrak{sp}(P_2)}1} \,\,  ...\overset{\mathfrak{so}_{2k}}4 \,\, ... \overset{\mathfrak{so}_{2k}}4 \,\, \overset{\mathfrak{sp}_{k-4}}1 \,\,[SO(2k)],
$$
then $N$ is constrained by $N \geq P_1, M_1 + 2 N \leq 4 P_1 + 16$ with the one exception of $N=4, P_1=0$, $M_1 \leq 10$.  Using the rules of \cite{Mekareeya:2016yal}, we can express these conditions on $N$ in terms of the partition $\mu$:
\begin{align}
&\frac{1}{2}(\mu_1^T -\mu_2^T)\geq N \geq \frac{1}{2}(\mu_1^T - 8) \\ \mbox { or } &N=4, \mu_1^T=8, \mu_2^T \leq 2.
\end{align}
The exceptional case in which $P_1=0$, $M_1 =9, 10$, $N = 4$ is due to the fact that $\mathfrak{su}_4 \oplus \mathfrak{so}_{10} \subset \mathfrak{so}_{16} \subset \mathfrak{e}_8$.  One might initially have thought that $P_1=0$, $M_1=13$, $N = 2$ would be allowed for the same reason: $\mathfrak{su}_2 \oplus \mathfrak{so}_{13} \subset \mathfrak{so}_{16} \subset \mathfrak{e}_8$.  However, there are a couple of ways to argue that such a theory is not allowed.  First off, the analysis of \cite{miranda1990persson} and \cite{Morrison:2016djb} (see also \cite{miranda1986extremal,persson1990configurations,global-symmetries}) indicates that it is impossible to construct such a configuration in F-theory.  Furthermore, if such a configuration were possible, one would expect that the configuration
$$
[Sp(5)] \,\, \overset{\mathfrak{so_{13}}}4 \,\,1 \,\, {\overset{\mathfrak{su_2}}2} \,\,\overset{\mathfrak{so_7}}3 \,\, {\overset{\mathfrak{su_2}}2}  \,\, 1\,\, {\overset{\mathfrak{e_7}}{8}} \,\, ...[E_7].
$$
should also be allowed, representing a homomorphism from $\Gamma_{E_7}$ into $E_8$, since it features a similar case of $\mathfrak{so}_{13}$ and $\mathfrak{su}_2$ meeting an unpaired $-1$ curve.  But it is also clear that such a homomorphism cannot exist because $Sp(5)$ is not a subgroup of $E_8$!  This gives us a second reason for doubting the existence of this particular theory.  We interpret the nonexistence of this theory as evidence that the global structure of the $\mathfrak{su}_2$ gauge algebra must in fact be $SU(2)$ rather than $SO(3)$.  Although $SO(3) \times SO(13)$ is a subgroup of $E_8$, $SU(2) \times SO(13)$ is not.  Thus, the $E_8$ gauging condition is satisfied at the level of algebras, but violated at the level of groups.

If we next attempt to stick $\mathfrak{so}(M)$ gauge algebra on the $-2$ curve, $M \geq 7$, we must have $P_1 \leq M-6$ if $13 \geq M \geq 9$, $P_1 \leq 2$ if $M=8$, or $P_1 \leq  4$ if $M=7$.  We must also have $M + \delta_{M,7} + M_1 \leq 4 P_1 + 16$ if $P_1 \geq 1$ and $M + M_1 \leq 16$ if $P_1 = 0$.  For the special case of $M = 7$, $P_1=1$, there is an additional possibility: rather than a spinor transforming as a mixed representation under $\mf{so}_7 \oplus \mf{sp}_1$ as $\frac{1}{2}(\bf{8}, \bf{2})$, we may have a fundamental $\frac{1}{2}(\bf{7}, \bf{2})$. In this case, we must have $7 + M_1 \leq 4 P_1 + 16 =20$.  These possibilities can be expressed in terms of the partition $\mu$ as follows:
\begin{align}
M&=7, ~16 \geq \mu_1^T \geq \mu_2^T + 8 - \delta_{\mu_1^T,8} \\ \mbox{ or } M&=7, ~\mu_1^T = 10, ~ \mu_2^T \leq 3 \\ \mbox{ or } M&=8, ~12 \geq \mu_1^T \geq \mu_2^T + 8\\  \mbox{ or } 13 &\geq M \geq 9, ~2M-4 \geq \mu_1^T \geq \mu_2^T + M.
\end{align}

Putting $\mathfrak{g}_2$ on the $-2$ curve, we have $P_1 \leq 4$, $4 P_1 + 9 \geq M_1$.  In terms of $\mu$, this is simply the condition
\begin{align}
 16 \geq \mu_1^T \geq \mu_2^T + 7.
\end{align}
If this condition is met, $\mathfrak{g}_2$ is a possible decoration.  Otherwise, it is not allowed.

Finally, we consider the case in which the $-2$ curve is left devoid of any gauge algebra, which for most practical purposes can be thought of as the case $\mathfrak{su}(N)$ with $N=1$ \cite{Apruzzi:2017iqe}.  In this case, there are two possibilities: $P_1 =0, M_1 \leq 16$ or else $P_1 =1, M_1 \leq 19$.
Expressing these conditions in terms of $\mu$, we have that for a $-2$ curve without a gauge algebra,
\begin{align}
\mu^T_1 &= 8 \\
\mbox{ or } \mu^T_1 &= 10, \mu_2^T \leq 9.
\end{align}
For $P_1=0$, it is not hard to see why $M_1$ must be smaller than $16$: $\mathfrak{so}_{16} \subset \mathfrak{e}_8$, while $\mathfrak{so}_{17} \not\subset \mathfrak{e}_8$ \cite{Morrison:2016djb}.  On the other hand, for $P_1=1$, we have a $-1$ curve with $\mathfrak{sp}_1$ gauge algebra and 20 half-hypermultiplets.  One half-hyper lives at the intersection with the $-2$ curve, and the remaining $19$ are free to transform under the $\mathfrak{so}_{19}$ gauge algebra of the adjacent $-4$ curve.  Indeed, by the analysis of \cite{global-symmetries}, one may verify that a Weierstrass model of the form
$$
\overset{II}2 \,\, \overset{I_3^{ns}}1 \,\,  \overset{I_6^{*,ns}}4 \,\, ...
$$
is allowed in F-theory and gives rise to the desired quiver
$$
2 \,\, \overset{\mathfrak{sp}_1}1 \,\,  \overset{\mathfrak{so}_{19}}4 \,\, ...
$$
Further evidence for the constraint $M_1 \leq 19$ comes from the match with the known homomorphisms from Dic$_3$ into $E_8$.  As shown in \cite{FREY}, there is one such homomorphism with centralizer $SO(9)$, which implies an associated 6D SCFT with flavor symmetry $SO(9)$.  The only possibility is
$$
2 \,\,      \underset{[SO(9)]}{\overset{\mathfrak{sp_1}}1} \,\,  \overset{\mathfrak{so_{10}}}4 \,\, ... [SO(10)],
$$
which indicates that there must indeed be a single half-hypermultiplet of $\mathfrak{sp}_1$ localized at the intersection with the $-2$ curve, leaving $19$ to transform as vectors under the $\mathfrak{so}_{10}$ gauge symmetry and the $SO(9)$ flavor symmetry.

Moving on to case (b), we now want to add a $-1$ curve on top of the $-4$ curve, resulting in a tree-like quiver.  This curve may be decorated with an $\mathfrak{sp}(P)$ gauge algebra, $P \geq 0$, subject to the constraints $M_1 -8 \geq P + P_1 + P_2$, $4 P + 16 \geq M_1$.  In terms of the partition $\mu$, this is simply
\begin{align}
\frac{1}{2}(\mu_2^T-\mu_3^T) \geq P \geq \frac{1}{4}(\mu_1^T+\mu_2^T) -4.
\end{align}

This concludes our discussion of case (i).

\vspace{.2cm}
\noindent
\textbf{Case (ii): $\mu^T_1 < 8, \mu_1^T+\mu_2^T \geq 6$.}

D-partitions satisfying $\mu^T_1 < 8, \mu_1^T+\mu_2^T \geq 6$ give rise to 6D SCFTs of the form,
\cite{Mekareeya:2016yal}
\begin{equation}
\overset{\mathfrak{g}}3 \,\,    {\overset{\mathfrak{sp}(P_2)}1} \,\,  ...\overset{\mathfrak{so}_{2k}}4 \,\, ... \overset{\mathfrak{so}_{2k}}4 \,\, \overset{\mathfrak{sp}_{k-4}}1 \,\,[SO(2k)].
\end{equation}
Here, $\mathfrak{g}$ can be $\mathfrak{so}(M), 12 \geq M \geq 7$, $\mathfrak{g}_2$, or $\mathfrak{su}_3$.  We may affinize this quiver by adding a $-1$ curve carrying gauge algebra $\mathfrak{sp}(P)$, $P \geq 0$ at the far left,
\begin{equation}
\overset{\mathfrak{sp}(P)}1 \,\, \overset{\mathfrak{g}}3 \,\,    {\overset{\mathfrak{sp}(P_2)}1} \,\,  ...\overset{\mathfrak{so}_{2k}}4 \,\, ... \overset{\mathfrak{so}_{2k}}4 \,\, \overset{\mathfrak{sp}_{k-4}}1 \,\,[SO(2k)].
\end{equation}
The choices for $P$ are dictated by the partition $\mu$.  We discuss the possibilities in turn:
\begin{enumerate}[(a)]
\item We may have $\mu_1^T = 4$, $\mu_2^T =2$.  This corresponds to $\mathfrak{g} = \mathfrak{su}_3$.  In this case, we must have $P=0$.

\item We may have $\mu_1^T =4, \mu_2^T = 3$.  In this case, we have $\mathfrak{g} = \mathfrak{g}_2$, and the constraint on $P$ is simply $P \leq  \frac{1}{2}(3-\mu_3^T)$.

\item We may have $\mu_1^T =4, \mu_2^T = 4$.  This gives $\mathfrak{g} = \mathfrak{so}_7$, and $P \leq 1$.

\item We may have $\mu_1^T =6, \mu_2^T = 1$.  This gives $\mathfrak{g} = \mathfrak{so}_7$, and $P \leq 2$.

We may have $\mu_1^T =6, \mu_2^T = 2$.  This gives $\mathfrak{g} = \mathfrak{so}_8$, and $P \leq 1$.

\item We may have $\mu_1^T =6, \mu_2^T \geq 2$.  This gives $\mathfrak{g} = \mathfrak{so}_{6 +\mu_2^T}$, and $P$ satisfies  $P \leq \frac{1}{2}( \mu_1^T+\mu_2^T-\mu_3^T) - 3$.
\end{enumerate}

This concludes our study of case (ii).

\vspace{.2cm}
\noindent
\textbf{Case (iii): $\mu_1^T+\mu_2^T < 6$.}

Finally, we have the simplest case, in which $\mu_1^T+\mu_2^T < 6$.  These theories are in 1-1 correspondence with nilpotent orbits: no decoration is allowed at all.  To get from the SCFT for the nilpotent orbit to the SCFT for the homomorphism Dic$_{k-2} \rightarrow E_8$, one simply adds a $-1$ curve to the far left of the quiver.  Thus, the theory for a partition with $\mu_1^T  =\mu_2^T= \mu_3^T=\mu_4^T= 2$ is
\begin{equation}
[E_7] \,\, 1 \,\, \overset{\mathfrak{su}_2}2   \,\,    \overset{\mathfrak{so}_7}3 \,\,    {\overset{\mathfrak{sp}_1}1} \,\,  ...\overset{\mathfrak{so}_{2k}}4 \,\, ... \overset{\mathfrak{so}_{2k}}4 \,\, \overset{\mathfrak{sp}_{k-4}}1 \,\,[SO(2k)].
\end{equation}
The theory for a partition with $\mu_1^T  = 4, \mu_2^T = 1$ is
\begin{equation}
[E_7] \,\, 1 \,\, \overset{\mathfrak{su}_2}2   \,\,   \underset{[SU(2)]}{\overset{\mathfrak{so}_7}3} \,\,    1 \,\,  ...\overset{\mathfrak{so}_{2k}}4 \,\, ... \overset{\mathfrak{so}_{2k}}4 \,\, \overset{\mathfrak{sp}_{k-4}}1 \,\,[SO(2k)].
\end{equation}
The theory for a partition with $\mu_1^T  =\mu_2^T= \mu_3^T=2$, $\mu_4^T=1$ is
\begin{equation}
[E_7] \,\, 1 \,\, \overset{\mathfrak{su}_2}2   \,\,    \overset{\mathfrak{g}_2}3 \,\,    1 \,\,  ...\overset{\mathfrak{so}_{2k}}4 \,\, ... \overset{\mathfrak{so}_{2k}}4 \,\, \overset{\mathfrak{sp}_{k-4}}1 \,\,[SO(2k)].
\end{equation}
And finally, the theory for a partition with $\mu_1^T=2, \mu_2^T =1$ is the theory of the trivial homomorphism in (\ref{eq:Dicstart}):
\begin{equation}
[E_8] \,\, 1 \,\, 2 \,\, \overset{\mathfrak{su}_2}2   \,\,    \overset{\mathfrak{g}_2}3 \,\,    1 \,\,  ...\overset{\mathfrak{so}_{2k}}4 \,\, ... \overset{\mathfrak{so}_{2k}}4 \,\, \overset{\mathfrak{sp}_{k-4}}1 \,\,[SO(2k)].
\end{equation}

The global symmetries for all of these theories, corresponding to the subgroup of $E_8$ left unbroken by the corresponding homomorphism, can be computed straightforwardly using the rules of \ref{ssec:GLOBAL}.  In appendix \ref{sec:LISTS}, we carry out the classification and work out the global symmetries explicitly for the cases $k=4$, $k=5$.

\subsubsection{Caveat: Outer Automorphisms}\label{ssec:CAVEAT}

In the case of Hom($SL(2,5),E_8$), we saw that two homomorphisms are sometimes exchanged under a $\mathbb{Z}_2$ outer automorphism. Fortunately, these two distinct automorphisms always lead to 6D SCFTs with distinct quivers. Or, to be more precise, the 6D SCFTs associated with these homomorphisms flow to distinct theories under a tensor branch flow.

This is not the case for nilpotent orbits of $D_k$. Here, there are two distinct orbits associated with a single ``very even" partition--a partition consisting of only even numbers--and these orbits are exchanged under outer automorphism. Thus, for $k=6$, we have two distinct orbits for the partition $[4^2, 2^2]$, exchanged by outer automorphism:
\begin{equation}
 [4^2, 2^2]_I~\leftrightarrow~ [4^2, 2^2]_{II}.
\end{equation}
In the special case of $k=4$, the outer automorphism group enlarges to $S_3$. Here, an additional nilpotent orbit is exchanged under this ``triality" automorphism:
\begin{align}
[2^4]_I~\leftrightarrow~&[2^4]_{II}~\leftrightarrow~[3,1^5], \nonumber \\
[4^2]_I~\leftrightarrow~&[4^2]_{II}~\leftrightarrow~[5,1^3].
\end{align}
As shown in Figure 4 of \cite{Heckman:2016ssk}, the 6D SCFTs associated with these nilpotent orbits have the \emph{same} quiver! In our estimation, the most likely interpretation of this fact is that these 6D SCFTs, while distinct at their superconformal fixed point, flow to the same free theories in the infrared.

Given this subtlety, a couple of caveats regarding our conjectured classification of homomorphisms are in order. First, we should note that nilpotent orbits of $D_k$ related by outer automorphism give rise to the same quiver. Appending an additional node to this quiver gives a unique quiver for a homomorphism in Hom(Dic$_{k-2}, E_8$). Therefore, the above classification of Hom(Dic$_{k-2}, E_8$) in terms of nilpotent orbits of $D_k$ should likely be modded out by outer automorphisms of $D_k$: we should only consider a single nilpotent orbit in each equivalence class.

Furthermore, the identification of multiple nilpotent orbits of $D_k$ with a single 6D SCFT quiver should make us wary of the same phenomenon occurring here for Hom(Dic$_{k-2}, E_8$). It is possible that the above classification fails to distinguish homomorphisms in Hom(Dic$_{k-2}, E_8$) that are related by outer automorphism. Therefore, we should be careful to simply conjecture a classification of Hom(Dic$_{k-2}, E_8$) \emph{up to outer automorphism}. Future progress in the study of these homomorphisms will hopefully shed light on this issue.

\subsubsection{Summary of Hom(Dic$_{k-2}, E_8$)}

Let us summarize the classification results of the last few pages.  Any homomorphism Dic$_{k-2} \cong \Gamma_{D_k} \rightarrow E_8$ is labeled by D-partition $\mu$ of $2k$ (that is, a partition of $2k$ in which every even number shows up an even number of times) along with a choice of Lie algebra $\mathfrak{g}$.  In 6D SCFT terms, this Lie algebra is a gauge algebra that ``affinizes" the quiver of the theory corresponding to the nilpotent orbit labeled by $\mu$ (see \cite{Heckman:2016ssk,Mekareeya:2016yal} for details).  Note that the allowed set of affinizing Lie algebras is heavily constrained according to the choice of partition $\mu$.  The possibilities are as follows:

\begin{enumerate}[(i)]
\item $\mu_1^T \geq 8$

\begin{enumerate}[(a)]

\item $\mathfrak{g}=\mathfrak{su}(N), N \geq 2, \frac{1}{2}(\mu_1^T -\mu_2^T)\geq N \geq \frac{1}{2}(\mu_1^T - 8) \mbox { or } N=4, \mu_1^T=8, \mu_2^T \leq 2$.

\item $\mathfrak{g}=\mathfrak{su}(N), N  = 1, \mu^T_1 = 8 \mbox{ or } \mu^T_1 = 10, \mu_2^T \leq 9.$

\item $\mathfrak{g}=\mathfrak{so}(M), M  = 7, ~16 \geq \mu_1^T \geq \mu_2^T + 8 - \delta_{\mu_1^T,8}$.

\item $\mathfrak{g}=\mathfrak{so}(M), M  = 7, ~\mu_1^T =10,~ \mu_2^T \leq 3$.

\item $\mathfrak{g}=\mathfrak{so}(M), M=8,~12 \geq \mu_1^T \geq \mu_2^T + 8$.

\item $\mathfrak{g}=\mathfrak{so}(M), 13 \geq M \geq 9, ~2M-4 \geq \mu_1^T \geq \mu_2^T + M$.

\item $\mathfrak{g}=\mathfrak{g}_2, 16 \geq \mu_1^T \geq \mu_2^T + 7.$

\item $\mathfrak{g}=\mathfrak{sp}(P), P \geq 0, \frac{1}{2}(\mu_2^T-\mu_3^T) \geq P \geq \frac{1}{4}(\mu_1^T+\mu_2^T) -4.$

\end{enumerate}

\item $8 > \mu_1^T , \mu_1^T+ \mu_2^T \geq 6 $.

\begin{enumerate}[(a)]

\item $\mathfrak{g}= \mathfrak{sp}(P), P=0, \mu_1^T = 4, \mu_2^T =2$. 

\item $\mathfrak{g}= \mathfrak{sp}(P), 0 \leq P \leq  \frac{1}{2}(3-\mu_3^T), \mu_1^T = 4, \mu_2^T =3$.

\item $\mathfrak{g}= \mathfrak{sp}(P),0 \leq  P \leq  1, \mu_1^T = 4, \mu_2^T =4$.

\item $\mathfrak{g}= \mathfrak{sp}(P), 0 \leq P \leq  2, \mu_1^T = 6, \mu_2^T =1$.

\item $\mathfrak{g}= \mathfrak{sp}(P), 0 \leq P \leq  1, \mu_1^T = 6, \mu_2^T =2$.

\item $\mathfrak{g}= \mathfrak{sp}(P), 0 \leq P \leq \frac{1}{2}( \mu_1^T+\mu_2^T-\mu_3^T) - 3, \mu_1^T = 6, \mu_2^T \geq 2$.

\end{enumerate}

\item $\mu_1^T + \mu_2^T  < 6$
\begin{enumerate}[(a)]
\item $\mathfrak{g} $ is trivial.
\end{enumerate}

\end{enumerate} 

Note than in case $(i)$, we have been careful to distinguish $\mf{g} = \mathfrak{sp}_0$ from $\mf{g} = \mathfrak{su}_1$, in accordance with our earlier discussion.  In 6D SCFT language, the former corresponds to an unpaired $-1$ curve, while the second corresponds to an unpaired $-2$ curve.

\subsection{Classification of Hom($\Gamma_{E_{6,7}}, E_8$) }

The classification of homomorphisms $\Gamma_{E_6} \cong SL(2,3) \rightarrow E_8$ and $\Gamma_{E_7} \rightarrow E_8$ can be performed in an analogous manner.  We present the results of this analysis in appendix \ref{sec:LISTS}.

\section{Renormalization Group Flows}\label{sec:RGFLOWS}

Renormalization group (RG) flows between different theories are among the central aspects of quantum field theory.  6D SCFTs admit no supersymmetry-preserving relevant or marginal deformations \cite{Cordova:2015fha}, so any RG flow from one 6D SCFT to another requires giving a vev to an operator of the theory.  All known interacting 6D SCFTs admit a ``tensor branch," whereby a scalar field in a tensor multiplet aquires a vev.  Many 6D SCFTs also feature a ``Higgs branch," in which a scalar field in a hypermultiplet aquires a vev.  Tensor branch flows are easy to understand in our F-theory framework and act only on the base $B_2$ of the Calabi-Yau three-fold.  Higgs branch flows, on the other hand, can act on both the elliptic fiber and the base of the F-theory geometry and are relatively poorly understood. 

We have seen that there is a 1-1 correspondence between 6D SCFTs and homomorphisms $\Gamma \rightarrow E_8$.  In \cite{Heckman:2016ssk}, a similar correspondence was observed for homomorphisms $\mf {su}_2 \rightarrow \mf{g}$ (or equivalently, for nilpotent orbits of $\mf g$).  In that case, the correspondence could actually be pushed beyond the classification of theories to the classification of RG flows: given two nilpotent orbits $\mu_1, \mu_2$ and corresponding 6D SCFTs $\mathcal{T}_1, \mathcal{T}_2$, there is a Higgs branch flow from $\mathcal{T}_1$ to $\mathcal{T}_2$ if and only if $\mu_1 \succ \mu_2$ in the usual ordering of nilpotent orbits.  Said differently, the Hasse diagram of nilpotent orbits matches the RG flow hierarchy.  

\begin{figure}
\begin{tikzpicture}[node distance=2cm]

\node (1) [startstop] {
$
1+1+1+1:~~~ 
[E_8] \,\, 1 \,\,2 \,\, \overset{\mathfrak{su}_2}2 \,\, \overset{\mathfrak{su}_3}2  \,\, {\overset{\mathfrak{su}_4}2} \,\,... [SU(4)]
$};

\node (2) [startstop, below of=1] {
$
1+1+2:~~~
[E_7] \,\, 1  \,\,  \underset{[N_f=1]}{\overset{\mathfrak{su}_2}2} \,\, \overset{\mathfrak{su}_3}2  \,\, {\overset{\mathfrak{su}_4}2} \,\,... [SU(4)]
$};

\node (3) [startstop, below of=2] {
$
1+1+2' :~~~~
[SO(14)] \,\,  \overset{\mathfrak{sp}_1}1  \,\,  \overset{\mathfrak{su}_3}2  \,\, {\overset{\mathfrak{su}_4}2} \,\,... [SU(4)]
$};

\node (4) [startstop, below of=3] {
$
1+3:~~~
[E_6] \,\,  1  \,\,  \underset{[  SU(2) ]}{\overset{\mathfrak{su}_3}2 } \,\, {\overset{\mathfrak{su}_4}2} \,\,... [SU(4)]
$};

\node (4b) [startstop, below of=4, xshift=4.3cm] {
$
1+3' :~~~~
[SU(8)] \,\,   {\overset{\mathfrak{su}_3}1 }  \,\,  {\overset{\mathfrak{su}_4}2 } \,\, {\overset{\mathfrak{su}_4}2} \,\,... [SU(4)]
$};

\node (5) [startstop, below of=4, xshift=-4.3cm] {
$
2+2 :~~~
[E_7] \,\,   1 \,\,  {\overset{\mathfrak{su}_2}2 } \,\, \underset{[SU(2)]}{\overset{\mathfrak{su}_4}2 }  \,\,... [SU(4)]
$};

\node (6) [startstop, below of=4b, xshift=-4.3cm] {
$
2+2':~~~~
[SO(12)] \,\,   \overset{\mathfrak{sp}_1}1 \,\,  \underset{[SU(2)]}{\overset{\mathfrak{su}_4}2 } \,\, {\overset{\mathfrak{su}_4}2 }  \,\,... [SU(4)]
$};

\node (7) [startstop, below of=6] {
$
4:~~~
[SO(10)] \,\,   1 \,\,  \underset{[SU(4)]}{\overset{\mathfrak{su}_4}2 } \,\, {\overset{\mathfrak{su}_4}2 }  \,\,... [SU(4)]
$};

\node (7b) [startstop, below of=7] {
$
4' :~~~~
[SU(8) \times SU(2)] \,\,    {\overset{\mathfrak{su}_4}1 } \,\,  {\overset{\mathfrak{su}_4}2 } \,\, {\overset{\mathfrak{su}_4}2 }  \,\,... [SU(4)]
$
};

\node (8) [startstop, below of=7b] {
$
2'+2' :~~~
[SO(16)] \,\,   \overset{\mathfrak{sp}_2}1 \,\,  {\overset{\mathfrak{su}_4}2 } \,\, {\overset{\mathfrak{su}_4}2 }  \,\,... [SU(4)]
$};

\draw [arrow, color=blue] (1) -- (2);
\draw [arrow] (2) -- (3);
\draw [arrow] (3) -- (4);
\draw [arrow] (4) -- (5);
\draw [arrow] (4) -- (4b);
\draw [arrow] (4b) -- (6);
\draw [arrow] (5) -- (6);
\draw [arrow] (6) -- (7);
\draw [arrow] (7) -- (7b);
\draw [arrow] (7b) -- (8);
\end{tikzpicture}

\caption{RG flow hierarchy for 6D SCFTs representing Hom($\mathbb{Z}_4, E_8$).}
\label{fig:RG}
\end{figure}
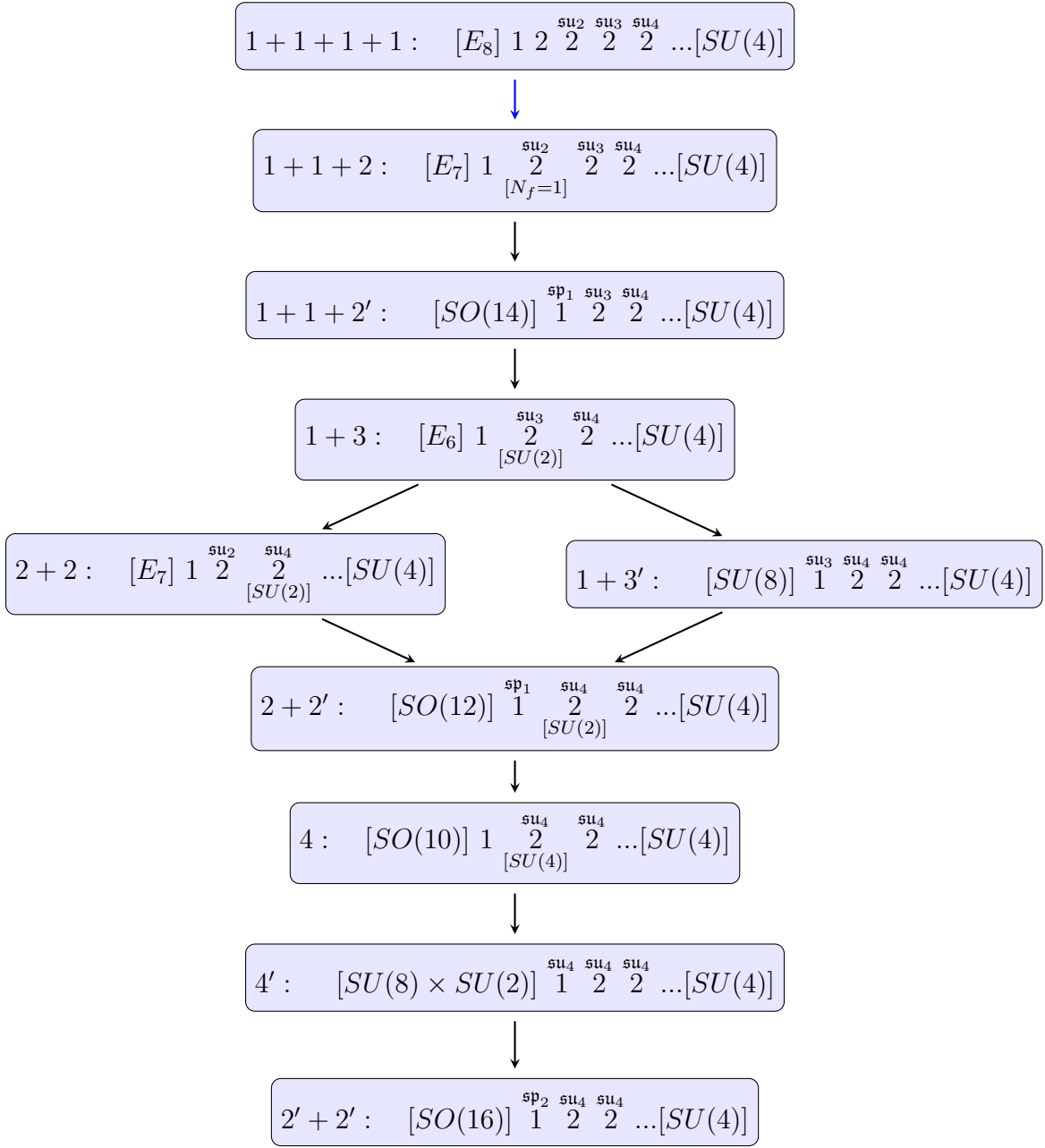

In our present case of homomorphisms $\Gamma \rightarrow E_8$, a similar relation should hold.  Namely, we may define a partial ordering on two homomorphisms $\rho_1$, $\rho_2$ by saying $\rho_1 \succ \rho_2$ if and only if there is a Higgs branch flow $\mathcal{T}_1 \rightarrow \mathcal{T}_2$ between the corresponding SCFTs.  In Figure \ref{fig:RG}, we demonstrate this hierarchy for the particular case of $\mathbb{Z}_4 \rightarrow E_8$ flows.  We do not yet have a mathematical interpretation for this formally-defined ordering, but it would be very interesting to explore further.

The hierarchy between these 6D SCFTs is also motivated by the results of \cite{Heckman:2015ola}, which studied some of the flows in the $SL(2,5) \rightarrow E_8$ hierarchy from the perspective of both F-theory geometry and 't Hooft anomaly matching.  However, the present work partially revises our understanding of these flows: we now see that \emph{every} theory in the hierarchy represents a distinct homomorphism $\Gamma \rightarrow E_8$, so every RG flow visible as a deformation of the F-theory geometry is also visible as a deformation of the heterotic M-theory setup of section \ref{sec:HOMS}.  Thus, the ordering of homomorphisms exactly matches the web of RG flows.

\section{Conclusions}\label{sec:CONC}

We have explicitly verified the correspondence between 6D SCFTs and Hom($\Gamma, E_8$) predicted by the string duality web.  This matching was facilitated by recent progress in classifying 6D SCFTs and understanding their global symmetries, which enabled us to address various subtleties and correct both the mathematics and physics literature.  Quite remarkably, this connection between 6D SCFTs and group homomorphisms has allowed us to classify homomorphisms from dicyclic groups into $E_8$, a task that has not been achieved to date from a purely mathematical perspective.  The result may be stated in a very simple way: homomorphisms from Dic$_{k-2}$ into $E_8$ are labeled by nilpotent orbits of $D_k$ supplemented with an appropriate choice of simple Lie algebra.

This work suggests a number of interesting future directions.  We have classified 6D SCFTs by relating them to elliptically-fibered Calabi-Yau three-folds, and we have classified homomorphisms by relating them to 6D SCFTs.  This shows that there is a correspondence between a particular class of elliptically-fibered Calabi-Yau three-folds and homomorphisms $\Gamma \rightarrow E_8$, and it would be interesting to understand this correspondence from a purely mathematical perspective.

On the physics side, the current work is yet another example of the power of group theory in understanding and classifying 6D SCFTs.  Indeed, as shown in \cite{Heckman:2018pqx}, for a fixed maximal gauge algebra, \emph{any} sufficiently long 6D SCFT quiver can be classified in terms of a pair of homomorphisms $(\rho_L, \rho_R)$, where $\rho_R$ is a homomorphism from $\mathfrak{su}_2$ into some simple gauge algebra (i.e. a nilpotent orbit) and $\rho_L$ is either a nilpotent orbit or a homomorphism $\Gamma \rightarrow E_8$.  It is tempting to conjecture that all 6D SCFTs, as well as the web of RG flows between them, can be given a simple group-theoretic interpretation.  Finally, this correspondence between 6D SCFTs and group theory may prove useful for understanding the compactification of general 6D SCFTs to lower dimensions.

\section*{Acknowledgements}

We thank Jonathan Heckman, David Morrison, Kantaro Ohmori, Cumrun Vafa, and Edward Witten for useful discussions. T.R. is supported by the Carl P. Feinberg Founders
Circle Membership and by NSF grant PHY-1606531.

\appendix

\section{On the Mathematical Classification of Homomorphisms }\label{sec:MATHEMATICAL}

In this appendix, we elaborate on the mathematical classification of Hom($SL(2,5), E_8$) performed in \cite{FREY}.  We further explain why the analogous classification of Hom(Dic$_2, E_8$) was more complicated than we originally thought it would be (so is not included in this article).

\subsection{Hom($SL(2,5), E_8$)}

The original intent in \cite{FREY} was to classify $Alt_5$ (the alternating group of degree 5) and $SL(2,5)$ (the binary icosahedral group) subgroups of $E_8(\bC)$ up to conjugacy (rather than homomorphisms $SL(2,5)\to E_8(\bC)$).  These two groups were classified together because the quotient of $SL(2,5)$ by its center (of order 2) is $Alt_5$.  The interest of this paper is to classify homomorphisms from $SL(2,5)$ (and $Alt_5$, as it is a quotient of $SL(2,5)$) into $E_8(\bC)$, and the number of such classes of homomorphisms is different than the number of classes of subgroups.  However, the two problems are closely related, and in fact the present question about homomorphisms is also addressed in \cite{FREY}.

The approach taken in \cite{FREY} was three-fold, and it centered around the notion of a fusion pattern.  A {\bf fusion pattern} from a group $A$ to a group $B$ is a function $f$ from the set of conjugacy classes of $A$ to the set of conjugacy classes of $B$ such that\footnote{This definition is essentially \cite[1.1]{Fr2}.}
\begin{enumerate}
\item if $K$ is a class in $A$ of elements of order $n$, then its image $f(K)$ is a class in $B$ of elements of order $n$.
\item $f$ commutes with power maps.  That is, if $K^{(m)}$ denotes the conjugacy class of all $m$th powers of elements of $K$, a conjugacy class in $A$, then $f(K^{(m)})=f(K)^{(m)}$ for every class $K$ in $A$ and all integers $m$.
\end{enumerate}

A list of conjugacy classes of elements of small order in $E_8$ can be found in a number of places, including \cite[Table 4]{CoGr} and \cite[Table 1.16]{FREY}, though \cite[Table 4]{CoGr} does not include classes of elements of order 10.  The classes of order 10 were calculated in \cite[Table 1.16]{FREY} using the approach of \cite{Kac90}.  

We typically denote fusion patterns by listing the images of the classes for the respective group.  For example, 

$2A, 3B, 5G$ for $Alt_5$

$2A, 3B, 4A, 5B, 6F, 10Z$ for $SL(2,5)$, though we often only list $4A, 6F, 10Z$, since these classes determine the others.

This means that the elements of order 2 in our $Alt_5$ subgroup of $E_8$ (which are all conjugate to each other in $Alt_5$) come from the $E_8$ conjugacy class labeled 2A in \cite[Table 4]{CoGr} (and in \cite[Table 1.16]{FREY}).  The elements of order 3 in our $Alt_5$ subgroup (which also are all conjugate to one another in $Alt_5$) come from the $E_8$ class labeled 3B, and the elements of order 5 come from the $E_8$ class labeled $5G$.  This fusion pattern represents a somewhat special case because generally, elements of order 5 in $E_8$ (and $Alt_5$) are not conjugate to their squares, but those in class 5G \emph{are} conjugate to their squares in $E_8$ (but not in $Alt_5$, of course).  Such a class in $E_8$ is called {\bf rational}.  There are two rational classes of elements of order 5 in $E_8$, namely 5C and 5G.  The other classes are not rational, but are {\bf real} (meaning that each element is conjugate in $E_8$ to its inverse).  In fact, all elements of $E_8$ are real.  Non-rationality for elements of order 5 is indicated in \cite[Table 4]{CoGr} and in \cite[Table 1.16]{FREY} with a ``[2]" next to the label name.  Such entries actually represent two conjugacy classes of elements rather than just one.  The ``[2]" indicates that the square of a given element is in a different conjugacy class.  So if $z$ is a non-rational element of order 5, its primitive powers fall into two classes, namely $\{z,z^{-1}\}$ and $\{z^2, z^3\}$. 

Elements of order 10 can similarly be rational or not in $E_8$.   Elements of the class 10Z, for example, are not rational because they are not conjugate to their cubes.  Thus, the primitive powers of an element $x$ of type 10Z similarly fall into two classes, namely $\{x,x^{-1}\}$ and $\{x^3,x^7\}$.  Non-rational classes of order 10 in \cite[Table 1.16]{FREY} are indicated with a ``[3]."  There are rational classes of elements of order 10 in $E_8$, namely 10FF, 10GG, 10OO, 10SS, 10TT, and 10FFF, but most are non-rational.

Two of the three parts of the three-fold approach in \cite{FREY} involved complex character theory, both for the finite groups $Alt_5$ and $SL(2,5)$, and for complex Lie groups.  One can find the elementary theory of characters for finite groups in \cite{Is} and \cite{JaLi}.  For the representation theory involved for Lie groups, consider \cite{FuHa}. The three-fold approach in \cite{FREY} consisted of the following:

\textbf{(1)} In the first stage, we use complex character theory to eliminate fusion pattern possibilities for $Alt_5$ and $SL(2,5)$ by eliminating those fusion patterns which did not yield nonnegative integer values for $$\left(\chi_L,\eta\right)=\frac1{|L|}\sum_{g\in L}\chi(g)\overline{\eta(g)}$$ where $\chi_L$ represents the adjoint character for $E_8$ restricted to the proposed subgroup $L$ with the given fusion pattern, and $\eta$ runs through the irreducible characters for $L$. (A character is {\bf irreducible} if it is afforded by an irreducible representation.  More simply a character $\eta$ is irreducible if and only if $(\eta,\eta)=1$ \cite[Corollary 2.17]{Is}. This part of the argument ended up being most important for eliminating possible fusion patterns with 0-dimensional centralizers.  In fact, there was only one conjugacy class of $Alt_5$ subgroups and there were no classes of $SL(2,5)$ subgroups with 0-dimensional centralizer.  Recall that the {\bf centralizer} of a subgroup $H$ in a group $G$ is the subgroup $$C_G(H)=\{g\in G\mid g^{-1}hg=h, \ \forall h\in H\}.$$ This part of the argument is also important for establishing the dimension of the connected component of the centralizer that contains the identity (the {\bf connected centralizer} for short) since that is measured by $\left(\chi|_L,\iota\right)$ where $\iota$ is the trivial character for $L$.

For example, consider the fusion pattern 2A, 3A, 5F.  How do we know there is no $Alt_5$ subgroup of $E_8$ with this fusion pattern?  If we assume that there is such a subgroup, say $L$, and we restrict the adjoint character $\chi$ for $E_8$ to $L$, it can be decomposed into irreducible characters for $Alt_5$.  According to character theory (see, for example, \cite[Theorem 14.17]{JaLi}), the multiplicity of each irreducible character $\eta$ for $Alt_5$ in the decomposition of $\chi_L$ can be determined by calculating the inner product $$\left(\chi_L,\eta\right)=\frac1{|L|}\sum_{g\in L}\chi(g)\overline{\eta(g)}.$$  In this case, the sum for $\eta=\iota$, the trivial character, looks like \begin{eqnarray*}
\left(\chi_L,\iota\right)&=&\frac1{60}[248\cdot 1+15\cdot 24\cdot 1+20\cdot (-4)\cdot 1+12\cdot(28+50\tau)\cdot 1+12\cdot (78-50\tau)\cdot 1]\\
&=&\frac1{60}[528+1,272]\\
&=&30
\end{eqnarray*}
Here, $\tau=\D\frac{1+\sqrt{5}}2$.  This calculation tells us that the trivial character (the character whose value on each element of $L$ is 1) occurs with multiplicity 30 in the decomposition of $\chi_L$ over $L$.  (It also tells us that the connected centralizer of $L$ is 30-dimensional.)  The values of $\chi$ on the elements of $L$ are given in \cite[Table 4]{CoGr} and \cite[Table 1.16]{FREY}.  However, if $\eta=3_a$ (the first degree 3 character mentioned in \cite{Atlas}), we also have 

\begin{align}
\left(\chi_L,3_a\right)&=\frac1{60}\left[248\cdot 3+15\cdot 24\cdot (-1)+20\cdot (-4)\cdot 0+12\cdot(28+50\tau)\cdot\left(\frac{1-\sqrt{5}}2\right)\right. \nonumber\\
& \left.+12\cdot (78-50\tau)\cdot \left(\frac{1+\sqrt{5}}2\right)\right] \nonumber
\end{align}
\begin{align}
&=\frac1{60}\left[384+12\left(14\left(1-\sqrt{5}\right)+\frac{25}2\left(1-\sqrt{5}\right)\left(1+\sqrt{5}\right)\right)\right. \nonumber\\
& \left.+12\left(39\left(1+\sqrt{5}\right)-\frac{25}2\left(1+\sqrt{5}\right)^2\right)\right]\nonumber
\end{align}
\begin{align}
&=\frac1{60}\left[384+12\left(14\left(1-\sqrt{5}\right)-50\right)\phantom{\frac12}\right.\nonumber\\
& \left.+12\left(39\left(1+\sqrt{5}\right)-\frac{25}2\left(6+2\sqrt{5}\right)\right)\right]\nonumber\\
&=\frac1{60}\left[-480\right]\nonumber\\
&=-8\nonumber
\end{align}

Since we are getting a negative multiplicity for this character, an $Alt_5$ subgroup with these character values cannot occur in $E_8$, so this fusion pattern does not correspond to an $Alt_5$ subgroup of $E_8$.  The character values for $\eta$ come from \cite{Atlas}.

\textbf{(2)} In the second stage, we use complex character theory to construct copies of $Alt_5$ and $SL(2,5)$ in various classical subgroups of $E_8$, particularly the centralizers of elements of orders 2 and 3 and certain elementary abelian subgroups with exponent 2 or 3, namely the $A_8$ subgroup (the centralizer of an element of type 3A), the $D_8$ subgroup (the centralizer of an element of type 2B), the $A_2E_6$ subgroup (the centralizer of an element of type 3B), the $A_2^4$ subgroup (the centralizer of a nine-element subgroup consisting of eight elements of type 3B in addition to the identity) and the $A_1^8$ subgroup (the centralizer of a 2A$^8$2B$^7$ elementary abelian group of order 16).  

For the construction of subgroups of the $A_8$ subgroup $\mathcal A$ (which is isomorphic to $SL(9,\bC)/\bZ_3$ by \cite[Lemma 3.3(iii)]{CoGr}), one uses straightforward character theory (with appropriate accommodations for the fact that the embedding is in a quotient group) and an observation that since 9 is odd, conjugacy in $GL(9,\bC)$ is equivalent to conjugacy in $SL(9,\bC)$ by \cite[4.12 and 4.13]{FREY}.  The values of the adjoint character on elements of such groups is ascertained using the formula $$\chi_{\mathcal A}=char\left(adj(\mathcal A)+\bigwedge\nolimits^3 V+\bigwedge\nolimits^3 V^*\right)$$  (see \cite[page 361]{FuHa}) where $adj(\mathcal A)$ is the adjoint character for $\mathcal A$, $V$ is the natural 9-dimensional module for $SL(9,\bC)$ and $V^*$ is its dual. (The dual has eigenvalues that are the inverses of the eigenvalues of $V$.)

As an example of such a calculation, consider the 9-dimensional $Alt_5$ character $3_a+6\cdot \iota$ (i.e. the sum of the first 3-dimensional irreducible character listed in \cite{Atlas} and six copies of the trivial character).  This character corresponds to a representation $\rho: Alt_5\to SL(9,\bC)$.  Let's consider an element $x$ of $Alt_5$ of order 2.  The eigenvalues of this matrix can be ascertained by the trace of this matrix, which will be $-1+6\cdot 1=5$ using the values in \cite{Atlas}.  Since $x$ has order 2, the eigenvalues must be square roots of 1, so must be $\pm 1$.  Thus, the eigenvalues of $x$ are $-1$ with multiplicity 2 and 1 with multiplicity 7.  This gives us the correct trace (which is the sum of the eigenvalues).  The adjoint representation for $\mathcal A$ is 80-dimensional and can be calculated by conjugating each of the basis vectors $E_{ij}$ for $i\neq j$ by $diag(-1,-1,1^7)$, where $E_{ij}$ is the matrix with all 0's except for a 1 at the $i,j$ position.  (These are the root vectors for the Lie algebra $\mf{a}_8$.  Note that there are $81-9=72$ of them.)  Then the multiplicity of the eigenvalue 1 should be increased by 8 to account for the (trivial) action on the Cartan subalgebra.  A shortcut to this calculation is to simply calculate the eigenvalues of $V\otimes V^*$ and subtract 1 from the multiplicity of the eigenvalue 1.  This is because $V \otimes V^*$ decomposes as $adj(\mathcal A)+\iota$.

In the case of $x$, the eigenvalues are: \begin{itemize}
\item  1, with multiplicity $2\cdot 2+7\cdot 7-1=52$ and 
\item $-1$, with multiplicity $2\cdot 7+7\cdot 2=28$.
\end{itemize}

For the exterior cubes, we simply calculate all products $\lambda_i\lambda_j\lambda_k$ for $i<j<k$ to get the eigenvalues of $\bigwedge^3V$, where the $\lambda_i$'s are the eigenvalues of the corresponding $9\times 9$ matrix for $x$.  Thus, $\bigwedge^3V$ is $\D{9\choose 3}=84$-dimensional.  In the case of $x$, we get the eigenvalue 1 with multiplicity $7+\D{7\choose 3}=42$ and the eigenvalue $-1$ with multiplicity $2\D{7\choose 2}=42$.  The eigenvalues of $V^*$ are simply the inverses of the eigenvalues of $V$, which in the case of $x$ does not change the eigenvalues, so we get the same results for $\bigwedge^3V^*$.  Thus, for $\chi(x)$, we get eigenvalue 1 with multiplicity $52+42+42=136$ and eigenvalue $-1$ with multiplicity $28+42+42=112$.  That matches the multiplicities for elements of type 2A in $E_8$.  

Similarly, if $y$ is an element of order 3 in $\mathcal A$ corresponding to the $Alt_5$ character $3_a+6\cdot\iota$, we see that the trace of such a matrix in $SL(9,\bC)$ is $0+6=6$, so we can use the diagonal matrix $diag(\omega,\overline\omega,1^7)$ where $\omega=e^{\frac{2\pi i}3}$.  For the adjoint character, we have 1 with multiplicity $2+7\cdot 7-1=50$, $\omega$ with multiplicity $1+2\cdot 7=15$ and $\overline\omega$ with multiplicity $1+2\cdot 7=15$.  For $\bigwedge^3V$ we get 1 with multiplicity $7+\D{7\choose 3}=42$, $\omega$ with multiplicity $\D{7\choose 2}=21$ and $\overline\omega$ with multiplicity $\D{7\choose 2}=21$.  We get the same results for $\bigwedge^3V^*$.  So for $y$, the multiplicities of the eigenvalues for $\chi$ are:
\begin{itemize}
\item $50+2\cdot 42=134$, for 1, 
\item $15+2\cdot 21=57$, for $\omega$ and $\overline\omega$. 
\end{itemize}
 This matches the multiplicities for elements of type 3D in $E_8$.

Similarly, we get class 5H for an element of order 5 corresponding to the $Alt_5$ character $3_a+6\cdot\iota$.  Thus, the $Alt_5$ subgroup of $A_8$ corresponding to the character $3_a+1^6$ will have fusion pattern $2A, 3D, 5H$.  The results of these calculations are given in \cite[Table 4.16 and Table 4.18]{FREY}.

The construction of subgroups of the $D_8$ subgroup $\mathcal D$ (which is isomorphic to $HSpin(16,\bC)$ by \cite[Lemma 3.3(i)]{CoGr}) is a little more subtle.  Since the smallest representation for $HSpin(16,\bC)$ is 120-dimensional, the constructions were performed using $SO(16,\bC)$.  Both $HSpin(16,\bC)$ and $SO(16,\bC)$ are quotients of $Spin(16,\bC)$ by central involutions, so one can move from one to the other by lifting to $Spin(16,\bC)$ and then performing a quotient by a different central involution.  By \cite[Lemma 5.4]{FREY}, conjugacy is preserved when going back and forth between these groups (although an $Alt_5$ can be converted to an $SL(2,5)$ and vice versa).  Of course, the 16-dimensional characters only determine conjugacy in $GL(16,\bC)$, but because the characters for $Alt_5$ 
have positive indicator (see \cite[p. xxviii]{Atlas}), and $Alt_5$ is simple, we may assume that our image is in $SO(16,\bC)$.  Furthermore, a theorem of Tits (\cite[7.3]{Ti}) tells us that any two such subgroups that are conjugate in $GL(16,\bC)$ are conjugate in $O(16,\bC)$.  If we can then find an element of $O(16,\bC)$ of determinant $-1$ that normalizes one of the two subgroups, the two must be conjugate in $SO(16,\bC)$.  A sufficient condition for such an element to exist is that one of the constituents of the 16-dimensional character is odd.  Thus, in most cases, the character is enough to determine conjugacy in $\mathcal{D}$.  

To determine the value of the adjoint character of $E_8$ for a given element of $\mathcal D$, we write our 16-dimensional matrix in the form $diag(A,A^{-t})$ where $A\in GL(8,\bC)$.  Our element $x\in\mathcal D$ is also conjugate to an element of the maximal torus of $\mathcal D$, so it corresponds to an element $a$ in the Cartan subalgebra, whose action on the root vectors $e_r$ of the Lie algebra $\mathfrak{e}_8$ (using the notation of \cite{Carter}) is to simply multiply $e_r$ by the scalar $(a,r)$ so that $x.e_r=e^{2\pi i(a,r)}e_r$.  Thus, we can find the eigenvalues of $x$ and determine the value of the adjoint character on $x$.  Of course, there are two such elements in $\mathcal D$ for each element in $SO(16,\bC)$, so there can be some ambiguity in the fusion pattern of the image, though this was not really a problem for $Alt_5$ characters.  

For example, consider the 16-dimensional matrix $diag(\omega,\omega,\overline{\omega},\overline{\omega},1^{12})$.  (This would be a matrix that would be determined by the 16-dimensional $Alt_5$ character $2\cdot 3_a+10\cdot\iota$.)  We write the roots of the Lie algebra $\mathfrak{e}_8$ in terms of a standard basis $\D\{e_i\}_{i=1}^8$, so the roots of $\mathfrak{e}_8$ are all vectors of the form $\pm e_i\pm e_j$ for $i\neq j$ or $\D\frac12\sum_{i=1}^8\epsilon_ie_i$, where $\epsilon_i=\pm 1$ and $\D\prod_{i=1}^8\epsilon_i=1$ (see \cite[p. 48]{Carter}).  We write our diagonal matrix as an 8-dimensional vector by splitting the eigenvalues into inverse pairs and choosing one eigenvalue from each pair, and we construct the corresponding vector in the Cartan subalgebra from the powers of $\omega$.  This gives us three choices for our vector, namely $(1,1,0^6)$, $(1,2,0^6)$ and $(2,2,0^6)$.  Let us consider the first.  We think of our vector $(1,1,0^6)$ as $e_1+e_2$ and calculate the inner product of our vector with each of the root vectors of $E_8$.  We find \begin{itemize}
\item 2 (or $-1$) for \begin{enumerate}
\item $e_1+e_2$, $-e_1\pm e_j$ and $-e_2\pm e_j$ for $j>2$ (25 vectors) 
\item  $\D\frac12\sum_{i=1}^8\epsilon_ie_i$ when $\epsilon_1=\epsilon_2=-1$ (32 vectors), \end{enumerate}
\item  1 (or $-2$) for 
\begin{enumerate}
\item $-e_1-e_2$, $e_1\pm e_j$ and $e_2\pm e_j$ for $j>2$ (25 vectors) 
\item $\D\frac12\sum_{i=1}^8\epsilon_ie_i$ when $\epsilon_1=\epsilon_2=1$ (32 vectors),\end{enumerate}
\item and 0 for 
\begin{enumerate}
\item $e_1-e_2, e_2-e_1$, $\pm e_i\pm e_j$ where $i,j>2$, $\D\left(2+4\cdot {6\choose 2}= 62 \mbox{ vectors}\right)$
\item $\D\frac12\sum_{i=1}^8\epsilon_ie_i$ when $\epsilon_1=-\epsilon_2$ (64 vectors).  \end{enumerate}\end{itemize}
So for the action of $x$ on the root vectors of $\mathfrak{e}_8$, \begin{itemize} 
\item we get the eigenvalue 1 with multiplicity $8+62+64=134$, (recall we get 1 eight times from the (trivial) action of $x$ on the Cartan subalgebra),
\item we get the eigenvalue $\omega$ with multiplicity $25+32=57$, and 
\item we get the eigenvalue $\overline\omega$ with multiplicity $25+32=57$.
\end{itemize}
These results match the multiplicities of a 3D element in $E_8$.

Note that we get the same result if we had chosen $(2,2,0^6)$ instead of $(1,1,0^6)$.  However, we do not get the same result if we choose $(1,2,0^6)$.  In this case, we actually get an element of order 6 because we get half-integers as inner products.  In this case we get inner products \begin{itemize}
\item 0 (or $\pm3$) for $e_1+e_2, -e_1-e_2, \pm e_i\pm e_j$ for $i,j>2$, $\D\left(2+4\cdot {6\choose 2}=62 \mbox{ vectors}\right)$
\item 1 (or $-2$) for $-e_1+e_2$, $e_1\pm e_j$ for $j>2$, $-e_2\pm e_j$ for $j>2$ (25 vectors) 
\item 2 (or $-1$) for $e_1-e_2$, $e_2\pm e_j$ for $j>2$, $-e_1\pm e_j$ for $j>2$ (25 vectors)
\item $\D\frac32\ \left(\mbox{or }-\frac32\right)$ for $\D\frac12\sum_{i=1}^8\epsilon_ie_i$ with $\varepsilon_1=\varepsilon_2$, (64 vectors)
\item $\D\frac12$ for $\D\frac12\sum_{i=1}^8\epsilon_ie_i$ with $\epsilon_1=-1, \epsilon_2=1$ (32 vectors)
\item $\D-\frac12$ for $\D\frac12\sum_{i=1}^8\epsilon_ie_i$ with $\epsilon_1=1, \epsilon_2=-1$ (32 vectors).
\end{itemize}

If we thought of our original matrix as a matrix with powers of $\alpha=e^{\frac{2\pi i}6}$ instead of powers of $\omega$, then our Cartan algebra vector would be $(2,4,0^6)$, and all the inner products would be twice as large.  This yields \begin{itemize}
\item eigenvalue 1 with multiplicity $62+8=70$
\item eigenvalue $\alpha$ with multiplicity 32 (see the $\D\frac12$ inner products)
\item eigenvalue $\alpha^2=\omega$ with multiplicity 25 (see the 1 inner products)
\item eigenvalue $\alpha^3=-1$ with multiplicity 64 (see the $\D\frac32$ and $\D-\frac32$ inner products)
\item eigenvalue $\alpha^4=\overline\omega$ with multplicity 25
\item eigenvalue $\overline\alpha$ with multiplicity 32.
\end{itemize}
This matches the action of an element of type 6M.  We expect to sometimes get $SL(2,5)$ subgroups in $HSpin(16,\bC)$ instead of $Alt_5$ subgroups because sometimes elements of $SO(16,\bC)$ of order 2 get lifted to elements of order 4 in $Spin(16,\bC)$ rather than elements of order 2.  One can decide which of the two cases is realized by looking at the multiplicity of the eigenvalue $-1$ in the elements of order 2 in the $Alt_5$ subgroup of $SO(16,\bC)$. In practice, for $Alt_5$ characters, this is determined by the parity of the number of nontrivial irreducible constituents (see \cite[Lemma 5.11]{FREY}).  The results of these calculations with $Alt_5$ characters are given in \cite[Tables 5.13 and 5.14]{FREY}.

On the other hand, the irreducible representations for $SL(2,5)$ are not real (although their characters are).  So an $SL(2,5)$ 16-dimensional character will only have an image in $O(16,\bC)$ if the faithful constituents have even multiplicity, by \cite[Corollary 5.16]{FREY}. (A character $\chi$ is {\bf faithful} if the set $\rm Ker\chi=\{g\in G\mid \chi(g)=\chi(1)\}=\{1\}$.  In these groups, the faithful characters correspond to faithful representations (i.e. representations with trivial kernel).) Since $L\cong SL(2,5)$ is {\bf perfect} (i.e. the commutator subgroup $\langle x^{-1}y^{-1}xy\mid x,y\in L\rangle$ is the whole group $L$), any image in $O(16,\bC)$ will in fact be in $SO(16,\bC)$.  And, as before, a sufficient condition for two subgroups with the same 16-dimensional character to be conjugate in $SO(16,\bC)$ is that one of the irreducible constituents has odd dimension.  In the case of an $SL(2,5)$ character, the problem of ambiguity really does require a solution, and a fair amount of time is spent in \cite[Chapter 5]{FREY} to address it.

\textbf{(3)} In the third stage, we use elements in the centralizer of an $Alt_5$ or $SL(2,5)$ subgroup of $E_8$ to force such a subgroup to have a conjugate in one of the subgroups where we've already constructed a complete list of $E_8$-conjugacy classes.  In particular, if the connected centralizer of a subgroup $L$ has rank at least 3 (2 in the case of $SL(2,5)$), then $L$ is conjugate to a subgroup of $\mathcal D$ by \cite[Lemma 5.24]{FREY}.  This does not always settle things immediately, since there are ambiguities in some $\mathcal D$- fusion patterns, and still conjugacy questions to settle. (For example, there are often multiple characters that yield the same fusion pattern, so one has to determine whether two such images are conjugate in $E_8$.)  These questions are almost all settled in \cite{FREY}, and a list of conjugacy classes of $Alt_5$ (resp. $SL(2,5)$) is given in Table 7.6 (resp. Table 8.2) of \cite{FREY}.  

As an example, consider the $SL(2,5)$ fusion pattern 4D, 6O, 10AAA, which was given the number 786 in \cite{FREY}.  Since $(\chi_L,\iota)=9$ (where $L$ is such an $SL(2,5)$ subgroup), the connected centralizer is 9-dimensional, and by \cite[Lemma 5.35 and Table 5.36]{FREY} has rank 3.  Thus, by \cite[Lemma 5.24]{FREY}, $L$ is conjugate to a subgroup of $\mathcal D$.  By \cite[Table 5.36]{FREY}, there are two characters that correspond to embeddings (i.e. injective homomorphisms), namely $2\cdot 2_a+2\cdot 5+2\cdot\iota$ and $2\cdot 2_a+2\cdot 3_a+2\cdot 3_b$, where $2_a$ is the first degree two character for $SL(2,5)$ in \cite{Atlas}, and $3_a$ and $3_b$ are the two degree three characters for $Alt_5$.  Since both characters have odd degree constituents, we know that these represent single $\cD$-classes.  Since all $SL(2, 5)$ subgroups with this fusion pattern are forced into $\cD$, there must be either one or two classes in $E_8$.  By \cite[Lemmas 5.32-5.34]{FREY}, one can ``see" the centralizing tori and can calculate the classes of elements in the tori using the technique described above.  In particular, we can find a 9-element elementary abelian group with 8 3B elements in the torus centralizing the group corresponding to the $2\cdot 2_a+2\cdot 3_a+2\cdot 3_b$ embedding so this group is conjugate to a subgroup of the $A_2^4$ subgroup which is the centralizer of such a 9-element subgroup of $E_8$.  On the other hand, the group corresponding to $2\cdot 2_a+2\cdot 5+2\cdot\iota$ has elements from all the classes of elements of order 3, but the centralizing torus does not have a 9-element subgroup with all 3B elements. Thus, these two classes in $\cD$ do not fuse in $E_8$, so there are two $E_8$ classes of $SL(2, 5)$ subgroups with fusion pattern 786.

This classification was not quite completed in \cite{FREY} as there were five fusion patterns where the number of classes of subgroups was not determined, although in three cases, the groups were proven to have conjugates inside the centralizer of an element of type 3B.  Those three cases were later resolved in \cite{Fr2}.  The remaining two cases were resolved in \cite{Lu}.

When considering the question of conjugacy classes of homomorphisms -- the interest of this paper -- it is important to keep in mind that in most cases, a single class of groups yields two classes of homomorphisms.  The reason is that both $Alt_5$ and $SL(2,5)$ have an outer automorphism group of order 2. For any homomorphism $\phi:L\to E_8$, there is another homomorphism $\eta\circ\phi$ where $\eta$ is an outer automorphism of $L$. (Here, we are assuming that functions act on the right, so that $\eta\circ\phi$ means we apply $\eta$ first and then $\phi$.)  An outer automorphism of $L\cong SL(2,5)$ will interchange the two classes of elements of order 5 as well as the two classes of elements of order 10  \cite{Atlas}.  The elements of order 10 in $L$ are in a different class than their cubes, while the elements of order 5 are in a different class than their squares.  The same is true of most elements of order 10 and 5 in $E_8$, so in most cases, the new homomorphism $\eta\circ\phi$ is not conjugate to the homomorphism $\phi$, i.e. there is no element $g\in N_{E_8}(L)$ such that $\eta\circ\phi=\phi\circ i_g$, where $i_g$ is the outer automorphism of $\phi(L)$ induced by conjugation by $g$.  (Of course $i_g$ is an inner automorphism of $E_8$, but induces an outer automorphism of $\phi(L)$.) For most $SL(2,5)$ fusion patterns, one can simply look at the elements of order 10 to see if the elements of order 10 are conjugate to their cubes.  This can be ascertained by looking at \cite[Table 1.16]{FREY}.  Those that are not conjugate in $E_8$ to their cubes have a ``[3]" at the end of their label.  Similarly, for $Alt_5$ fusion patterns, one can simply look at the elements of order 5 to see if they are conjugate in $E_8$ to their squares.  If not, then there cannot be an inner automorphism $i_g$ of $E_8$ that makes $\eta\circ\phi=\phi\circ i_g$, since then the elements of order 10 (5) in fact would be conjugate in $E_8$ to their cubes (squares).  But this is only a sufficient condition for there to be two homomorphisms for a given class of $SL(2,5)$ or $Alt_5$ subgroups.  In the cases where the classes of order 5 and 10 are rational (i.e. conjugate to their squares or cubes respectively), more thought needs to be applied to determine whether there is one or two classes of homomorphisms per class of subgroups.  

Now consider $L\cong SL(2,5)$ where the elements of order 10 are rational.  (This forces the elements of order 5 to also be rational.)  If a homomorphism $\phi$ of $L$ corresponds to a character, the homomorphism $\eta\circ\phi$ (where $\eta$ is an outer automorphism of $L$) will be represented by the same character, except the two characters of degree 2 and the two characters of degree 3 will be interchanged.  This was called an {\bf outer twist} in \cite{FREY}.  Suppose the character of $\phi$ is a 16-dimensional character in which the multiplicities of the faithful characters of $L$ have even multiplicity.  Then the image of $L$ is conjugate to a subgroup of $SO(16,\bC)$.  If the outer twist results in the same character, then by ordinary character theory, there is an element $g\in GL(16,\bC)$ such that $\eta\circ\phi=\phi\circ i_g$.  By \cite[7.3]{Ti}, we may assume that $g\in O(16,\bC)$.  But then, by \cite[5.9]{FREY}, we may assume that $g\in SO(16,\bC)$ if there is an element of $O(16,\bC)$ of determinant $-1$ that normalizes $\eta\circ\phi(G)$.  But there is such an element if at least one of the constituents of the character is odd, by \cite[5.10]{FREY}.  Since the element $g$ is in $SO(16,\bC)$, when we lift to $Spin(16,\bC)$ and project to $HSpin(16,\bC)$ (the $D_8$-subgroup of $E_8$), our element $g$ comes along for the ride, so it is available to perform the outer automorphism on the resulting subgroup of $\mathcal D$. So for each $SL(2,5)$ fusion pattern, if an embedding in $SO(16,\bC)$ can be found that is unchanged by an outer twist and has at least one odd constituent, then there is only one class of homomorphisms corresponding to that class of groups rather than the usual two.  A very similar argument works for $Alt_5$, and such characters are listed in \cite[7.7 and 8.3]{FREY}.

It should be noted that \cite{FREY} was considering homomorphisms into $E_8(\bC)$, while most physicists are interested in $E_8(\bR)$.  However, it turns out that all of the homomorphisms discussed in \cite{FREY} have images in $E_8(\bR)$.  To see this, we note that by \cite[p. xxviii]{Atlas}, the {\bf Frobenius-Schur indicator function} for a character $\chi$ of a group $G$ defined by $$\rm ind\chi=\frac1{|G|}\sum_{g\in G}\chi(g^2)$$ takes on three possible values: 1, if $\chi$ is afforded by a real representation, $-1$, if the character only has real values, but there are no real representations that afford $\chi$, and 0, if $\chi$ has nonreal values.  On page 2 of \cite{Atlas}, we see that all of the characters for $Alt_5$ have positive indicator, so are afforded by real representations.  Hence, any embedding of $Alt_5$ in $E_8(\bC)$ is also an embedding in $E_8(\bR)$.  However, the indicators for the faithful characters of $SL(2,5)$ are negative, meaning that they are not afforded by a real representation.  But by \cite[23.6]{JaLi}, the character $\chi+\overline\chi$ \emph{can} be afforded by a real representation.  In the case of characters with negative indicators, since $\overline\chi=\chi$, we see that $\chi+\chi$ can be afforded by a real representation.  Thus, if we can decompose the adjoint character for $E_8$ over a given $SL(2,5)$ subgroup, and all of the faithful irreducible constituents have even multiplicity, then that particular subgroup appears in $E_8(\bR)$.  This calculation is done for $SL(2,5)$ in \cite[Table 4.9]{FREY}, but there are several errors in that table.  So we did the calculations again, and discovered that each entry that involves an $SL(2,5)$ subgroup of $E_8(\bC)$ has even multiplicities for each of the faithful $SL(2,5)$ characters.  These corrected multiplicities are listed in Table~\ref{tab:mults}.

\begin{center}
\begin{longtable}{|l|p{2in}|p{2in}|} \hline \label{evenmult}
Fusion pattern & Multiplicities of Nonfaithful Characters 1, $3_a, 3_b$, 4, 5& Multiplicities of Faithful Characters $2_a, 2_b, 4_f$, 6\\\hline
3 4A, 6C, 10N & 11, 13, 9, 6, 7 & 0, 16, 2, 12\\ 
19 4A, 6C, 10D & 17, 0, 28, 0, 7 & 0, 28, 14, 0\\
22 4A, 6C, 10Z & 7, 8, 10, 10, 7 & 8, 10, 4, 10\\
37 4A, 6C, 10U & 9, 5, 15, 8, 7 & 6, 16, 8, 6\\
57 4A, 6C, 10HH & 6, 9, 8, 11, 7 & 6, 10, 2, 12\\
152 4D, 6G, 10L & 9, 15, 3, 2, 13 & 0, 8, 6, 12\\
170 4D, 6G, 10P & 9, 2, 16, 2, 13 & 0, 14, 12, 6\\
174 4D, 6G, 10JJ & 3, 6, 6, 8, 13 & 4, 4, 6, 12\\
188 4D, 6G, 10KK & 3, 5, 7, 8, 13 & 4, 6, 8, 10\\
210 4D, 6G, 10AAA & 6, 7, 8, 5, 13 & 2, 8, 8, 10\\
598 4A, 6F, 10Z & 4, 8, 10, 7, 10 & 2, 4, 10, 10\\
613 4A, 6F, 10U & 6, 5, 15, 5, 10 & 0, 10, 14, 6\\
633 4A, 6F, 10HH & 3, 9, 8, 8, 10 & 0, 4, 8, 12\\
750 4D, 6O, 10JJ & 6, 6, 6, 11, 10 & 4, 4, 6, 12\\
764 4D, 6O, 10KK & 6, 5, 7, 11, 10 & 4, 6, 8, 10\\
785 4D, 6O, 10ZZ & 17, 23, 0, 0, 10 & 18, 0, 16, 2\\
786 4D, 6O, 10AAA & 9, 7, 8, 8, 10 & 2, 8, 8, 10\\
800 4D, 6O, 10WW & 17, 22, 1, 0, 10 & 0, 4, 2, 16 \\
934 4E, 6P, 10YY & 39, 1, 10, 16, 0 & 18, 32, 0, 2 \\
951 4E, 6P, 10BBB & 55, 0, 27, 0, 0 & 0, 52, 2, 0\\
1310 4D, 6R, 10L & 21, 15, 3, 14, 1 & 6, 14, 0, 12\\
1328 4D, 6R 10P & 21, 2, 16, 14, 1 & 6, 20, 6, 6\\
1368 4D, 6R, 10AAA & 18, 7, 8, 17, 1 & 8, 14, 2, 10\\
1401 4D, 6R, 10EEE & 35, 31, 1, 0, 1 & 0, 20, 0, 12 \\
1419 4D 6R, 10DDD & 35, 0, 32, 0, 1 & 0, 32, 12, 0\\
1504 4E, 6L, 10YY & 24, 1, 10, 1, 15 & 0, 14, 18, 2\\
1556 4E, 6L, 10TT & 17, 2, 2, 8, 15 & 0, 0, 4, 16\\
2294 4G, 6S, 10CCC & 133, 0, 1, 0, 0 & 0, 56, 0, 0\\
2305 4B, 6A, 10A & 36, 0, 28, 0, 0 & 0, 48, 8, 0\\
2324 4B, 6A, 10O & 24, 1, 15, 12, 0 & 14, 28, 2, 6\\
2342 4B, 6A, 10FF & 20, 6, 6, 16, 0 & 20, 20, 0, 8\\
2458 4C, 6I, 10EE & 6, 14, 2, 4, 10 & 0, 8, 4, 16\\
2475 4C, 6I, 10AA & 6, 5, 11, 4, 10 & 4, 10, 10, 10\\
2476 4C, 6I, 10II & 4, 5, 9, 6, 10 & 4, 8, 8, 12\\
2491 4C, 6I, 10T & 10, 0, 20, 0, 10 & 0, 20, 16, 4\\
2493 4C, 6I, 10OO & 2, 6, 6, 8, 10 & 6, 6, 8, 12\\
2511 4C, 6I, 10MM & 3, 8, 5, 7, 10 & 4, 6, 6, 14\\
2900 4B, 6H, 10O & 13, 1, 15, 1, 11 & 0, 14, 16, 6\\
2918 4B, 6H, 10FF & 9, 6, 6, 5, 11 & 6, 6, 14, 8\\
2937 4B, 6H, 10MM & 6, 6, 3, 8, 11 & 2, 4, 8, 14\\
3052 4C, 6K, 10II & 11, 5, 9, 13, 3 & 8, 12, 4, 12\\
3063 4C, 6H, 10OO & 3, 6, 6, 9, 9 & 4, 4, 10, 12\\
3069 4C, 6K, 10OO & 9, 6, 6, 15, 3 & 10, 10, 4, 12\\
3088 4C, 6K, 10PP & 24, 0, 27, 0, 3 & 0, 32, 16, 0\\
3089 4C, 6K, 10QQ & 16, 16, 3, 8, 3 & 16, 8, 8, 8\\
3105 4C, 6K, 10XX & 22, 17, 8, 2, 3 & 0, 16, 0, 16\\
3141 4C, 6K, 10SS & 15, 9, 9, 9, 3 & 8, 8, 0, 16\\
3500 4B, 6J, 10FF & 10, 6, 6, 6, 10 & 4, 4, 16, 8\\
3628 4C, 6I, 10II & 6, 5, 9, 8, 8 & 0, 4, 12, 12\\
3645 4C, 6J, 10OO & 4, 6, 6, 10, 8 & 2, 2, 12, 12\\
3665 4C, 6J, 10QQ & 11, 16, 3, 3, 8 & 8, 0, 16, 8\\
3717 4C, 6J, 10SS & 10, 9, 9, 4, 8 & 0, 0, 8, 16\\
3847 4F, 6Q, 10B & 78, 0, 14, 0, 0 & 0, 64, 0, 0\\
3868 4F, 6Q, 10FFF & 66, 1, 1, 12, 0 & 32, 32, 0, 0\\
4438 4F, 6M, 10FFF & 55, 1, 1, 1, 11 & 0, 0, 32, 0\\\hline
\caption{Multiplicities of irreducible $SL(2,5)$ constituents of the adjoint character for $E_8$ when restricted to $SL(2,5)$ subgroups of each fusion pattern.}
\label{tab:mults}
\end{longtable}
\end{center}

Also, we should note that we can clean up a logic error in \cite[Lemma 5.43]{FREY}.  The proof erroneously says that an element of order 4 in an $SL(2,5)$ subgroup $M$ with one of the fusion patterns listed, would force some conjugate of $M$ to be in the $E_7$ subgroup.  The proof is correct in saying that subgroups with these fusion patterns cannot live in the $E_7$ subgroup and also that any such subgroup would have to live in the $A_1E_7$ centralizer of an element of type 2A.  However, there is a list of possible fusion patterns for $SL(2,5)$ subgroups of $A_1E_7$ in \cite[Table XIX]{Fr3}, and none of the fusion patterns mentioned in this Lemma appear on that list, so they were correctly eliminated from consideration.

\subsection{Hom(Dic$_{k-2}, E_8$)}

We attempted to do a mathematical classification of $Dic_2 \rightarrow E_8$ homomorphisms as well.  One might have expected that this task would be easier than the $Alt_5$ and $SL(2,5)$ problem, since $Dic_2$ is much smaller than either $Alt_5$ or $SL(2,5)$. However, other subtleties arise in the dicyclic case, significantly complicating the analysis.  These subtleties arise in the consideration of dicyclic subgroups of the $D_8$ subgroup because of our use of $SO(16,\bC)$ to build embeddings rather than building them directly in $HSpin(16,\bC)$.  The lifting of groups to $Spin(16,\bC)$ produces different behavior in the solvable dicyclic groups than it did in the perfect alternating and $SL(2,5)$ groups, the extent of which we didn't fully realize initially.  Progress has recently been made, some results have been achieved, and we expect to achieve and publish more results in the near future.

\section{Lists of Homomorphisms $\Gamma \rightarrow E_8$}\label{sec:LISTS}

In this appendix, we list the 6D SCFTs corresponding to homomorphisms $\Gamma_G \rightarrow E_8$ for $G=D_4$, $D_5$, $E_6$, $E_7$, and $E_8$. Each homomorphism is labeled by its centralizer, which shows up as the global symmetry of the 6D SCFT.

\subsection{$\Gamma_{D_4} \cong $ Dic$_2$ \label{ssec:D4list}}

$Sp(4) $:
$$
[Sp(4)  ] \,\,  {\overset{\mathfrak{g_2}}2} \,\, 1 \,\,  \overset{\mathfrak{so_8}}4 \,\,    1 \,\, {\overset{\mathfrak{so_8}}4} \,\,  ... [SO(8)]
$$

$SU(6) \times U(1) \times U(1) $:
$$
[SU(6)  ] \,\,  {\overset{\mathfrak{su_3}}2} \,\, \underset{[U(1) \times U(1)]}1 \,\,  \overset{\mathfrak{so_8}}4 \,\,    1 \,\, {\overset{\mathfrak{so_8}}4} \,\,  ... [SO(8)]
$$

$SO(7) \times Sp(2) \text{ or } SO(7) \times SU(2) \times SU(2) \times SU(2)$:
$$
[SO(7)  ] \,\,  {\overset{\mathfrak{su_2}}2} \,\, \underset{[Sp(2) \text{ or } SU(2) \times SU(2) \times SU(2)]}1 \,\,  \overset{\mathfrak{so_8}}4 \,\,    1 \,\, {\overset{\mathfrak{so_8}}4} \,\,  ... [SO(8)]
$$

$SO(8) \times SU(2)$:
$$
[SU(2) ] \,\, 2 \,\,  \underset{[SO(8)]}1 \,\,  \overset{\mathfrak{so_8}}4 \,\,    1 \,\, {\overset{\mathfrak{so_8}}4} \,\,  ... [SO(8)]
$$

$Sp(4) \times SU(2) $:
$$
[Sp(4) \times SU(2) ] \,\,  {\overset{\mathfrak{so_7}}2} \,\, 1 \,\,  \overset{\mathfrak{so_8}}4 \,\,    1 \,\, {\overset{\mathfrak{so_8}}4} \,\,  ... [SO(8)]
$$

$F_4  \times SU(2)$:
$$
[ F_4  ] \,\,  1 \,\, \underset{[SU(2)]}{\overset{\mathfrak{g_2}}3} \,\, 1 \,\,  \overset{\mathfrak{so_8}}4 \,\,    1 \,\, {\overset{\mathfrak{so_8}}4} \,\,  ... [SO(8)]
$$

$SO(13) $:
$$
[SO(13) ] \,\, {\overset{\mathfrak{sp_1}}1}    \,\,   {\overset{\mathfrak{g_2}}3} \,\, 1 \,\,  \overset{\mathfrak{so_8}}4 \,\,    1 \,\, {\overset{\mathfrak{so_8}}4} \,\,  ... [SO(8)]
$$

$SO(9) \times Sp(2) $:
$$
[SO(9) ] \,\, 1  \,\,  \underset{[Sp(2)]}{\overset{\mathfrak{so_7}}3} \,\, 1 \,\,  \overset{\mathfrak{so_8}}4 \,\,    1 \,\, {\overset{\mathfrak{so_8}}4} \,\,  ... [SO(8)]
$$

$Sp(2) \times Sp(2) \times Sp(2)$:
$$
[Sp(2) \times Sp(2) \times Sp(2) ] \,\,  {\overset{\mathfrak{so_8}}2} \,\,1 \,\,  \overset{\mathfrak{so_8}}4 \,\,    1 \,\, {\overset{\mathfrak{so_8}}4} \,\,  ... [SO(8)]
$$

$E_6$:
$$
[ E_6  ] \,\,  1 \,\, {\overset{\mathfrak{su_3}}3} \,\, 1 \,\,  \overset{\mathfrak{so_8}}4 \,\,    1 \,\, {\overset{\mathfrak{so_8}}4} \,\,  ... [SO(8)]
$$

$SO(8) \times SU(2) \times SU(2) \times SU(2) $:
$$
[SO(8) ] \,\, 1  \,\,  \underset{[SU(2) \times SU(2) \times SU(2)]}{\overset{\mathfrak{so_8}}3} \,\, 1 \,\,  \overset{\mathfrak{so_8}}4 \,\,    1 \,\, {\overset{\mathfrak{so_8}}4} \,\,  ... [SO(8)]
$$

$SO(12) \times SU(2) $:
$$
[SO(12) ] \,\, {\overset{\mathfrak{sp_1}}1}    \,\,   \underset{[SU(2)]}{\overset{\mathfrak{so_7}}3} \,\, 1 \,\,  \overset{\mathfrak{so_8}}4 \,\,    1 \,\, {\overset{\mathfrak{so_8}}4} \,\,  ... [SO(8)]
$$

$SO(12) \times SU(2) \times SU(2) $:
$$
[SO(12) ] \,\, {\overset{\mathfrak{sp_1}}1}    \,\,  \underset{[SU(2) \times SU(2)]}{\overset{\mathfrak{so_8}}3} \,\, 1 \,\,  \overset{\mathfrak{so_8}}4 \,\,    1 \,\, {\overset{\mathfrak{so_8}}4} \,\,  ... [SO(8)]
$$

$SU(8)  \times U(1)$:
$$
[SU(8)  ] \,\,  {\overset{\mathfrak{su_4}}2} \,\, \underset{[U(1)]}1 \,\,  \overset{\mathfrak{so_8}}4 \,\,    1 \,\, {\overset{\mathfrak{so_8}}4} \,\,  ... [SO(8)]
$$

$E_7 $:
$$
[E_7] \,\, 1   \,\, {\overset{\mathfrak{su_2}}2} \,\,  {\overset{\mathfrak{g_2}}3} \,\, 1 \,\,  \overset{\mathfrak{so_8}}4 \,\,    1 \,\, {\overset{\mathfrak{so_8}}4} \,\,  ... [SO(8)]
$$

$SO(8) \times SO(8)$:
$$
[SO(8) ] \,\,  1  \,\, \underset{[SO(8)]}{ \underset{1}{\overset{\mathfrak{so_8}}4}} \,\,    1 \,\, {\overset{\mathfrak{so_8}}4} \,\,  ... [SO(8)]
$$

$SO(16) $:
$$
[SO(16) ] \,\, {\overset{\mathfrak{sp_2}}1}    \,\,  {\overset{\mathfrak{so_7}}3} \,\, 1 \,\,  \overset{\mathfrak{so_8}}4 \,\,    1 \,\, {\overset{\mathfrak{so_8}}4} \,\,  ... [SO(8)]
$$

$E_7 \times SU(2)$:
$$
[E_7] \,\, 1   \,\, {\overset{\mathfrak{su_2}}2} \,\, \underset{[SU(2)]}{\overset{\mathfrak{so_7}}3} \,\, 1 \,\,  \overset{\mathfrak{so_8}}4 \,\,    1 \,\, {\overset{\mathfrak{so_8}}4} \,\,  ... [SO(8)]
$$

$E_8$:
$$
[E_8] \,\, 1 \,\, 2  \,\, {\overset{\mathfrak{su_2}}2} \,\,\overset{\mathfrak{g_2}}3 \,\, 1 \,\,  \overset{\mathfrak{so_8}}4 \,\,    1 \,\, {\overset{\mathfrak{so_8}}4} \,\,  ... [SO(8)]
$$

%%%%%%%%%%%%%%%%%%%%%%%%%%%%%%%%%%%%%%%%%%%%%%

\subsection{$\Gamma_{D_5} \cong $ Dic$_3$ \label{ssec:D5list}}

$SO(7) \times SU(2)$:
$$
[SU(2) ] \,\, 2 \,\,  \underset{[SO(7)]}1 \,\,  \overset{\mathfrak{so_9}}4 \,\,    \underset{[N_f=1/2]}{\overset{\mathfrak{sp_1}}1} \,\, {\overset{\mathfrak{so_{10}}}4} \,\,  ... [SO(10)]
$$

$Sp(3) \times SU(2)$:

$$
[Sp(3)] \,\, {\overset{\mathfrak{g_{2}}}2} \,\,   \underset{[SU(2)]}{\overset{\mathfrak{sp_1}}1} \,\,  \overset{\mathfrak{so_{10}}}4 \,\, ... [SO(10)]
$$

$Sp(3) \times SU(2) \times U(1)$:

$$
[Sp(3) \times SU(2)] \,\, {\overset{\mathfrak{so_{7}}}2} \,\,   \underset{[U(1)]}{\overset{\mathfrak{sp_1}}1} \,\,  \overset{\mathfrak{so_{10}}}4 \,\, ... [SO(10)]
$$

$Sp(4) $:
$$
[Sp(4)  ] \,\,  {\overset{\mathfrak{g_2}}2} \,\, 1 \,\,  \overset{\mathfrak{so_9}}4 \,\,    \underset{[N_f=1/2]}{\overset{\mathfrak{sp_1}}1} \,\, {\overset{\mathfrak{so_{10}}}4} \,\,  ... [SO(10)]
$$

$SO(9)$:
$$
2 \,\,      \underset{[SO(9)]}{\overset{\mathfrak{sp_1}}1} \,\,  \overset{\mathfrak{so_{10}}}4 \,\, ... [SO(10)]
$$

$SO(7)  \times SU(2) \times SU(2) $:
$$
[SO(7)  ] \,\,  {\overset{\mathfrak{su_2}}2} \,\, \underset{[ SU(2) \times SU(2)]}1 \,\,  \overset{\mathfrak{so_9}}4 \,\,    \underset{[N_f=1/2]}{\overset{\mathfrak{sp_1}}1} \,\, {\overset{\mathfrak{so_{10}}}4} \,\,  ... [SO(10)]
$$

$$
[SO(7)] \,\,  1   \,\,   \underset{[SU(2) \times SU(2)]}{\overset{\mathfrak{so_{9}}}3} \,\,    \underset{[N_f=1/2]}{\overset{\mathfrak{sp_1}}1} \,\,  \overset{\mathfrak{so_{10}}}4 \,\, ... [SO(10)]
$$

$SO(7)  \times SU(2) \times SU(2) \times U(1) $:

$$
[SO(7) ] \,\,  {\overset{\mathfrak{su_2}}2} \,\,  \underset{[SU(2) \times U(1)]}1 \,\, \underset{[SU(2)]}{\overset{\mathfrak{so_{10}}}4} \,\,  ... [SO(10)]
$$

$SU(4) \times SU(2) \times SU(2)$:
$$
[SU(4)] \,\, {\overset{\mathfrak{su_{3}}}2} \,\,      \underset{[SU(2) \times SU(2)]}{\overset{\mathfrak{sp_1}}1} \,\,  \overset{\mathfrak{so_{10}}}4 \,\, ... [SO(10)]
$$

$$
[SU(2) \times SU(2)] \,\, {\overset{\mathfrak{su_{2}}}2} \,\,      \underset{[SU(4)]}{\overset{\mathfrak{sp_1}}1} \,\,  \overset{\mathfrak{so_{10}}}4 \,\, ... [SO(10)]
$$

$$
[SU(2) ] \,\,  2 \,\,  \underset{[SU(4)]}1 \,\, \underset{[SU(2)]}{\overset{\mathfrak{so_{10}}}4} \,\,  ... [SO(10)]
$$

$SU(4) \times Sp(2)$:
$$
[SU(4)] \,\, 1    \,\,   \underset{[Sp(2)]}{\overset{\mathfrak{so_{10}}}3} \,\,   {\overset{\mathfrak{sp_1}}1} \,\,  \overset{\mathfrak{so_{10}}}4 \,\, ... [SO(10)]
$$

$SU(6) $:
$$
[SU(6)  ] \,\,  {\overset{\mathfrak{su_3}}2} \,\, 1 \,\,  \overset{\mathfrak{so_9}}4 \,\,    \underset{[N_f=1/2]}{\overset{\mathfrak{sp_1}}1} \,\, {\overset{\mathfrak{so_{10}}}4} \,\,  ... [SO(10)]
$$

$SU(6) \times U(1)$:

$$
[SU(6)] \,\, {\overset{\mathfrak{su_{4}}}2} \,\,      \underset{[U(1)]}{\overset{\mathfrak{sp_1}}1} \,\,  \overset{\mathfrak{so_{10}}}4 \,\, ... [SO(10)]
$$

$Sp(4) \times SU(2) $:
$$
[Sp(4) \times SU(2) ] \,\,  {\overset{\mathfrak{so_7}}2} \,\, 1 \,\,  \overset{\mathfrak{so_9}}4 \,\,    \underset{[N_f=1/2]}{\overset{\mathfrak{sp_1}}1} \,\, {\overset{\mathfrak{so_{10}}}4} \,\,  ... [SO(10)]
$$

$$
[Sp(4) \times SU(2)] \,\, {\overset{\mathfrak{so_{10}}}2} \,\,   {\overset{\mathfrak{sp_1}}1} \,\,  \overset{\mathfrak{so_{10}}}4 \,\, ... [SO(10)]
$$

$$
[Sp(4)] \,\, {\overset{\mathfrak{so_{7}}}2} \,\,   \underset{[SU(2)]}{\overset{\mathfrak{sp_1}}1} \,\,  \overset{\mathfrak{so_{10}}}4 \,\, ... [SO(10)]
$$

$Sp(3) \times Sp(2)$:
$$
[Sp(3) \times Sp(2)] \,\, {\overset{\mathfrak{so_{9}}}2} \,\,   \underset{[N_f=1/2]}{\overset{\mathfrak{sp_1}}1} \,\,  \overset{\mathfrak{so_{10}}}4 \,\, ... [SO(10)]
$$

$Sp(2) \times Sp(2) \times SU(2)$:
$$
[Sp(2) \times Sp(2) \times SU(2)] \,\, {\overset{\mathfrak{so_{8}}}2} \,\,   \underset{[N_f=1]}{\overset{\mathfrak{sp_1}}1} \,\,  \overset{\mathfrak{so_{10}}}4 \,\, ... [SO(10)]
$$

$SO(9) \times SU(2) \times U(1)$:
$$
[SO(9)] \,\,  1  \,\,  \underset{[SU(2)]}{\overset{\mathfrak{so_{7}}}3} \,\,   \underset{[U(1)]}{\overset{\mathfrak{sp_1}}1} \,\,  \overset{\mathfrak{so_{10}}}4 \,\, ... [SO(10)]
$$

$F_4  \times SU(2)$:
$$
[ F_4  ] \,\,  1 \,\, \underset{[SU(2)]}{\overset{\mathfrak{g_2}}3} \,\, 1 \,\,  \overset{\mathfrak{so_9}}4 \,\,    \underset{[N_f=1/2]}{\overset{\mathfrak{sp_1}}1} \,\, {\overset{\mathfrak{so_{10}}}4} \,\,  ... [SO(10)]
$$

$$
[F_4] \,\,  1  \,\,  {\overset{\mathfrak{g_{2}}}3} \,\,   \underset{[SU(2)]}{\overset{\mathfrak{sp_1}}1} \,\,  \overset{\mathfrak{so_{10}}}4 \,\, ... [SO(10)]
$$

$SO(11) \times SU(2)$:
$$
[SO(11)] \,\,  {\overset{\mathfrak{sp_1}}1}    \,\,   \underset{[SU(2)]}{\overset{\mathfrak{so_{9}}}3} \,\,    \underset{[N_f=1/2]}{\overset{\mathfrak{sp_1}}1} \,\,  \overset{\mathfrak{so_{10}}}4 \,\, ... [SO(10)]
$$

$SO(8) \times SU(2) \times SU(2) \times U(1)$
$$
[SO(8)] \,\, 1    \,\,   \underset{[SU(2) \times SU(2)]}{\overset{\mathfrak{so_{8}}}3} \,\,   \underset{[U(1)]}{\overset{\mathfrak{sp_1}}1} \,\,  \overset{\mathfrak{so_{10}}}4 \,\, ... [SO(10)]
$$

$SO(12)$:
$$
[SO(12)] \,\,  {\overset{\mathfrak{sp_1}}1}    \,\,  {\overset{\mathfrak{so_{7}}}3} \,\,   \underset{[N_f=1]}{\overset{\mathfrak{sp_1}}1} \,\,  \overset{\mathfrak{so_{10}}}4 \,\, ... [SO(10)]
$$

$SO(13) $:
$$
[SO(13) ] \,\, {\overset{\mathfrak{sp_1}}1}    \,\,   {\overset{\mathfrak{g_2}}3} \,\, 1 \,\,  \overset{\mathfrak{so_9}}4 \,\,    \underset{[N_f=1/2]}{\overset{\mathfrak{sp_1}}1} \,\, {\overset{\mathfrak{so_{10}}}4} \,\,  ... [SO(10)]
$$

$SO(9) \times Sp(2) $:
$$
[SO(9) ] \,\, 1  \,\,  \underset{[Sp(2)]}{\overset{\mathfrak{so_7}}3} \,\, 1 \,\,  \overset{\mathfrak{so_9}}4 \,\,    \underset{[N_f=1/2]}{\overset{\mathfrak{sp_1}}1} \,\, {\overset{\mathfrak{so_{10}}}4} \,\,  ... [SO(10)]
$$

$SO(7) \times SO(7)$:
$$
[SO(7) ] \,\,  1  \,\, \underset{[SO(7)]}{ \underset{1}{\overset{\mathfrak{so_9}}4}} \,\,    \underset{[N_f=1/2]}{\overset{\mathfrak{sp_1}}1} \,\, {\overset{\mathfrak{so_{10}}}4} \,\,  ... [SO(10)]
$$

$SO(10) \times SU(2)$:
$$
[SO(10)] \,\, \overset{\mathfrak{sp_1}}1    \,\,   \underset{[SU(2)]}{\overset{\mathfrak{so_{10}}}3} \,\,   {\overset{\mathfrak{sp_1}}1} \,\,  \overset{\mathfrak{so_{10}}}4 \,\, ... [SO(10)]
$$

$E_6$:
$$
[ E_6  ] \,\,  1 \,\, {\overset{\mathfrak{su_3}}3} \,\, 1 \,\,  \overset{\mathfrak{so_9}}4 \,\,    \underset{[N_f=1/2]}{\overset{\mathfrak{sp_1}}1} \,\, {\overset{\mathfrak{so_{10}}}4} \,\,  ... [SO(10)]
$$

$SU(6) \times SU(2) $:
$$
[SU(6) ] \,\,  {\overset{\mathfrak{su_3}}2} \,\,  1 \,\, \underset{[SU(2)]}{\overset{\mathfrak{so_{10}}}4} \,\,  ... [SO(10)]
$$

$SO(12) \times SU(2) $:
$$
[SO(12) ] \,\, {\overset{\mathfrak{sp_1}}1}    \,\,   \underset{[SU(2)]}{\overset{\mathfrak{so_7}}3} \,\, 1 \,\,  \overset{\mathfrak{so_9}}4 \,\,    \underset{[N_f=1/2]}{\overset{\mathfrak{sp_1}}1} \,\, {\overset{\mathfrak{so_{10}}}4} \,\,  ... [SO(10)]
$$

$SO(12) \times SU(2) \times U(1)$:

$$
[SO(12)] \,\,  {\overset{\mathfrak{sp_1}}1}    \,\,   \underset{[SU(2)]}{\overset{\mathfrak{so_{8}}}3} \,\,   \underset{[U(1)]}{\overset{\mathfrak{sp_1}}1} \,\,  \overset{\mathfrak{so_{10}}}4 \,\, ... [SO(10)]
$$

$SU(4) \times SU(4) \times SU(2)$:
$$
[SU(4) ] \,\,  1  \,\, \underset{[SU(2)]}{\overset{[SU(4)]}{ \overset{1}{\overset{\mathfrak{so_{10}}}4}}} \,\,    1 \,\, {\overset{\mathfrak{so_{10}}}4} \,\,  ... [SO(10)]
$$

$SU(8) $:
$$
[SU(8)  ] \,\,  {\overset{\mathfrak{su_4}}2} \,\, 1 \,\,  \overset{\mathfrak{so_9}}4 \,\,    \underset{[N_f=1/2]}{\overset{\mathfrak{sp_1}}1} \,\, {\overset{\mathfrak{so_{10}}}4} \,\,  ... [SO(10)]
$$

$$
[SU(8)] \,\, {\overset{\mathfrak{su_{5}}}2} \,\,   {\overset{\mathfrak{sp_1}}1} \,\,  \overset{\mathfrak{so_{10}}}4 \,\, ... [SO(10)]
$$

$SO(14) \times U(1)$:
$$
[SO(14)] \,\, \overset{\mathfrak{sp_2}}1    \,\,   \underset{[U(1)]}{\overset{\mathfrak{so_{10}}}3} \,\,   {\overset{\mathfrak{sp_1}}1} \,\,  \overset{\mathfrak{so_{10}}}4 \,\, ... [SO(10)]
$$

$SO(10) \times SU(4)$:
$$
[SO(10) ] \,\,  \overset{\mathfrak{sp_1}}1  \,\, \underset{[SU(4)]}{ \underset{1}{\overset{\mathfrak{so_{10}}}4}} \,\,    1 \,\, {\overset{\mathfrak{so_{10}}}4} \,\,  ... [SO(10)]
$$

$SO(16) $:
$$
[SO(16) ] \,\, {\overset{\mathfrak{sp_2}}1}    \,\,  {\overset{\mathfrak{so_7}}3} \,\, 1 \,\,  \overset{\mathfrak{so_9}}4 \,\,    \underset{[N_f=1/2]}{\overset{\mathfrak{sp_1}}1} \,\, {\overset{\mathfrak{so_{10}}}4} \,\,  ... [SO(10)]
$$

$E_6 \times SU(2) \times U(1)$:
$$
[E_6] \,\,  1  \,\,  {\overset{\mathfrak{su_{3}}}3} \,\,  \underset{[U(1)]}1 \,\,   \underset{[SU(2)]}{\overset{\mathfrak{so_{10}}}4} \,\, ... [SO(10)]
$$

$E_7$:
$$
[E_7] \,\, 1    \,\, {\overset{\mathfrak{su_2}}2} \,\,\overset{\mathfrak{g_2}}3 \,\, 1 \,\,  \overset{\mathfrak{so_9}}4 \,\,    \underset{[N_f=1/2]}{\overset{\mathfrak{sp_1}}1} \,\, {\overset{\mathfrak{so_{10}}}4} \,\,  ... [SO(10)]
$$

$E_7 \times U(1)$:

$$
[E_7] \,\, 1    \,\, {\overset{\mathfrak{su_2}}2} \,\,\overset{\mathfrak{so_7}}3 \,\, \underset{[U(1)]}{\overset{\mathfrak{sp_1}}1} \,\,  \overset{\mathfrak{so_{10}}}4 \,\, ... [SO(10)]
$$

$SU(8) \times SU(2) $:
$$
[SU(8) ] \,\,  {\overset{\mathfrak{su_4}}2} \,\,  1 \,\, \underset{[SU(2)]}{\overset{\mathfrak{so_{10}}}4} \,\,  ... [SO(10)]
$$

$E_7 \times SU(2)$:
$$
[E_7] \,\, 1   \,\, {\overset{\mathfrak{su_2}}2} \,\, \underset{[SU(2)]}{\overset{\mathfrak{so_7}}3} \,\, 1 \,\,  \overset{\mathfrak{so_9}}4 \,\,    \underset{[N_f=1/2]}{\overset{\mathfrak{sp_1}}1} \,\, {\overset{\mathfrak{so_{10}}}4} \,\,  ... [SO(10)]
$$

$E_8$:
$$
[E_8] \,\, 1 \,\, 2  \,\, {\overset{\mathfrak{su_2}}2} \,\,\overset{\mathfrak{g_2}}3 \,\, 1 \,\,  \overset{\mathfrak{so_9}}4 \,\,    \underset{[N_f=1/2]}{\overset{\mathfrak{sp_1}}1} \,\, {\overset{\mathfrak{so_{10}}}4} \,\,  ... [SO(10)]
$$

%%%%%%%%%%%%%%%%%%%%%%%%%%%%%%%%%%%%%%%%%%%%%%

\subsection{$\Gamma_{E_6} \cong SL(2,3)$ \label{ssec:E6list}}

$SU(3)$:
$$
[SU(3)] \,\,  {\overset{\mathfrak{e_6}}3} \,\,    1 \,\, {\overset{\mathfrak{su_3}}3} \,\, 1\,\, \overset{\mathfrak{e_6}}{6} \,\, ...[E_6]
$$

$Sp(2)$:
$$
[Sp(2)] \,\,  {\overset{\mathfrak{f_4}}3} \,\,    1 \,\, {\overset{\mathfrak{su_3}}3} \,\, 1\,\, \overset{\mathfrak{e_6}}{6} \,\, ...[E_6]
$$

$G_2 \times U(1)$:
$$
  2 \,\, \underset{[G_2]}{\overset{\mathfrak{su_2}}2} \,\, \underset{[U(1)]}1 \,\, \overset{\mathfrak{e_6}}{6} \,\, ...[E_6]
$$

$SU(2) \times SU(2) \times SU(2) \times U(1)$:

$$
[SU(2) \times SU(2)] \,\,{\overset{\mathfrak{su_2}}2} \,\, \underset{[SU(2)]}{\overset{\mathfrak{su_2}}2} \,\,  \underset{[U(1)]}1 \,\, \overset{\mathfrak{e_6}}{6} \,\, ...[E_6]
$$

$SU(2) \times SU(2) \times SU(2) \times U(1) \times U(1)$:
$$
[SU(2) \times SU(2) \times SU(2)] \,\,  {\overset{\mathfrak{so_{8}}}3} \,\,     \underset{[U(1) \times U(1)]}1 \,\, {\overset{\mathfrak{su_3}}3} \,\, 1\,\, \overset{\mathfrak{e_6}}{6} \,\, ...[E_6]
$$

$SU(3) \times SU(2)$:

$$
[SU(2)] \,\, {\overset{\mathfrak{g_{2}}}3} \,\,     \underset{[SU(3)]}1 \,\, {\overset{\mathfrak{su_3}}3} \,\, 1\,\, \overset{\mathfrak{e_6}}{6} \,\, ...[E_6]
$$

$$
[SU(2)] \,\, 2 \,\,  2  \,\,  \underset{[SU(3)]}1 \,\, \overset{\mathfrak{e_6}}{6} \,\, ...[E_6]
$$

$SU(3) \times SU(2) \times U(1)$:

$$
[SU(2)] \,\, 2\,\,  \underset{[SU(3)]}1 \,\,  \underset{[U(1)]}{\overset{\mathfrak{e_6}}5} \,\,    1 \,\, {\overset{\mathfrak{su_3}}3} \,\, 1\,\, \overset{\mathfrak{e_6}}{6} \,\, ...[E_6]
$$

$$
[SU(3)] \,\, 1 \,\,  \underset{[SU(2) \times U(1)]}{\overset{\mathfrak{e_6}}4} \,\,    1 \,\, {\overset{\mathfrak{su_3}}3} \,\, 1\,\, \overset{\mathfrak{e_6}}{6} \,\, ...[E_6]
$$

$Sp(2)\times SU(2) \times U(1)$:
$$
[Sp(2) \times SU(2)] \,\,  {\overset{\mathfrak{so_{9}}}3} \,\,  \underset{[U(1)]}1   \,\, {\overset{\mathfrak{su_3}}3} \,\, 1\,\, \overset{\mathfrak{e_6}}{6} \,\, ...[E_6]
$$

$$
[Sp(2)] \,\, {\overset{\mathfrak{so_{7}}}3} \,\,     \underset{[SU(2) \times U(1)]}1 \,\, {\overset{\mathfrak{su_3}}3} \,\, 1\,\, \overset{\mathfrak{e_6}}{6} \,\, ...[E_6]
$$

$Sp(3) \times U(1) \times U(1)$:
$$
[Sp(3) \times U(1)] \,\,  {\overset{\mathfrak{so_{10}}}3} \,\,    \underset{[U(1)]}1 \,\, {\overset{\mathfrak{su_3}}3} \,\, 1\,\, \overset{\mathfrak{e_6}}{6} \,\, ...[E_6]
$$

$G_2 \times SU(2)$:
$$
[SU(2)] \,\, 2 \,\, \underset{[G_2]}1 \,\,  \overset{\mathfrak{f_4}}5 \,\,    1 \,\, {\overset{\mathfrak{su_3}}3} \,\, 1\,\, \overset{\mathfrak{e_6}}{6} \,\, ...[E_6]
$$

$$
[G_2] 1 \,\,   \underset{[SU(2)]}{\overset{\mathfrak{f_4}}4} \,\,    1 \,\, {\overset{\mathfrak{su_3}}3} \,\, 1\,\, \overset{\mathfrak{e_6}}{6} \,\, ...[E_6]
$$

$SU(4) \times U(1)$:
$$
[U(1)] \,\, {\overset{\mathfrak{su_2}}2} \,\, \underset{[SU(4)]}{\overset{\mathfrak{su_3}}2} \,\, 1\,\, \overset{\mathfrak{e_6}}{6} \,\, ...[E_6]
$$

$SU(4) \times U(1) \times U(1)$:

$$
[SU(4)] \,\,{\overset{\mathfrak{su_3}}2} \,\, \underset{[U(1)]}{\overset{\mathfrak{su_2}}2} \,\, \underset{[U(1)]}1\,\, \overset{\mathfrak{e_6}}{6} \,\, ...[E_6]
$$

$SO(7) \times U(1) \times U(1)$:
$$
[SO(7)] \,\,  {\overset{\mathfrak{su_2}}2} \,\,  \underset{[U(1)]}1 \,\,  \underset{[U(1)]}{\overset{\mathfrak{e_6}}5} \,\,    1 \,\, {\overset{\mathfrak{su_3}}3} \,\, 1\,\, \overset{\mathfrak{e_6}}{6} \,\, ...[E_6]
$$

$SO(7) \times SU(2)$:
$$
[SO(7)] \,\,  {\overset{\mathfrak{su_2}}2} \,\, \underset{[SU(2)]}1 \,\,  \overset{\mathfrak{f_4}}5 \,\,    1 \,\, {\overset{\mathfrak{su_3}}3} \,\, 1\,\, \overset{\mathfrak{e_6}}{6} \,\, ...[E_6]
$$

$SO(7) \times SU(2) \times U(1)$:

$$
[SO7)] \,\,  1  \,\, \underset{[SU(2)]}{\overset{\mathfrak{so_{9}}}4} \,\,  \underset{[U(1)]}1   \,\, {\overset{\mathfrak{su_3}}3} \,\, 1\,\, \overset{\mathfrak{e_6}}{6} \,\, ...[E_6]
$$

$SU(3) \times SU(3) \times U(1)$:
$$
[SU(3)] 1 \,\,   \underset{[U(1)]}{\overset{[SU(3)]}{\overset{1}{\overset{\mathfrak{e_6}}5}}} \,\,    1 \,\, {\overset{\mathfrak{su_3}}3} \,\, 1\,\, \overset{\mathfrak{e_6}}{6} \,\, ...[E_6]
$$

$$
[SU(3) \times U(1)] \,\,  {\overset{\mathfrak{su_3}}2} \,\, \underset{[SU(3)]}{\overset{\mathfrak{su_3}}2} \,\, 1\,\, \overset{\mathfrak{e_6}}{6} \,\, ...[E_6]
$$

$SU(3)\times SU(3)$ or $G_2$:\footnote{Note that $SU(3) \times SU(3) \times SU(3)$ and $SU(3) \times G_2$ are both maximal subgroups of $E_6$, so it is unclear which one will be leftover after gauging $SU(3) \times SU(3) \subset SU(3) \times E_6 \subset E_8$).}
$$
 {\overset{\mathfrak{su_{3}}}3} \,\,    \underset{[SU(3) \times SU(3) \mbox{ or } G_2]}1 \,\, {\overset{\mathfrak{su_3}}3} \,\, 1\,\, \overset{\mathfrak{e_6}}{6} \,\, ...[E_6]
$$

$SO(8) \times U(1) \times U(1)$:
$$
[SO(8)] 1 \,\,  {\overset{\mathfrak{so_8}}4} \,\,    \underset{[U(1) \times U(1)]}1  \,\, {\overset{\mathfrak{su_3}}3} \,\, 1\,\, \overset{\mathfrak{e_6}}{6} \,\, ...[E_6]
$$

$Sp(4)$:
$$
[Sp(4)] \,\,  {\overset{\mathfrak{g_2}}2} \,\, 1 \,\,  \overset{\mathfrak{f_4}}5 \,\,    1 \,\, {\overset{\mathfrak{su_3}}3} \,\, 1\,\, \overset{\mathfrak{e_6}}{6} \,\, ...[E_6]
$$

$G_2 \times SU(3)$:
$$
[G_2] \,\,  {\overset{\mathfrak{su_2}}2} \,\,  2  \,\,  \underset{[SU(3)]}1 \,\, \overset{\mathfrak{e_6}}{6} \,\, ...[E_6]
$$

$G_2 \times G_2$:
$$
[G_2] 1 \,\,   \overset{[G_2]}{\overset{1}{\overset{\mathfrak{f_4}}5}} \,\,    1 \,\, {\overset{\mathfrak{su_3}}3} \,\, 1\,\, \overset{\mathfrak{e_6}}{6} \,\, ...[E_6]
$$

$F_4 \times U(1)$:
$$
[F_4] \,\,1  \,\,  {\overset{\mathfrak{g_2}}3} \,\, {\overset{\mathfrak{su_2}}2}  \,\, \underset{[U(1)]}1 \,\, \overset{\mathfrak{e_6}}{6} \,\, ...[E_6]
$$

$SU(4) \times Sp(2)$:
$$
[SU(4))] \,\, 1 \,\, \underset{[Sp(2)]}{\overset{\mathfrak{so_{10}}}4} \,\,    1 \,\, {\overset{\mathfrak{su_3}}3} \,\, 1\,\, \overset{\mathfrak{e_6}}{6} \,\, ...[E_6]
$$

$SU(6)$:

$$
[SU(6)] \,\,  {\overset{\mathfrak{su_3}}2} \,\, 1 \,\,  \overset{\mathfrak{f_4}}5 \,\,    1 \,\, {\overset{\mathfrak{su_3}}3} \,\, 1\,\, \overset{\mathfrak{e_6}}{6} \,\, ...[E_6]
$$

$SU(6) \times U(1)$:

$$
[SU(6)] \,\,  {\overset{\mathfrak{su_4}}2} \,\, {\overset{\mathfrak{su_2}}2} \,\, \underset{[U(1)]}1 \,\, \overset{\mathfrak{e_6}}{6} \,\, ...[E_6]
$$

$$
[SU(6)] \,\,  {\overset{\mathfrak{su_3}}2} \,\, 1 \,\,   \underset{[U(1)]}{\overset{\mathfrak{e_6}}5}  \,\,    1 \,\, {\overset{\mathfrak{su_3}}3} \,\, 1\,\, \overset{\mathfrak{e_6}}{6} \,\, ...[E_6]
$$

$SU(5) \times SU(2) \times U(1)$:
$$
[SU(5) \times U(1)] \,\,  {\overset{\mathfrak{su_4}}2} \,\, \underset{[SU(2)]}{\overset{\mathfrak{su_3}}2} \,\, 1\,\, \overset{\mathfrak{e_6}}{6} \,\, ...[E_6]
$$

$SO(7) \times SU(3) \times U(1)$:
$$
[SO(7)] \,\,  {\overset{\mathfrak{su_2}}2}  \,\,  \underset{[U(1)]}1 \,\,   \overset{[SU(3)]}{\overset{1}{\overset{\mathfrak{e_6}}{6}}} \,\, ...[E_6]
$$

$SO(9) \times SU(2) \times U(1)$:
$$
[SO(9)] \,\,1  \,\, \underset{[SU(2)]}{\overset{\mathfrak{so_7}}3} \,\, {\overset{\mathfrak{su_2}}2}  \,\,  \underset{[U(1)]}1 \,\, \overset{\mathfrak{e_6}}{6} \,\, ...[E_6]
$$

$SU(3) \times SU(3) \times SU(2)$:
$$
[SU(2)] \,\,  2 \,\,  \underset{[SU(3)]}1 \,\,   \overset{[SU(3)]}{\overset{1}{\overset{\mathfrak{e_6}}{6}}} \,\, ...[E_6]
$$

$SO(11)$:
$$
[SO(11)] \,\, {\overset{\mathfrak{sp_1}}1} \,\, {\overset{\mathfrak{so_{9}}}4} \,\,    1 \,\, {\overset{\mathfrak{su_3}}3} \,\, 1\,\, \overset{\mathfrak{e_6}}{6} \,\, ...[E_6]
$$

$SO(10) \times SU(2)$:
$$
[SO(10)] \,\, {\overset{\mathfrak{sp_1}}1} \,\, \underset{[SU(2)]}{\overset{\mathfrak{so_{10}}}4} \,\,    1 \,\, {\overset{\mathfrak{su_3}}3} \,\, 1\,\, \overset{\mathfrak{e_6}}{6} \,\, ...[E_6]
$$

$F_4 \times SU(2)$:
$$
[F_4] \,\, 1  \,\, \underset{[SU(2)]}{\overset{\mathfrak{g_2}}3} \,\, 1 \,\,  \overset{\mathfrak{f_4}}5 \,\,    1 \,\, {\overset{\mathfrak{su_3}}3} \,\, 1\,\, \overset{\mathfrak{e_6}}{6} \,\, ...[E_6]
$$

$SU(3) \times SU(3) \times SU(3)$:
$$
[SU(3))]\,\, 1 \,\,   \underset{[SU(3)]}{\underset{1}{\overset{[SU(3)]}{\overset{1}{\overset{\mathfrak{e_6}}{6}}}}} \,\, ...[E_6]
$$

$SU(7) \times U(1)$:
$$
[SU(7)] \,\,  {\overset{\mathfrak{su_5}}2} \,\, \underset{[U(1)]}{\overset{\mathfrak{su_3}}2} \,\, 1\,\, \overset{\mathfrak{e_6}}{6} \,\, ...[E_6]
$$

$SO(12) \times U(1)$:
$$
[SO(12)] \,\, {\overset{\mathfrak{sp_1}}1}  \,\, {\overset{\mathfrak{so_7}}3} \,\, {\overset{\mathfrak{su_2}}2}  \,\, \underset{[U(1)]}1  \,\, \overset{\mathfrak{e_6}}{6} \,\, ...[E_6]
$$

$E_6$:
$$
[E_6] \,\, 1  \,\,\overset{\mathfrak{su_3}}3 \,\, 1 \,\,  \overset{\mathfrak{f_4}}5 \,\,    1 \,\, {\overset{\mathfrak{su_3}}3} \,\, 1\,\, \overset{\mathfrak{e_6}}{6} \,\, ...[E_6]
$$

$E_6 \times U(1)$:

$$
[E_6] \,\, 1  \,\,\overset{\mathfrak{su_3}}3 \,\, 1 \,\,  \underset{[U(1)]}{\overset{\mathfrak{e_6}}5} \,\,    1 \,\, {\overset{\mathfrak{su_3}}3} \,\, 1\,\, \overset{\mathfrak{e_6}}{6} \,\, ...[E_6]
$$

$SO(14)$:
$$
[SO(14)] \,\, {\overset{\mathfrak{sp_2}}1} \,\, {\overset{\mathfrak{so_{10}}}4} \,\,    \underset{[U(1)]}1 \,\, {\overset{\mathfrak{su_3}}3} \,\, 1\,\, \overset{\mathfrak{e_6}}{6} \,\, ...[E_6]
$$

$SU(6) \times SU(3)$:
$$
[SU(6)] \,\,  {\overset{\mathfrak{su_3}}2}  \,\,  1 \,\,   \overset{[SU(3)]}{\overset{1}{\overset{\mathfrak{e_6}}{6}}} \,\, ...[E_6]
$$

$E_7$:
$$
[E_7] \,\, 1 \,\, {\overset{\mathfrak{su_2}}2}  \,\,\overset{\mathfrak{g_2}}3 \,\, 1 \,\,  \overset{\mathfrak{f_4}}5 \,\,    1 \,\, {\overset{\mathfrak{su_3}}3} \,\, 1\,\, \overset{\mathfrak{e_6}}{6} \,\, ...[E_6]
$$

$SU(9)$:
$$
[SU(9)] \,\,  {\overset{\mathfrak{su_6}}2} \,\, {\overset{\mathfrak{su_3}}2} \,\, 1\,\, \overset{\mathfrak{e_6}}{6} \,\, ...[E_6]
$$

$E_6 \times SU(3)$:
$$
[E_6] \,\, 1 \,\,  {\overset{\mathfrak{su_3}}3} \,\, 1 \,\,   \overset{[SU(3)]}{\overset{1}{\overset{\mathfrak{e_6}}{6}}} \,\, ...[E_6]
$$

$E_8$:
$$
[E_8] \,\, 1 \,\, 2  \,\, {\overset{\mathfrak{su_2}}2} \,\,\overset{\mathfrak{g_2}}3 \,\, 1 \,\,  \overset{\mathfrak{f_4}}5 \,\,    1 \,\, {\overset{\mathfrak{su_3}}3} \,\, 1\,\, \overset{\mathfrak{e_6}}{6} \,\, ...[E_6]
$$

%%%%%%%%%%%%%%%%%%%%%%%%%%%%%%%%%%%%%%%%%%%%%

\subsection{$\Gamma_{E_7} $ \label{ssec:E7list}}

$SU(2) \times U(1) \times U(1)$:
$$
[U(1)]\,\,{\overset{\mathfrak{su_2}}2} \,\, \underset{[SU(2)]}{\overset{\mathfrak{su_3}}2} \,\, \underset{[U(1)]}{\overset{\mathfrak{su_2}}2} \,\,  1\,\, \overset{\mathfrak{e_7}}{8} \,\, ...[E_7]
$$

$$
[SU(2) \times U(1)]  \,\, \overset{\mathfrak{e_6}}4 \,\, \underset{[U(1)]}1 \,\, {\overset{\mathfrak{su_2}}2} \,\,\overset{\mathfrak{so_7}}3 \,\, {\overset{\mathfrak{su_2}}2}  \,\, 1\,\, {\overset{\mathfrak{e_7}}{8}} \,\, ...[E_7]
$$

$SU(2) \times SU(2)$:

$$
 [SU(2)]  \,\, {\overset{\mathfrak{g_{2}}}3} \,\,  \underset{[SU(2)]}1 \,\, {\overset{\mathfrak{g_2}}3} \,\, {\overset{\mathfrak{su_2}}2}  \,\, 1\,\, \overset{\mathfrak{e_7}}{8} \,\, ...[E_7]
$$

$$
2 \,\, \overset{\mathfrak{su_2}}2 \,\, \underset{[SU(2) \times SU(2)]}{\overset{\mathfrak{su_2}}2} \,\,  1\,\, \overset{\mathfrak{e_7}}{8} \,\, ...[E_7]
$$

$$
[SU(2)]  \,\,  2 \,\, 2 \,\, 2 \,\,  \underset{[SU(2)]}1\,\, \overset{\mathfrak{e_7}}{8} \,\, ...[E_7]
$$

$$
[SU(2)] \,\,   2 \,\, 2 \,\,  \underset{[SU(2)]}1\,\,  \underset{[N_f=1/2]}{\overset{\mathfrak{e_7}}{7}}  \,\, ...[E_7]
$$

$$
[SU(2)  ]  \,\, 1 \,\, \underset{[SU(2)]}{\overset{\mathfrak{e_7}}{5}} \,\, ...[E_7]
$$

$$
[SU(2)  \times SU(2)]   \overset{\mathfrak{e_7}}{4} \,\, ...[E_7]
$$

$$
[SU(2)]  \,\,  \overset{\mathfrak{f_4}}4 \,\,\underset{[SU(2)]}1 \,\, {\overset{\mathfrak{su_2}}2} \,\,\overset{\mathfrak{so_7}}3 \,\, {\overset{\mathfrak{su_2}}2}  \,\, 1\,\, {\overset{\mathfrak{e_7}}{8}} \,\, ...[E_7]
$$

$SU(2) \times SU(2) \times U(1)$:

$$
[SU(2)]  \,\,   {\overset{\mathfrak{g_{2}}}3} \,\,   \underset{[U(1)]}1 \,\,\underset{[SU(2)]}{\overset{\mathfrak{so_7}}3} \,\, {\overset{\mathfrak{su_2}}2}  \,\, 1\,\, \overset{\mathfrak{e_7}}{8} \,\, ...[E_7]
$$

$$
  {\overset{\mathfrak{su_{3}}}3} \,\,   \underset{[SU(2) \times U(1)]}1 \,\,\underset{[SU(2)]}{\overset{\mathfrak{so_7}}3} \,\, {\overset{\mathfrak{su_2}}2}  \,\, 1\,\, \overset{\mathfrak{e_7}}{8} \,\, ...[E_7]
$$

$$
[SU(2)  ]  \,\, 1 \,\,\overset{[SU(2)]}{\overset{1}{\underset{[U(1)]}{\overset{\mathfrak{e_7}}{6}}}} \,\, ...[E_7]
$$

$SU(3)$:
$$
  {\overset{\mathfrak{su_{3}}}3} \,\,  \underset{[SU(3)]}1 \,\, {\overset{\mathfrak{g_2}}3} \,\, {\overset{\mathfrak{su_2}}2}  \,\, 1\,\, \overset{\mathfrak{e_7}}{8} \,\, ...[E_7]
$$

$SU(3) \times U(1) \times U(1)$:

$$
[SU(3)]  \,\, {\overset{\mathfrak{su_3}}2} \,\, \underset{[U(1)]}{\overset{\mathfrak{su_3}}2} \,\, \underset{[U(1)]}{\overset{\mathfrak{su_2}}2} \,\,  1\,\, \overset{\mathfrak{e_7}}{8} \,\, ...[E_7]
$$

$$
[SU(3)]  \,\, 1\,\, {\underset{[U(1)]}{\overset{\mathfrak{e_6}}5}} \,\, \underset{[U(1)]}1  \,\, {\overset{\mathfrak{su_2}}2} \,\,\overset{\mathfrak{so_7}}3 \,\, {\overset{\mathfrak{su_2}}2}  \,\, 1\,\, {\overset{\mathfrak{e_7}}{8}} \,\, ...[E_7]
$$

$Sp(2)$:

$$
  {\overset{\mathfrak{g_{2}}}3} \,\,    \underset{[Sp(2)]}{\overset{\mathfrak{sp_1}}1} \,\,{\overset{\mathfrak{so_7}}3} \,\, {\overset{\mathfrak{su_2}}2}  \,\, 1\,\, \overset{\mathfrak{e_7}}{8} \,\, ...[E_7]
$$

$$
 [Sp(2)] \,\, {\overset{\mathfrak{f_{4}}}3} \,\,  1 \,\, {\overset{\mathfrak{g_2}}3} \,\, {\overset{\mathfrak{su_2}}2}  \,\, 1\,\, \overset{\mathfrak{e_7}}{8} \,\, ...[E_7]
$$

$$
 {\overset{\mathfrak{su_2}}2} \,\, \underset{[Sp(2) ]}{\overset{\mathfrak{g_2}}2} \,\, \overset{\mathfrak{su_2}}2 \,\,  1\,\, \overset{\mathfrak{e_7}}{8} \,\, ...[E_7]
$$

$Sp(2)  \times U(1)$:
$$
[Sp(2)]  \,\,   {\overset{\mathfrak{so_{10}}}3} \,\,    \underset{[U(1)]}{\overset{\mathfrak{sp_1}}1} \,\,{\overset{\mathfrak{so_7}}3} \,\, {\overset{\mathfrak{su_2}}2}  \,\, 1\,\, \overset{\mathfrak{e_7}}{8} \,\, ...[E_7]
$$

$$
 [Sp(2)]  \,\, {\overset{\mathfrak{so_{7}}}3} \,\,   \underset{[U(1)]}1 \,\, {\overset{\mathfrak{g_2}}3} \,\, {\overset{\mathfrak{su_2}}2}  \,\, 1\,\, \overset{\mathfrak{e_7}}{8} \,\, ...[E_7]
$$

$G_2$:
$$
  2 \,\,  \underset{[G_2]}{\overset{\mathfrak{su_2}}2} \,\, 1\,\,  \underset{[N_f=1/2]}{\overset{\mathfrak{e_7}}{7}} \,\, ...[E_7]
$$

$SU(2) \times SU(2) \times SU(2)$:
$$
[SU(2) \times SU(2)]  \,\,   {\overset{\mathfrak{so_{9}}}3} \,\,    \underset{[SU(2)]}{\overset{\mathfrak{sp_1}}1} \,\,{\overset{\mathfrak{so_7}}3} \,\, {\overset{\mathfrak{su_2}}2}  \,\, 1\,\, \overset{\mathfrak{e_7}}{8} \,\, ...[E_7]
$$

$$
[SU(2)]  \,\,   {\overset{\mathfrak{so_{7}}}3} \,\,    \underset{[SU(2) \times SU(2)]}{\overset{\mathfrak{sp_1}}1} \,\,{\overset{\mathfrak{so_7}}3} \,\, {\overset{\mathfrak{su_2}}2}  \,\, 1\,\, \overset{\mathfrak{e_7}}{8} \,\, ...[E_7]
$$

$$
 [SU(2) \times SU(2) \times SU(2)] \,\, {\overset{\mathfrak{so_{8}}}3} \,\,  1 \,\, {\overset{\mathfrak{g_2}}3} \,\, {\overset{\mathfrak{su_2}}2}  \,\, 1\,\, \overset{\mathfrak{e_7}}{8} \,\, ...[E_7]
$$

$$
[SU(2)\times SU(2)]  \,\, {\overset{\mathfrak{su_2}}2} \,\, {\overset{\mathfrak{su_2}}2} \,\, \underset{[SU(2)]}{\overset{\mathfrak{su_2}}2} \,\,  1\,\, \overset{\mathfrak{e_7}}{8} \,\, ...[E_7]
$$

$$
[SU(2)\times SU(2)]  \,\, \overset{\mathfrak{su_2}}2 \,\, \overset{\mathfrak{su_2}}2 \,\, 2 \,\,  \underset{[SU(2)]}1\,\, \overset{\mathfrak{e_7}}{8} \,\, ...[E_7]
$$

$$
 [SU(2)] \,\, 2 \,\,  2 \,\, \underset{[SU(2)]}1\,\, \underset{[SU(2)]}{\underset{1}{\overset{\mathfrak{e_7}}{8}}} \,\, ...[E_7]
$$

$$
[SU(2) \times SU(2) ]  \,\,   {\overset{\mathfrak{su_2}}2} \,\, \underset{[SU(2)]}{\overset{\mathfrak{su_2}}2}  \,\, 1\,\, \underset{[N_f=1/2]}{\overset{\mathfrak{e_7}}{7}}\,\, ...[E_7]
$$

$$
[SU(2) ]  \,\,   2  \,\, \underset{[SU(2)]}1 \,\, \underset{[SU(2)]}{\underset{1}{\overset{\mathfrak{e_7}}{7}}} \,\, ...[E_7]
$$

$$
[SU(2)  ]  \,\, 1 \,\,\overset{[SU(2)]}{\overset{1}{\underset{[SU(2)]}{\underset{1}{\overset{\mathfrak{e_7}}{7}}}}} \,\, ...[E_7]
$$

$$
[SU(2)]  \,\,  \overset{\mathfrak{so_{9}}}4 \,\,\underset{[SU(2) \times SU(2)]}1 \,\, {\overset{\mathfrak{su_2}}2} \,\,\overset{\mathfrak{so_7}}3 \,\, {\overset{\mathfrak{su_2}}2}  \,\, 1\,\, {\overset{\mathfrak{e_7}}{8}} \,\, ...[E_7]
$$

$Sp(2) $ or $SU(2) \times SU(2) \times SU(2)$:
$$
 \overset{\mathfrak{so_{8}}}4 \,\,\underset{[Sp(2) \mbox{ or } SU(2) \times SU(2) \times SU(2)]}1 \,\, {\overset{\mathfrak{su_2}}2} \,\,\overset{\mathfrak{so_7}}3 \,\, {\overset{\mathfrak{su_2}}2}  \,\, 1\,\, {\overset{\mathfrak{e_7}}{8}} \,\, ...[E_7]
$$

$SU(3) \times SU(2)$:

$$
 2 \,\, \underset{[SU(3)]}{\overset{\mathfrak{su_2}}2} \,\, 2 \,\,  \underset{[SU(2)]}1\,\, \overset{\mathfrak{e_7}}{8} \,\, ...[E_7]
$$

$SU(3) \times SU(2) \times U(1)$:

$$
[SU(2) \times U(1)]  \,\,  {\overset{\mathfrak{su_3}}2} \,\, \underset{[SU(3)]}{\overset{\mathfrak{su_4}}2} \,\, \overset{\mathfrak{su_2}}2 \,\,  1\,\, \overset{\mathfrak{e_7}}{8} \,\, ...[E_7]
$$

$$
[SU(2)] 2 \,\, \underset{[SU(3)]}1\,\, \overset{\mathfrak{e_6}}6 \,\,\underset{[U(1)]}1 \,\, {\overset{\mathfrak{su_2}}2} \,\,\overset{\mathfrak{so_7}}3 \,\, {\overset{\mathfrak{su_2}}2}  \,\, 1\,\, {\overset{\mathfrak{e_7}}{8}} \,\, ...[E_7]
$$

$SU(4)$:
$$
 {\overset{\mathfrak{su_2}}2} \,\, \underset{[SU(4)]}{\overset{\mathfrak{su_4}}2} \,\, \overset{\mathfrak{su_2}}2 \,\,  1\,\, \overset{\mathfrak{e_7}}{8} \,\, ...[E_7]
$$

$SU(4) \times U(1)$:

$$
[SU(4)] \,\,  {\overset{\mathfrak{su_3}}2} \,\, \underset{[U(1)]}{\overset{\mathfrak{su_2}}2}  \,\, 1\,\, \underset{[N_f=1/2]}{\overset{\mathfrak{e_7}}{7}} \,\, ...[E_7]
$$

$Sp(2) \times SU(2)$:
$$
 [Sp(2) \times SU(2)] \,\, {\overset{\mathfrak{so_{9}}}3} \,\,  1 \,\, {\overset{\mathfrak{g_2}}3} \,\, {\overset{\mathfrak{su_2}}2}  \,\, 1\,\, \overset{\mathfrak{e_7}}{8} \,\, ...[E_7]
$$

$$
 {\overset{\mathfrak{su_2}}2} \,\, \underset{[Sp(2) \times SU(2)]}{\overset{\mathfrak{so_7}}2} \,\, \overset{\mathfrak{su_2}}2 \,\,  1\,\, \overset{\mathfrak{e_7}}{8} \,\, ...[E_7]
$$

$Sp(2) \times SU(2) \times U(1)$:

$$
[Sp(2)]  \,\,   {\overset{\mathfrak{so_{7}}}3} \,\,   \underset{[U(1)]}1 \,\,\underset{[SU(2)]}{\overset{\mathfrak{so_7}}3} \,\, {\overset{\mathfrak{su_2}}2}  \,\, 1\,\, \overset{\mathfrak{e_7}}{8} \,\, ...[E_7]
$$

$$
[Sp(2)]  \,\,  \overset{\mathfrak{so_{10}}}4 \,\,\underset{[SU(2) \times U(1)]}1 \,\, {\overset{\mathfrak{su_2}}2} \,\,\overset{\mathfrak{so_7}}3 \,\, {\overset{\mathfrak{su_2}}2}  \,\, 1\,\, {\overset{\mathfrak{e_7}}{8}} \,\, ...[E_7]
$$

$SO(7) \times U(1)$:
$$
[SO(7)] \overset{\mathfrak{su_2}}2 \,\, \underset{[U(1)]}1\,\, \overset{\mathfrak{e_6}}6 \,\,1 \,\, {\overset{\mathfrak{su_2}}2} \,\,\overset{\mathfrak{so_7}}3 \,\, {\overset{\mathfrak{su_2}}2}  \,\, 1\,\, {\overset{\mathfrak{e_7}}{8}} \,\, ...[E_7]
$$

$Sp(3) $:
$$
[Sp(3)]  \,\,   {\overset{\mathfrak{so_{11}}}3} \,\,    \underset{[N_f=1/2]}{\overset{\mathfrak{sp_1}}1} \,\,{\overset{\mathfrak{so_7}}3} \,\, {\overset{\mathfrak{su_2}}2}  \,\, 1\,\, \overset{\mathfrak{e_7}}{8} \,\, ...[E_7]
$$

$$
[Sp(3)] \,\,  {\overset{\mathfrak{g_2}}2} \,\, {\overset{\mathfrak{su_2}}2}  \,\, 1\,\,  \underset{[N_f=1/2]}{\overset{\mathfrak{e_7}}{7}}  \,\, ...[E_7]
$$

$G_2 \times SU(2)$:
$$
  2 \,\,  \underset{[G_2]}{\overset{\mathfrak{su_2}}2} \,\, 1\,\, \underset{[SU(2)]}{\underset{1}{\overset{\mathfrak{e_7}}{8}}} \,\, ...[E_7]
$$

$$
[G_2 ]  \,\,   {\overset{\mathfrak{su_2}}2} \,\,2  \,\, \underset{[SU(2)]}1\,\, \underset{[N_f=1/2]}{\overset{\mathfrak{e_7}}{7}}\,\, ...[E_7]
$$

$$
[G_2]  \,\, 1\,\,  \overset{\mathfrak{f_4}}5 \,\,\underset{[SU(2)]}1 \,\, {\overset{\mathfrak{su_2}}2} \,\,\overset{\mathfrak{so_7}}3 \,\, {\overset{\mathfrak{su_2}}2}  \,\, 1\,\, {\overset{\mathfrak{e_7}}{8}} \,\, ...[E_7]
$$

$$
[SU(2)] \,\, 2 \,\, \underset{[G_2]}1 \,\,  \overset{\mathfrak{f_4}}5 \,\,    1 \,\, {\overset{\mathfrak{g_2}}3}\,\, {\overset{\mathfrak{su_2}}2}  \,\, 1\,\, \overset{\mathfrak{e_7}}{8} \,\, ...[E_7]
$$

$$
[G_2]  \,\,  1 \,\,  \underset{[SU(2)]}{\overset{\mathfrak{f_{4}}}4} \,\,    1 \,\, {\overset{\mathfrak{g_2}}3}\,\, {\overset{\mathfrak{su_2}}2}  \,\, 1\,\, \overset{\mathfrak{e_7}}{8} \,\, ...[E_7]
$$

$SU(2) \times SU(2) \times SU(2) \times SU(2)$:
$$
[SU(2) \times SU(2)]  \,\,   {\overset{\mathfrak{so_{8}}}3} \,\,    \underset{[SU(2) \times SU(2)]}{\overset{\mathfrak{sp_1}}1} \,\,{\overset{\mathfrak{so_7}}3} \,\, {\overset{\mathfrak{su_2}}2}  \,\, 1\,\, \overset{\mathfrak{e_7}}{8} \,\, ...[E_7]
$$

$$
[SU(2) \times SU(2) \times SU(2)]  \,\,   {\overset{\mathfrak{so_{8}}}3} \,\,   1 \,\,\underset{[SU(2)]}{\overset{\mathfrak{so_7}}3} \,\, {\overset{\mathfrak{su_2}}2}  \,\, 1\,\, \overset{\mathfrak{e_7}}{8} \,\, ...[E_7]
$$

$$
[SU(2) \times SU(2) ]  \,\,   {\overset{\mathfrak{su_2}}2} \,\, \underset{[SU(2)]}{\overset{\mathfrak{su_2}}2}  \,\, 1\,\, \underset{[SU(2)]}{\underset{1}{\overset{\mathfrak{e_7}}{8}}} \,\, ...[E_7]
$$

$$
[SU(2) ]  \,\, 2  \,\, \underset{[SU(2)]}1\,\,\underset{[SU(2)]}{\underset{2}{\underset{[SU(2)]}{\underset{1}{\overset{\mathfrak{e_7}}{8}}}}} \,\, ...[E_7]
$$

$$
[SU(2)  ]  \,\, 2   \,\, \underset{[SU(2)]}1\,\,\overset{[SU(2)]}{\overset{1}{\underset{[SU(2)]}{\underset{1}{\overset{\mathfrak{e_7}}{8}}}}} \,\, ...[E_7]
$$

$$
[SU(2)  ]  \,\,  1\,\,  \overset{[SU(2)]\,[SU(2)]}{\overset{1\,\,\,\,\,1}{\underset{[SU(2)]}{\underset{1}{\overset{\mathfrak{e_7}}{8}}}}} \,\, ...[E_7]
$$

$SU(4) \times SU(2)$:

$$
[SU(4)]  \,\, {\overset{\mathfrak{su_3}}2} \,\, {\overset{\mathfrak{su_2}}2} \,\, 2 \,\,  \underset{[SU(2)]}1\,\, \overset{\mathfrak{e_7}}{8} \,\, ...[E_7]
$$

$SU(4) \times SU(2) \times U(1)$:
$$
[SU(4) \times U(1)]  \,\,  {\overset{\mathfrak{su_4}}2} \,\, \underset{[SU(2)]}{\overset{\mathfrak{su_4}}2} \,\, \overset{\mathfrak{su_2}}2 \,\,  1\,\, \overset{\mathfrak{e_7}}{8} \,\, ...[E_7]
$$

$$
[SU(4)] \,\,  {\overset{\mathfrak{su_3}}2} \,\, \underset{[U(1)]}{\overset{\mathfrak{su_2}}2}  \,\, 1\,\, \underset{[SU(2)]}{\underset{1}{\overset{\mathfrak{e_7}}{8}}} \,\, ...[E_7]
$$

$$
[SU(4)]  \,\,  1 \,\,   \underset{[SU(2)]}{\overset{\mathfrak{so_{10}}}4} \,\,    \underset{[U(1)]}{\overset{\mathfrak{sp_1}}1} \,\, {\overset{\mathfrak{so_7}}3}\,\, {\overset{\mathfrak{su_2}}2}  \,\, 1\,\, \overset{\mathfrak{e_7}}{8} \,\, ...[E_7]
$$

$SU(3) \times SU(3) \times U(1)$:
$$
[SU(3)]  \,\, 1\,\, \underset{[SU(3)]}{\underset{1}{\overset{\mathfrak{e_6}}6}} \,\,  \underset{[U(1)]}1 \,\, {\overset{\mathfrak{su_2}}2} \,\,\overset{\mathfrak{so_7}}3 \,\, {\overset{\mathfrak{su_2}}2}  \,\, 1\,\, {\overset{\mathfrak{e_7}}{8}} \,\, ...[E_7]
$$

$Sp(2) \times SU(2) \times SU(2)$:
$$
 [Sp(2) \times SU(2)]\,\, {\overset{\mathfrak{so_{9}}}3} \,\,  1 \,\, \underset{[SU(2)]}{\overset{\mathfrak{so_7}}3} \,\, {\overset{\mathfrak{su_2}}2}  \,\, 1\,\, \overset{\mathfrak{e_7}}{8} \,\, ...[E_7]
$$

$Sp(2) \times Sp(2)$:
$$
[Sp(2)]  \,\,  1 \,\,  \underset{[Sp(2)]}{\overset{\mathfrak{so_{11}}}4} \,\,    \underset{[N_f=1/2]}{\overset{\mathfrak{sp_1}}1} \,\, {\overset{\mathfrak{so_7}}3}\,\, {\overset{\mathfrak{su_2}}2}  \,\, 1\,\, \overset{\mathfrak{e_7}}{8} \,\, ...[E_7]
$$

$G_2 \times SU(2) \times SU(2)$:
$$
[G_2 ]  \,\,   {\overset{\mathfrak{su_2}}2} \,\,2  \,\, \underset{[SU(2)]}1\,\, \underset{[SU(2)]}{\underset{1}{\overset{\mathfrak{e_7}}{8}}} \,\, ...[E_7]
$$

$SU(5) \times U(1)$:
$$
[SU(5)]  \,\, {\overset{\mathfrak{su_4}}2} \,\, {\overset{\mathfrak{su_3}}2} \,\, \underset{[U(1)]}{\overset{\mathfrak{su_2}}2} \,\,  1\,\, \overset{\mathfrak{e_7}}{8} \,\, ...[E_7]
$$

$SO(7) \times SU(2)$:
$$
[SO(7)]  \,\,  1 \,\,  {\overset{\mathfrak{so_{9}}}4} \,\,    \underset{[SU(2)]}{\overset{\mathfrak{sp_1}}1} \,\, {\overset{\mathfrak{so_7}}3}\,\, {\overset{\mathfrak{su_2}}2}  \,\, 1\,\, \overset{\mathfrak{e_7}}{8} \,\, ...[E_7]
$$

$$
[SO(7)]  \,\, 1 \,\,   \underset{[SU(2)]}{\overset{\mathfrak{so_{9}}}4} \,\,    1 \,\, {\overset{\mathfrak{g_2}}3}\,\, {\overset{\mathfrak{su_2}}2}  \,\, 1\,\, \overset{\mathfrak{e_7}}{8} \,\, ...[E_7]
$$

$$
[SO(7) ]  \,\,   {\overset{\mathfrak{su_2}}2}   \,\, 1 \,\, \underset{[SU(2)]}{\underset{1}{\overset{\mathfrak{e_7}}{7}}} \,\, ...[E_7]
$$

$$
[SO(7)] \,\, {\overset{\mathfrak{su_2}}2} \,\, \underset{[SU(2)]}1 \,\,  \overset{\mathfrak{f_4}}5 \,\,    1 \,\, {\overset{\mathfrak{g_2}}3}\,\, {\overset{\mathfrak{su_2}}2}  \,\, 1\,\, \overset{\mathfrak{e_7}}{8} \,\, ...[E_7]
$$

$Sp(3) \times SU(2)$:
$$
[Sp(3)] \,\,  {\overset{\mathfrak{g_2}}2} \,\, {\overset{\mathfrak{su_2}}2}  \,\, 1\,\, \underset{[SU(2)]}{\underset{1}{\overset{\mathfrak{e_7}}{8}}} \,\, ...[E_7]
$$

$$
[Sp(3) \times SU(2)] \,\,   {\overset{\mathfrak{so_7}}2} \,\, {\overset{\mathfrak{su_2}}2}  \,\, 1\,\,  \underset{[N_f=1/2]}{\overset{\mathfrak{e_7}}{7}}  \,\, ...[E_7]
$$

$$
[Sp(3)]  \,\,  \overset{\mathfrak{so_{11}}}4 \,\,\underset{[SU(2)]}1 \,\, {\overset{\mathfrak{su_2}}2} \,\,\overset{\mathfrak{so_7}}3 \,\, {\overset{\mathfrak{su_2}}2}  \,\, 1\,\, {\overset{\mathfrak{e_7}}{8}} \,\, ...[E_7]
$$

$SO(8)$:
$$
[SO(8)]  \,\,  1 \,\,  {\overset{\mathfrak{so_{8}}}4} \,\,    1 \,\,{\overset{\mathfrak{g_2}}3} \,\, {\overset{\mathfrak{su_2}}2}  \,\, 1\,\, \overset{\mathfrak{e_7}}{8} \,\, ...[E_7]
$$

$Sp(4)$:
$$
[Sp(4)] \,\, {\overset{\mathfrak{g_2}}2} \,\, 1 \,\,  \overset{\mathfrak{f_4}}5 \,\,    1 \,\, {\overset{\mathfrak{g_2}}3}\,\, {\overset{\mathfrak{su_2}}2}  \,\, 1\,\, \overset{\mathfrak{e_7}}{8} \,\, ...[E_7]
$$

$$
[Sp(4)]  \,\,   {\overset{\mathfrak{so_{12}}}3} \,\,    {\overset{\mathfrak{sp_1}}1} \,\,{\overset{\mathfrak{so_7}}3} \,\, {\overset{\mathfrak{su_2}}2}  \,\, 1\,\, \overset{\mathfrak{e_7}}{8} \,\, ...[E_7]
$$

$G_2 \times G_2$:
$$
[G_2]  \,\,  1 \,\,  \underset{[G_2]}{\underset{1}{\overset{\mathfrak{f_4}}5}}  \,\,    1 \,\,{\overset{\mathfrak{g_2}}3} \,\, {\overset{\mathfrak{su_2}}2}  \,\, 1\,\, \overset{\mathfrak{e_7}}{8} \,\, ...[E_7]
$$

$F_4$:
$$
[F_4] \,\, 1 \,\,  {\overset{\mathfrak{g_2}}3} \,\, {\overset{\mathfrak{su_2}}2}  \,\, 1\,\, \underset{[N_f=1/2]}{\overset{\mathfrak{e_7}}{7}}  \,\, ...[E_7]
$$

$SU(6)$:

$$
[SU(6)] \,\, {\overset{\mathfrak{su_3}}2} \,\, 1 \,\,  \overset{\mathfrak{f_4}}5 \,\,    1 \,\, {\overset{\mathfrak{g_2}}3}\,\, {\overset{\mathfrak{su_2}}2}  \,\, 1\,\, \overset{\mathfrak{e_7}}{8} \,\, ...[E_7]
$$

$$
[SU(6)] \,\,  {\overset{\mathfrak{su_4}}2} \,\, {\overset{\mathfrak{su_2}}2}  \,\, 1\,\, \underset{[N_f=1/2]}{\overset{\mathfrak{e_7}}{7}} \,\, ...[E_7]
$$

$SU(6) \times U(1)$:

$$
[SU(6)]  \,\,  {\overset{\mathfrak{su_5}}2} \,\, \underset{[U(1)]}{\overset{\mathfrak{su_4}}2} \,\, \overset{\mathfrak{su_2}}2 \,\,  1\,\, \overset{\mathfrak{e_7}}{8} \,\, ...[E_7]
$$

$$
[SU(6)] \overset{\mathfrak{su_3}}2 \,\, 1\,\, \overset{\mathfrak{e_6}}6 \,\, \underset{[U(1)]}1 \,\, {\overset{\mathfrak{su_2}}2} \,\,\overset{\mathfrak{so_7}}3 \,\, {\overset{\mathfrak{su_2}}2}  \,\, 1\,\, {\overset{\mathfrak{e_7}}{8}} \,\, ...[E_7]
$$

$Sp(4) \times SU(2) $:
$$
[Sp(4)]  \,\,  \overset{\mathfrak{so_{12}}}4 \,\,\underset{[SU(2)]}1 \,\, {\overset{\mathfrak{su_2}}2} \,\,\overset{\mathfrak{so_7}}3 \,\, {\overset{\mathfrak{su_2}}2}  \,\, 1\,\, {\overset{\mathfrak{e_7}}{8}} \,\, ...[E_7]
$$

$SO(8) \times SU(2) $:
$$
[SO(8)]  \,\,  1 \,\,  {\overset{\mathfrak{so_{8}}}4} \,\,    1 \,\, \underset{[SU(2)]}{\overset{\mathfrak{so_7}}3}\,\, {\overset{\mathfrak{su_2}}2}  \,\, 1\,\, \overset{\mathfrak{e_7}}{8} \,\, ...[E_7]
$$

$SO(7) \times SU(2) \times SU(2)$:
$$
[SO(7)]  \,\,  1 \,\,  \underset{[SU(2)]}{\overset{\mathfrak{so_{9}}}4} \,\,    1 \,\, \underset{[SU(2)]}{\overset{\mathfrak{so_7}}3}\,\, {\overset{\mathfrak{su_2}}2}  \,\, 1\,\, \overset{\mathfrak{e_7}}{8} \,\, ...[E_7]
$$

$$
[SU(2) ]  \,\, 2  \,\, \underset{[SU(2)]}1\,\,\underset{[SO(7)]}{\underset{\mathfrak{su_2}}{\underset{2}{\underset{1}{\overset{\mathfrak{e_7}}{8}}}}} \,\, ...[E_7]
$$

$$
[SO(7) ]  \,\, {\overset{\mathfrak{su_2}}2}    \,\, 1\,\,\overset{[SU(2)]}{\overset{1}{\underset{[SU(2)]}{\underset{1}{\overset{\mathfrak{e_7}}{8}}}}} \,\, ...[E_7]
$$

$Sp(3) \times SU(2) \times SU(2)$:
$$
[SU(2) \times SU(2)]  \,\, 1 \,\,  \underset{[Sp(3)]}{\overset{\mathfrak{so_{12}}}4} \,\,    \overset{\mathfrak{sp_1}}1 \,\, {\overset{\mathfrak{so_7}}3}\,\, {\overset{\mathfrak{su_2}}2}  \,\, 1\,\, \overset{\mathfrak{e_7}}{8} \,\, ...[E_7]
$$

$$
[Sp(3) \times SU(2)] \,\,  {\overset{\mathfrak{so_7}}2} \,\, {\overset{\mathfrak{su_2}}2}  \,\, 1\,\, \underset{[SU(2)]}{\underset{1}{\overset{\mathfrak{e_7}}{8}}} \,\, ...[E_7]
$$

$SO(10) \times U(1)$:
$$
[SO(10)]  \,\,  \overset{\mathfrak{sp_1}}1 \,\,  {\overset{\mathfrak{so_{10}}}4} \,\,    \underset{[U(1)]}{\overset{\mathfrak{sp_1}}1} \,\, {\overset{\mathfrak{so_7}}3}\,\, {\overset{\mathfrak{su_2}}2}  \,\, 1\,\, \overset{\mathfrak{e_7}}{8} \,\, ...[E_7]
$$

$SO(11) $:
$$
[SO(11)]  \,\,  {\overset{\mathfrak{sp_1}}1} \,\,  {\overset{\mathfrak{so_{9}}}4} \,\,    1 \,\, {\overset{\mathfrak{g_2}}3}\,\, {\overset{\mathfrak{su_2}}2}  \,\, 1\,\, \overset{\mathfrak{e_7}}{8} \,\, ...[E_7]
$$

$SO(9) \times SU(2)$:
$$
[SO(9)]  \,\, \overset{\mathfrak{sp_1}}1 \,\,  \underset{[SU(2)]}{\overset{\mathfrak{so_{11}}}4} \,\,    \underset{[N_f=1/2]}{\overset{\mathfrak{sp_1}}1} \,\, {\overset{\mathfrak{so_7}}3}\,\, {\overset{\mathfrak{su_2}}2}  \,\, 1\,\, \overset{\mathfrak{e_7}}{8} \,\, ...[E_7]
$$

$$
[SO(9)] \,\, 1 \,\,  \underset{[SU(2)]}{\overset{\mathfrak{so_7}}3} \,\, {\overset{\mathfrak{su_2}}2}  \,\, 1\,\, \underset{[N_f=1/2]}{\overset{\mathfrak{e_7}}{7}} \,\, ...[E_7]
$$

$F_4 \times SU(2)$:
$$
[F_4] \,\, 1  \,\, \underset{[SU(2)]}{\overset{\mathfrak{g_2}}3} \,\, 1 \,\,  \overset{\mathfrak{f_4}}5 \,\,    1 \,\, {\overset{\mathfrak{g_2}}3}\,\, {\overset{\mathfrak{su_2}}2}  \,\, 1\,\, \overset{\mathfrak{e_7}}{8} \,\, ...[E_7]
$$

$$
[F_4] \,\, 1 \,\,  {\overset{\mathfrak{g_2}}3} \,\, {\overset{\mathfrak{su_2}}2}  \,\, 1\,\, \underset{[SU(2)]}{\underset{1}{\overset{\mathfrak{e_7}}{8}}} \,\, ...[E_7]
$$

$SU(6) \times SU(2)$:
$$
[SU(6)] \,\,  {\overset{\mathfrak{su_4}}2} \,\, {\overset{\mathfrak{su_2}}2}  \,\, 1\,\, \underset{[SU(2)]}{\underset{1}{\overset{\mathfrak{e_7}}{8}}} \,\, ...[E_7]
$$

$SO(9)\times SU(2) \times SU(2)$:
$$
[SO(9)] \,\, 1 \,\,  \underset{[SU(2)]}{\overset{\mathfrak{so_7}}3} \,\, {\overset{\mathfrak{su_2}}2}  \,\, 1\,\, \underset{[SU(2)]}{\underset{1}{\overset{\mathfrak{e_7}}{8}}} \,\, ...[E_7]
$$

$SO(7) \times SO(7) $
$$
[SO(7) ]  \,\, {\overset{\mathfrak{su_2}}2}  \,\, 1\,\, \underset{[SO(7)]}{\underset{\mathfrak{su_2}}{\underset{2}{\underset{1}{\overset{\mathfrak{e_7}}{8}}}}} \,\, ...[E_7]
$$

$SO(11) \times SU(2)$:
$$
[SO(11)]  \,\,  {\overset{\mathfrak{sp_1}}1} \,\,  {\overset{\mathfrak{so_{9}}}4} \,\,   1 \,\, \underset{[SU(2)]}{\overset{\mathfrak{so_7}}3}\,\, {\overset{\mathfrak{su_2}}2}  \,\, 1\,\, \overset{\mathfrak{e_7}}{8} \,\, ...[E_7]
$$

$SO(12)$:
$$
[SO(12)] \,\, \overset{\mathfrak{sp_1}}1 \,\,\overset{\mathfrak{so_7}}3 \,\, {\overset{\mathfrak{su_2}}2}  \,\, 1\,\, \underset{[N_f=1/2]}{\overset{\mathfrak{e_7}}{7}} \,\, ...[E_7]
$$

$SO(13)$:
$$
[SO(13)] \,\, \overset{\mathfrak{sp_1}}1  \,\, {\overset{\mathfrak{g_2}}3} \,\, 1 \,\,  \overset{\mathfrak{f_4}}5 \,\,    1 \,\, {\overset{\mathfrak{g_2}}3}\,\, {\overset{\mathfrak{su_2}}2}  \,\, 1\,\, \overset{\mathfrak{e_7}}{8} \,\, ...[E_7]
$$

$$
[SO(13)]  \,\, \overset{\mathfrak{sp_2}}1 \,\,  {\overset{\mathfrak{so_{11}}}4} \,\,    \underset{[N_f=1/2]}{\overset{\mathfrak{sp_1}}1} \,\, {\overset{\mathfrak{so_7}}3}\,\, {\overset{\mathfrak{su_2}}2}  \,\, 1\,\, \overset{\mathfrak{e_7}}{8} \,\, ...[E_7]
$$

$SO(8) \times Sp(2)$:
$$
[SO(8)]  \,\, \overset{\mathfrak{sp_1}}1 \,\,  \underset{[Sp(2)]}{\overset{\mathfrak{so_{12}}}4} \,\,    \overset{\mathfrak{sp_1}}1 \,\, {\overset{\mathfrak{so_7}}3}\,\, {\overset{\mathfrak{su_2}}2}  \,\, 1\,\, \overset{\mathfrak{e_7}}{8} \,\, ...[E_7]
$$

$E_6$:
$$
[E_6] \,\, 1  \,\, \overset{\mathfrak{su_3}}3 \,\, 1 \,\,  \overset{\mathfrak{f_4}}5 \,\,    1 \,\, {\overset{\mathfrak{g_2}}3}\,\, {\overset{\mathfrak{su_2}}2}  \,\, 1\,\, \overset{\mathfrak{e_7}}{8} \,\, ...[E_7]
$$

$$
[E_6] 1\,\, \overset{\mathfrak{su_3}}3 \,\, 1\,\, \overset{\mathfrak{e_6}}6 \,\,1 \,\, {\overset{\mathfrak{su_2}}2} \,\,\overset{\mathfrak{so_7}}3 \,\, {\overset{\mathfrak{su_2}}2}  \,\, 1\,\, {\overset{\mathfrak{e_7}}{8}} \,\, ...[E_7]
$$

$SU(8)$:
$$
[SU(8)]  \,\,  {\overset{\mathfrak{su_6}}2} \,\, {\overset{\mathfrak{su_4}}2} \,\, \overset{\mathfrak{su_2}}2 \,\,  1\,\, \overset{\mathfrak{e_7}}{8} \,\, ...[E_7]
$$

$SO(12) \times SU(2)$:
$$
[SO(12)]  \,\, \overset{\mathfrak{sp_2}}1 \,\,  \underset{[SU(2)]}{\overset{\mathfrak{so_{12}}}4} \,\,    \overset{\mathfrak{sp_1}}1 \,\, {\overset{\mathfrak{so_7}}3}\,\, {\overset{\mathfrak{su_2}}2}  \,\, 1\,\, \overset{\mathfrak{e_7}}{8} \,\, ...[E_7]
$$

$$
[SO(12)] \,\, \overset{\mathfrak{sp_1}}1 \,\,\overset{\mathfrak{so_7}}3 \,\, {\overset{\mathfrak{su_2}}2}  \,\, 1\,\, \underset{[SU(2)]}{\underset{1}{\overset{\mathfrak{e_7}}{8}}} \,\, ...[E_7]
$$

$E_7$:
$$
[E_7] \,\, 1 \,\, {\overset{\mathfrak{su_2}}2} \,\,\overset{\mathfrak{g_2}}3 \,\, 1 \,\,  \overset{\mathfrak{f_4}}5 \,\,    1 \,\, {\overset{\mathfrak{g_2}}3}\,\, {\overset{\mathfrak{su_2}}2}  \,\, 1\,\, \overset{\mathfrak{e_7}}{8} \,\, ...[E_7]
$$

$$
[E_7] \,\, 1 \,\, {\overset{\mathfrak{su_2}}2} \,\,\overset{\mathfrak{so_7}}3 \,\, {\overset{\mathfrak{su_2}}2}  \,\, 1\,\, \underset{[N_f=1/2]}{\overset{\mathfrak{e_7}}{7}} \,\, ...[E_7]
$$

$E_7 \times SU(2)$:
$$
[E_7] \,\, 1 \,\, {\overset{\mathfrak{su_2}}2} \,\,\overset{\mathfrak{so_7}}3 \,\, {\overset{\mathfrak{su_2}}2}  \,\, 1\,\, \underset{[SU(2)]}{\underset{1}{\overset{\mathfrak{e_7}}{8}}} \,\, ...[E_7]
$$

$SO(16)$:
$$
[SO(16)]  \,\, \overset{\mathfrak{sp_3}}1 \,\,  \overset{\mathfrak{so_{12}}}4 \,\,    \overset{\mathfrak{sp_1}}1 \,\, {\overset{\mathfrak{so_7}}3}\,\, {\overset{\mathfrak{su_2}}2}  \,\, 1\,\, \overset{\mathfrak{e_7}}{8} \,\, ...[E_7]
$$

$E_8$:
$$
[E_8] \,\, 1 \,\, 2  \,\, {\overset{\mathfrak{su_2}}2} \,\,\overset{\mathfrak{g_2}}3 \,\, 1 \,\,  \overset{\mathfrak{f_4}}5 \,\,    1 \,\, {\overset{\mathfrak{g_2}}3}\,\, {\overset{\mathfrak{su_2}}2}  \,\, 1\,\, \overset{\mathfrak{e_7}}{8} \,\, ...[E_7]
$$

%%%%%%%%%%%%%%%%%%%%%%%%%%%%%%%%%%%%%%%%%%%%%

\subsection{$\Gamma_{E_8} \cong SL(2,5)$ \label{ssec:E8list}}

$\emptyset:$

$$
\overset{\mathfrak{e_8}}{(6)} \,\, ...[E_8]
$$

$U(1) \times U(1)$:

$$
 [U(1)] \,\, {\overset{\mathfrak{su_2}}2} \,\, \underset{[U(1)]}{\overset{\mathfrak{su_3}}2} \,\, \underset{[N_f=1]}{\overset{\mathfrak{su_3}}2} \,\, \overset{\mathfrak{su_2}}2 \,\, 2 \,\, 1 \,\, \overset{\mathfrak{e_8}}{(12)} \,\, ...[E_8]
$$

$$
 \overset{\mathfrak{so_8}}4 \,\, \underset{[U(1) \times U(1)]}1 \,\, {\overset{\mathfrak{su_3}}3} \,\, 1 \,\, \overset{\mathfrak{f_4}}5 \,\, 1 \,\, \overset{\mathfrak{g_2}}3 \,\,{\overset{\mathfrak{su_2}}2} \,\, 2 \,\, 1 \,\, \overset{\mathfrak{e_8}}{(12)} \,\, ...[E_8]
$$

$$
\overset{\mathfrak{su_3}}3 \,\, \underset{[U(1) \times U(1)]}1 \,\, {\overset{\mathfrak{so_8}}4} \,\,  1 \,\, {\overset{\mathfrak{g_2}}3} \,\, \overset{\mathfrak{su_2}}2 \,\, 2 \,\, 1 \,\, \overset{\mathfrak{e_8}}{(12)} \,\, ...[E_8]
$$

$$
 [U(1)] \,\, {\overset{\mathfrak{e_6}}5} \,\,  \underset{[U(1)]}1 \,\, {\overset{\mathfrak{su_2}}2} \,\, \overset{\mathfrak{g_2}}3 \,\, 1 \,\, \overset{\mathfrak{f_4}}5 \,\, 1 \,\, \overset{\mathfrak{g_2}}3 \,\,{\overset{\mathfrak{su_2}}2} \,\, 2 \,\, 1 \,\, \overset{\mathfrak{e_8}}{(12)} \,\, ...[E_8]
$$

$SU(2)$:

$$
2 \,\, {\overset{\mathfrak{su_2}}2} \,\, \underset{[SU(2)]}{\overset{\mathfrak{su_3}}2} \,\, \overset{\mathfrak{su_2}}2 \,\, 2 \,\, 1 \,\, \overset{\mathfrak{e_8}}{(12)} \,\, ...[E_8]
$$

$$
[SU(2)] \,\, 2 \,\, 2 \,\, 2 \,\, 2 \,\, 2 \,\, 1 \,\, \overset{\mathfrak{e_8}}{(12)} \,\, ...[E_8]
$$

$$
2 \,\, \underset{[SU(2)]}{\overset{\mathfrak{su_2}}2} \,\, \overset{\mathfrak{su_2}}2 \,\, 2 \,\, 1 \,\, \overset{\mathfrak{e_8}}{(11)} \,\, ...[E_8]
$$

$$
[SU(2)] \,\, 2 \,\, 2 \,\, 2 \,\, 2 \,\, 1 \,\, \overset{\mathfrak{e_8}}{(11)} \,\, ...[E_8]
$$

$$
[SU(2)] \,\, 2 \,\, 2 \,\, 2 \,\, 1 \,\, \overset{\mathfrak{e_8}}{(10)} \,\, ...[E_8]
$$

$$
[SU(2)] \,\, 2 \,\, 2  \,\, 1 \,\, \overset{\mathfrak{e_8}}{(9)} \,\, ...[E_8]
$$

$$
[SU(2)] \,\, 2 \,\, 1 \,\, \overset{\mathfrak{e_8}}{(8)} \,\, ...[E_8]
$$

$$
[SU(2)] \,\,  \overset{\mathfrak{g_2}}3 \,\, 1 \,\, \overset{\mathfrak{so_8}}4 \,\,  1 \,\, {\overset{\mathfrak{g_2}}3} \,\, \overset{\mathfrak{su_2}}2 \,\, 2 \,\, 1 \,\, \overset{\mathfrak{e_8}}{(12)} \,\, ...[E_8]
$$

$$
\overset{\mathfrak{su_3}}3 \,\, 1 \,\, \underset{[SU(2)]}{\overset{\mathfrak{f_4}}4} \,\,  1 \,\, {\overset{\mathfrak{g_2}}3} \,\, \overset{\mathfrak{su_2}}2 \,\, 2 \,\, 1 \,\, \overset{\mathfrak{e_8}}{(12)} \,\, ...[E_8]
$$

$$
\overset{\mathfrak{so_8}}4 \,\, 1 \,\, \underset{[SU(2)]}{\overset{\mathfrak{g_2}}3} \,\, 1 \,\, \overset{\mathfrak{f_4}}5 \,\, 1 \,\, \overset{\mathfrak{g_2}}3 \,\,{\overset{\mathfrak{su_2}}2} \,\, 2 \,\, 1 \,\, \overset{\mathfrak{e_8}}{(12)} \,\, ...[E_8]
$$

$$
[SU(2)] \,\, {\overset{\mathfrak{f_4}}4} \,\, 1 \,\, {\overset{\mathfrak{su_3}}3} \,\, 1 \,\, \overset{\mathfrak{f_4}}5 \,\, 1 \,\, \overset{\mathfrak{g_2}}3 \,\,{\overset{\mathfrak{su_2}}2} \,\, 2 \,\, 1 \,\, \overset{\mathfrak{e_8}}{(12)} \,\, ...[E_8]
$$

$$
\overset{\mathfrak{g_2}}3 \,\, {\overset{\mathfrak{su_2}}2} \,\,  \underset{[SU(2)]}1 \,\, \overset{\mathfrak{f_4}}5 \,\, 1 \,\, \overset{\mathfrak{g_2}}3 \,\,{\overset{\mathfrak{su_2}}2} \,\, 2 \,\, 1 \,\, \overset{\mathfrak{e_8}}{(12)} \,\, ...[E_8]
$$

$$
{\overset{\mathfrak{f_4}}5} \,\, \underset{[SU(2)]}1 \,\, {\overset{\mathfrak{su_2}}2} \,\, \overset{\mathfrak{g_2}}3 \,\, 1 \,\, \overset{\mathfrak{f_4}}5 \,\, 1 \,\, \overset{\mathfrak{g_2}}3 \,\,{\overset{\mathfrak{su_2}}2} \,\, 2 \,\, 1 \,\, \overset{\mathfrak{e_8}}{(12)} \,\, ...[E_8]
$$

$$
[SU(2)]   \,\, {\overset{\mathfrak{e_7}}5} \,\, 1 \,\, {\overset{\mathfrak{su_2}}2} \,\, \overset{\mathfrak{g_2}}3 \,\, 1 \,\, \overset{\mathfrak{f_4}}5 \,\, 1 \,\, \overset{\mathfrak{g_2}}3 \,\,{\overset{\mathfrak{su_2}}2} \,\, 2 \,\, 1 \,\, \overset{\mathfrak{e_8}}{(12)} \,\, ...[E_8]
$$

$SU(2) \times U(1)$:

$$
2 \,\, \underset{[SU(2)]}{\overset{\mathfrak{su_2}}2} \,\, {\overset{\mathfrak{su_2}}2} \,\, \underset{[U(1)]}{\overset{\mathfrak{su_2}}2} \,\, 2 \,\, 1 \,\, \overset{\mathfrak{e_8}}{(12)} \,\, ...[E_8]
$$

$$
[U(1)] \,\, {\overset{\mathfrak{su_2}}2} \,\, \underset{[SU(2)]}{\overset{\mathfrak{su_3}}2} \,\, \overset{\mathfrak{su_2}}2 \,\, 2 \,\, 1 \,\, \overset{\mathfrak{e_8}}{(11)} \,\, ...[E_8]
$$

$$
\overset{\mathfrak{su_3}}3 \,\, \underset{[U(1)]}1 \,\, \underset{[SU(2)]}{\overset{\mathfrak{so_9}}4} \,\,  1 \,\, {\overset{\mathfrak{g_2}}3} \,\, \overset{\mathfrak{su_2}}2 \,\, 2 \,\, 1 \,\, \overset{\mathfrak{e_8}}{(12)} \,\, ...[E_8]
$$

$$
[SU(2)]\,\,  \overset{\mathfrak{so_9}}4 \,\, \underset{[U(1)]}1 \,\, {\overset{\mathfrak{su_3}}3} \,\, 1 \,\, \overset{\mathfrak{f_4}}5 \,\, 1 \,\, \overset{\mathfrak{g_2}}3 \,\,{\overset{\mathfrak{su_2}}2} \,\, 2 \,\, 1 \,\, \overset{\mathfrak{e_8}}{(12)} \,\, ...[E_8]
$$

$$
[SU(2) \times U(1)]\,\,  \overset{\mathfrak{e_6}}4 \,\, 1 \,\, {\overset{\mathfrak{su_3}}3} \,\, 1 \,\, \overset{\mathfrak{f_4}}5 \,\, 1 \,\, \overset{\mathfrak{g_2}}3 \,\,{\overset{\mathfrak{su_2}}2} \,\, 2 \,\, 1 \,\, \overset{\mathfrak{e_8}}{(12)} \,\, ...[E_8]
$$

$$
[SU(2)] \,\, 1\,\, \underset{[U(1)]}{\overset{\mathfrak{e_7}}6} \,\, 1 \,\, {\overset{\mathfrak{su_2}}2} \,\, \overset{\mathfrak{g_2}}3 \,\, 1 \,\, \overset{\mathfrak{f_4}}5 \,\, 1 \,\, \overset{\mathfrak{g_2}}3 \,\,{\overset{\mathfrak{su_2}}2} \,\, 2 \,\, 1 \,\, \overset{\mathfrak{e_8}}{(12)} \,\, ...[E_8]
$$

$$
 [U(1)] \,\,{\overset{\mathfrak{e_7}}6} \,\, \underset{[SU(2)]}1 \,\, 2 \,\, {\overset{\mathfrak{su_2}}2} \,\, \overset{\mathfrak{g_2}}3 \,\, 1 \,\, \overset{\mathfrak{f_4}}5 \,\, 1 \,\, \overset{\mathfrak{g_2}}3 \,\,{\overset{\mathfrak{su_2}}2} \,\, 2 \,\, 1 \,\, \overset{\mathfrak{e_8}}{(12)} \,\, ...[E_8]
$$

$SU(2) \times SU(2)$:

$$
[SU(2) \times SU(2)] \,\, \overset{\mathfrak{su_2}}2  \,\, {\overset{\mathfrak{su_2}}2} \,\, {\overset{\mathfrak{su_2}}2} \,\, \overset{\mathfrak{su_2}}2 \,\, 2 \,\, 1 \,\, \overset{\mathfrak{e_8}}{(12)} \,\, ...[E_8]
$$

$$
[SU(2) \times SU(2)]  \,\, {\overset{\mathfrak{su_2}}2} \,\, {\overset{\mathfrak{su_2}}2} \,\, \overset{\mathfrak{su_2}}2 \,\, 2 \,\, 1 \,\, \overset{\mathfrak{e_8}}{(11)} \,\, ...[E_8]
$$

$$
[SU(2) \times SU(2)]  \,\, {\overset{\mathfrak{su_2}}2} \,\, \overset{\mathfrak{su_2}}2 \,\, 2 \,\, 1 \,\, \overset{\mathfrak{e_8}}{(10)} \,\, ...[E_8]
$$

$$
[SU(2)] \,\, 2 \,\, 2 \,\, 2 \,\, 1 \,\, \overset{[SU(2)]}{\overset{2}{\overset{1}{\overset{\mathfrak{e_8}}{(12)}}}} \,\, ...[E_8]
$$

$$
[SU(2)] \,\, 2 \,\, 2  \,\, 1 \,\, \overset{[SU(2)]}{\overset{2}{\overset{2}{\overset{1}{\overset{\mathfrak{e_8}}{(12)}}}}} \,\, ...[E_8]
$$

$$
[SU(2)] \,\, 2 \,\, 2  \,\, 1 \,\, \overset{[SU(2)]}{\overset{2}{\overset{1}{\overset{\mathfrak{e_8}}{(11)}}}} \,\, ...[E_8]
$$

$$
[SU(2)] \,\, 2 \,\, 1 \,\, \overset{[SU(2)]}{\overset{2}{\overset{1}{\overset{\mathfrak{e_8}}{(10)}}}} \,\, ...[E_8]
$$

$$
[SU(2)] \,\, {\overset{\mathfrak{f_4}}4} \,\, 1 \,\, \underset{[SU(2)]}{\overset{\mathfrak{g_2}}3} \,\, 1 \,\, \overset{\mathfrak{f_4}}5 \,\, 1 \,\, \overset{\mathfrak{g_2}}3 \,\,{\overset{\mathfrak{su_2}}2} \,\, 2 \,\, 1 \,\, \overset{\mathfrak{e_8}}{(12)} \,\, ...[E_8]
$$

$$
[SU(2)]\,\,  \overset{\mathfrak{so_9}}4 \,\, 1 \,\, \underset{[SU(2)]}{\overset{\mathfrak{g_2}}3} \,\, 1 \,\, \overset{\mathfrak{f_4}}5 \,\, 1 \,\, \overset{\mathfrak{g_2}}3 \,\, {\overset{\mathfrak{su_2}}2} \,\, 2 \,\, 1 \,\, \overset{\mathfrak{e_8}}{(12)} \,\, ...[E_8]
$$

$$
[SU(2)] \,\,  \overset{\mathfrak{so_{10}}}4 \,\, \underset{[SU(2)]}{\overset{\mathfrak{sp_1}}1} \,\, {\overset{\mathfrak{g_2}}3} \,\, 1 \,\, \overset{\mathfrak{f_4}}5 \,\, 1 \,\, \overset{\mathfrak{g_2}}3 \,\,{\overset{\mathfrak{su_2}}2} \,\, 2 \,\, 1 \,\, \overset{\mathfrak{e_8}}{(12)} \,\, ...[E_8]
$$

$$
 \overset{\mathfrak{so_{9}}}4 \,\, \underset{[SU(2) \times SU(2)]}{\overset{\mathfrak{sp_1}}1} \,\, {\overset{\mathfrak{g_2}}3} \,\, 1 \,\, \overset{\mathfrak{f_4}}5 \,\, 1 \,\, \overset{\mathfrak{g_2}}3 \,\,{\overset{\mathfrak{su_2}}2} \,\, 2 \,\, 1 \,\, \overset{\mathfrak{e_8}}{(12)} \,\, ...[E_8]
$$

$$
 [SU(2)]\,\,\overset{\mathfrak{g_2}}3 \,\,  \underset{[SU(2)]}1 \,\, \overset{\mathfrak{g_2}}3 \,\,{\overset{\mathfrak{su_2}}2} \,\, 2 \,\, 1 \,\, \overset{\mathfrak{e_8}}{(11)} \,\, ...[E_8]
$$

$$
[SU(2)]\,\,  \overset{\mathfrak{so_7}}3 \,\, \overset{\mathfrak{su_2}}2 \,\,  \underset{[SU(2)]}1 \,\, \overset{\mathfrak{f_4}}5 \,\, 1 \,\, \overset{\mathfrak{g_2}}3 \,\,{\overset{\mathfrak{su_2}}2} \,\, 2 \,\, 1 \,\, \overset{\mathfrak{e_8}}{(12)} \,\, ...[E_8]
$$

$$
 \overset{\mathfrak{g_2}}3 \,\,  \underset{[SU(2) \times SU(2)]}{\overset{\mathfrak{sp_1}}1} \,\, {\overset{\mathfrak{so_9}}4} \,\,1 \,\, {\overset{\mathfrak{g_2}}3} \,\, \overset{\mathfrak{su_2}}2 \,\, 2 \,\, 1 \,\, \overset{\mathfrak{e_8}}{(12)} \,\, ...[E_8]
$$

$$
[SU(2)] \,\, \overset{\mathfrak{so_7}}3 \,\,  \underset{[SU(2)]}{\overset{\mathfrak{sp_1}}1} \,\, {\overset{\mathfrak{so_9}}4} \,\,1 \,\, {\overset{\mathfrak{g_2}}3} \,\, \overset{\mathfrak{su_2}}2 \,\, 2 \,\, 1 \,\, \overset{\mathfrak{e_8}}{(12)} \,\, ...[E_8]
$$

$$
[SU(2)]\,\,  \overset{\mathfrak{g_2}}3 \,\,  1 \,\, \underset{[SU(2)]}{\overset{\mathfrak{f_4}}4} \,\,1 \,\, {\overset{\mathfrak{g_2}}3} \,\, \overset{\mathfrak{su_2}}2 \,\, 2 \,\, 1 \,\, \overset{\mathfrak{e_8}}{(12)} \,\, ...[E_8]
$$

$$
[SU(2)]\,\,  \overset{\mathfrak{g_2}}3 \,\,  1 \,\, \underset{[SU(2)]}{\overset{\mathfrak{so_9}}4} \,\,1 \,\, {\overset{\mathfrak{g_2}}3} \,\, \overset{\mathfrak{su_2}}2 \,\, 2 \,\, 1 \,\, \overset{\mathfrak{e_8}}{(12)} \,\, ...[E_8]
$$

$$
[SU(2)]  \,\,  2 \,\, 2 \,\, \underset{[SU(2)]}1\,\, \overset{\mathfrak{e_7}}8 \,\,1 \,\, {\overset{\mathfrak{su_2}}2} \,\, \overset{\mathfrak{g_2}}3 \,\, 1 \,\, \overset{\mathfrak{f_4}}5 \,\, 1 \,\, \overset{\mathfrak{g_2}}3 \,\,{\overset{\mathfrak{su_2}}2} \,\, 2 \,\, 1 \,\, \overset{\mathfrak{e_8}}{(12)} \,\, ...[E_8]
$$

$$
[SU(2)]  \,\,  2  \,\, \underset{[SU(2)]}1\,\, \underset{[N_f=1/2]}{\overset{\mathfrak{e_7}}7} \,\,1 \,\, {\overset{\mathfrak{su_2}}2} \,\, \overset{\mathfrak{g_2}}3 \,\, 1 \,\, \overset{\mathfrak{f_4}}5 \,\, 1 \,\, \overset{\mathfrak{g_2}}3 \,\,{\overset{\mathfrak{su_2}}2} \,\, 2 \,\, 1 \,\, \overset{\mathfrak{e_8}}{(12)} \,\, ...[E_8]
$$

$$
[SU(2)] \,\, 1\,\, \underset{[N_f=1/2]}{\overset{\mathfrak{e_7}}7} \,\,\underset{[SU(2)]}1 \,\, 2 \,\, {\overset{\mathfrak{su_2}}2} \,\, \overset{\mathfrak{g_2}}3 \,\, 1 \,\, \overset{\mathfrak{f_4}}5 \,\, 1 \,\, \overset{\mathfrak{g_2}}3 \,\,{\overset{\mathfrak{su_2}}2} \,\, 2 \,\, 1 \,\, \overset{\mathfrak{e_8}}{(12)} \,\, ...[E_8]
$$

$$
[SU(2)] \,\, 1\,\, \overset{[SU(2)]}{\overset{1}{\underset{[N_f=1/2]}{\overset{\mathfrak{e_7}}7}}} \,\, 1 \,\, {\overset{\mathfrak{su_2}}2} \,\, \overset{\mathfrak{g_2}}3 \,\, 1 \,\, \overset{\mathfrak{f_4}}5 \,\, 1 \,\, \overset{\mathfrak{g_2}}3 \,\,{\overset{\mathfrak{su_2}}2} \,\, 2 \,\, 1 \,\, \overset{\mathfrak{e_8}}{(12)} \,\, ...[E_8]
$$

$SU(2) \times SU(2) \times U(1)$:

$$
[SU(2) \times U(1)] \,\, \overset{\mathfrak{su_3}}2  \,\, \underset{[SU(2)]}{\overset{\mathfrak{su_4}}2} \,\, {\overset{\mathfrak{su_3}}2} \,\, \overset{\mathfrak{su_2}}2 \,\, 2 \,\, 1 \,\, \overset{\mathfrak{e_8}}{(12)} \,\, ...[E_8]
$$

$$
[SU(2) \times SU(2)] \,\, \overset{\mathfrak{so_9}}3 \,\,  \underset{[U(1)]}{\overset{\mathfrak{sp_1}}1} \,\, {\overset{\mathfrak{so_9}}4} \,\,1 \,\, {\overset{\mathfrak{g_2}}3} \,\, \overset{\mathfrak{su_2}}2 \,\, 2 \,\, 1 \,\, \overset{\mathfrak{e_8}}{(12)} \,\, ...[E_8]
$$

$SU(3)$:

$$
\overset{\mathfrak{su_2}}2  \,\, \underset{[SU(3)]}{\overset{\mathfrak{su_4}}2} \,\, {\overset{\mathfrak{su_3}}2} \,\, \overset{\mathfrak{su_2}}2 \,\, 2 \,\, 1 \,\, \overset{\mathfrak{e_8}}{(12)} \,\, ...[E_8]
$$

$$
 2 \,\, \underset{[SU(3)]}{\overset{\mathfrak{su_2}}2} \,\, 2 \,\, 1 \,\, \overset{\mathfrak{e_8}}{(10)} \,\, ...[E_8]
$$

$$
\overset{\mathfrak{su_3}}3 \,\,  \underset{[SU(3)]}1 \,\, {\overset{\mathfrak{g_2}}3} \,\, {\overset{\mathfrak{su_2}}2} \,\, 2 \,\, 1 \,\, \overset{\mathfrak{e_8}}{(11)} \,\, ...[E_8]
$$

$$
 {\overset{\mathfrak{e_6}}6} \,\, \underset{[SU(3)]}1 \,\, 2 \,\, {\overset{\mathfrak{su_2}}2} \,\, \overset{\mathfrak{g_2}}3 \,\, 1 \,\, \overset{\mathfrak{f_4}}5 \,\, 1 \,\, \overset{\mathfrak{g_2}}3 \,\,{\overset{\mathfrak{su_2}}2} \,\, 2 \,\, 1 \,\, \overset{\mathfrak{e_8}}{(12)} \,\, ...[E_8]
$$

$SU(3) \times U(1)$:

$$
[SU(3)] \,\, \overset{\mathfrak{su_3}}2  \,\, {\overset{\mathfrak{su_3}}2} \,\, \underset{[U(1)]}{\overset{\mathfrak{su_3}}2} \,\, \overset{\mathfrak{su_2}}2 \,\, 2 \,\, 1 \,\, \overset{\mathfrak{e_8}}{(12)} \,\, ...[E_8]
$$

$$
[SU(3)] \,\,  {\overset{\mathfrak{su_3}}2} \,\, \underset{[U(1)]}{\overset{\mathfrak{su_3}}2} \,\, \overset{\mathfrak{su_2}}2 \,\, 2 \,\, 1 \,\, \overset{\mathfrak{e_8}}{(11)} \,\, ...[E_8]
$$

$$
[SU(3)] \,\, 1\,\, {\overset{\mathfrak{e_6}}6} \,\, \underset{[U(1)]}1 \,\, {\overset{\mathfrak{su_2}}2} \,\, \overset{\mathfrak{g_2}}3 \,\, 1 \,\, \overset{\mathfrak{f_4}}5 \,\, 1 \,\, \overset{\mathfrak{g_2}}3 \,\,{\overset{\mathfrak{su_2}}2} \,\, 2 \,\, 1 \,\, \overset{\mathfrak{e_8}}{(12)} \,\, ...[E_8]
$$

$$
[SU(3)] \,\, 1  \,\,  \underset{[U(1)]}{\overset{\mathfrak{e_6}}5} \,\, 1 \,\, \overset{\mathfrak{su_3}}3 \,\, 1 \,\, \overset{\mathfrak{f_4}}5 \,\, 1 \,\, \overset{\mathfrak{g_2}}3 \,\,{\overset{\mathfrak{su_2}}2} \,\, 2 \,\, 1 \,\, \overset{\mathfrak{e_8}}{(12)} \,\, ...[E_8]
$$

$SU(2) \times SU(2) \times SU(2)$:

$$
  [SU(2) \times SU(2)]   \,\, {\overset{\mathfrak{su_2}}2}  \,\, {\overset{\mathfrak{su_2}}2} \,\, 2 \,\, 1 \,\,  \overset{[SU(2)]}{\overset{2}{\overset{1}{\overset{\mathfrak{e_8}}{(12)}}}} \,\, ...[E_8]
$$

$$
[SU(2) \times SU(2) \times SU(2)]\,\,  \overset{\mathfrak{so_8}}3 \,\,  1 \,\, \overset{\mathfrak{g_2}}3 \,\,{\overset{\mathfrak{su_2}}2} \,\, 2 \,\, 1 \,\, \overset{\mathfrak{e_8}}{(11)} \,\, ...[E_8]
$$

$$
  [SU(2) ]   \,\,  2 \,\, 1 \,\, \underset{[SU(2)]}{\underset{2}{\underset{1}{\overset{[SU(2)]}{\overset{2}{\overset{1}{\overset{\mathfrak{e_8}}{(12)}}}}}}} \,\, ...[E_8]
$$

$$
[SU(2) \times SU(2)] \,\,  \overset{\mathfrak{so_8}}3 \,\, \underset{[SU(2)]}{\overset{\mathfrak{sp_1}}1} \,\, \overset{\mathfrak{so_9}}4 \,\,  1 \,\, {\overset{\mathfrak{g_2}}3} \,\, \overset{\mathfrak{su_2}}2 \,\, 2 \,\, 1 \,\, \overset{\mathfrak{e_8}}{(12)} \,\, ...[E_8]
$$

$$
[SU(2) \times SU(2) \times SU(2)] \,\,  \overset{\mathfrak{so_8}}3 \,\, 1 \,\, \overset{\mathfrak{so_8}}4 \,\,  1 \,\, {\overset{\mathfrak{g_2}}3} \,\, \overset{\mathfrak{su_2}}2 \,\, 2 \,\, 1 \,\, \overset{\mathfrak{e_8}}{(12)} \,\, ...[E_8]
$$

$$
[SU(2)]\,\, 2  \,\, \underset{[SU(2)]}1 \,\, \overset{[SU(2)]}{\overset{1}{\overset{\mathfrak{e_7}}8}} \,\,1 \,\, {\overset{\mathfrak{su_2}}2} \,\, \overset{\mathfrak{g_2}}3 \,\, 1 \,\, \overset{\mathfrak{f_4}}5 \,\, 1 \,\, \overset{\mathfrak{g_2}}3 \,\,{\overset{\mathfrak{su_2}}2} \,\, 2 \,\, 1 \,\, \overset{\mathfrak{e_8}}{(12)} \,\, ...[E_8]
$$

$$
[SU(2)]\,\, 1 \,\, \overset{[SU(2)]}{\overset{1}{\overset{\mathfrak{e_7}}8}} \,\,\underset{[SU(2)]}1 \,\,2 \,\, {\overset{\mathfrak{su_2}}2} \,\, \overset{\mathfrak{g_2}}3 \,\, 1 \,\, \overset{\mathfrak{f_4}}5 \,\, 1 \,\, \overset{\mathfrak{g_2}}3 \,\,{\overset{\mathfrak{su_2}}2} \,\, 2 \,\, 1 \,\, \overset{\mathfrak{e_8}}{(12)} \,\, ...[E_8]
$$

$$
[SU(2)] \,\, 2 \,\, \underset{[SU(2)]}1 \,\, \overset{\mathfrak{e_7}}8 \,\,\underset{[SU(2)]}1 \,\,2 \,\, {\overset{\mathfrak{su_2}}2} \,\, \overset{\mathfrak{g_2}}3 \,\, 1 \,\, \overset{\mathfrak{f_4}}5 \,\, 1 \,\, \overset{\mathfrak{g_2}}3 \,\,{\overset{\mathfrak{su_2}}2} \,\, 2 \,\, 1 \,\, \overset{\mathfrak{e_8}}{(12)} \,\, ...[E_8]
$$

$$
[SU(2) \times SU(2)] \,\, \overset{\mathfrak{su_2}}2 \,\, \underset{[SU(2)]}{\overset{\mathfrak{su_2}}2} \,\, 1 \,\, \overset{\mathfrak{e_7}}8 \,\, 1  \,\, {\overset{\mathfrak{su_2}}2} \,\, \overset{\mathfrak{g_2}}3 \,\, 1 \,\, \overset{\mathfrak{f_4}}5 \,\, 1 \,\, \overset{\mathfrak{g_2}}3 \,\,{\overset{\mathfrak{su_2}}2} \,\, 2 \,\, 1 \,\, \overset{\mathfrak{e_8}}{(12)} \,\, ...[E_8]
$$

$$
[SU(2)] \,\, 1\,\, \underset{[SU(2)]}{\underset{1}{\overset{[SU(2)]}{\overset{1}{\overset{\mathfrak{e_7}}8}}}} \,\,1 \,\, {\overset{\mathfrak{su_2}}2} \,\, \overset{\mathfrak{g_2}}3 \,\, 1 \,\, \overset{\mathfrak{f_4}}5 \,\, 1 \,\, \overset{\mathfrak{g_2}}3 \,\,{\overset{\mathfrak{su_2}}2} \,\, 2 \,\, 1 \,\, \overset{\mathfrak{e_8}}{(12)} \,\, ...[E_8]
$$

$SU(3) \times SU(2)$:
$$
 2 \,\, \underset{[SU(3)]}{\overset{\mathfrak{su_2}}2} \,\, 2 \,\, 1 \,\,  \overset{[SU(2)]}{\overset{2}{\overset{1}{\overset{\mathfrak{e_8}}{(12)}}}} \,\, ...[E_8]
$$

$$
[SU(2)] \,\, 2  \,\, \underset{[SU(3)]}1 \,\, {\overset{\mathfrak{e_6}}6} \,\, 1  \,\, \overset{\mathfrak{su_3}}3 \,\, 1 \,\, \overset{\mathfrak{f_4}}5 \,\, 1 \,\, \overset{\mathfrak{g_2}}3 \,\,{\overset{\mathfrak{su_2}}2} \,\, 2 \,\, 1 \,\, \overset{\mathfrak{e_8}}{(12)} \,\, ...[E_8]
$$

$Sp(2)$:

$$
[Sp(2)] \,\, \overset{\mathfrak{so_{7}}}3 \,\,1 \,\, \overset{\mathfrak{so_{8}}}4 \,\, 1 \,\, {\overset{\mathfrak{g_2}}3} \,\, \overset{\mathfrak{su_2}}2 \,\, 2 \,\, 1 \,\, \overset{\mathfrak{e_8}}{(12)} \,\, ...[E_8]
$$

$$
[Sp(2)] \,\, \overset{\mathfrak{f_{4}}}3  \,\, 1 \,\, {\overset{\mathfrak{g_2}}3} \,\, \overset{\mathfrak{su_2}}2 \,\, 2 \,\, 1 \,\, \overset{\mathfrak{e_8}}{(11)} \,\, ...[E_8]
$$

$$
 \overset{\mathfrak{su_2}}2 \,\, \underset{[Sp(2)]}{\overset{\mathfrak{g_2}}2} \,\, \overset{\mathfrak{su_2}}2 \,\, 2 \,\, 1 \,\, \overset{\mathfrak{e_8}}{(11)} \,\, ...[E_8]
$$

$$
 2 \,\, \overset{\mathfrak{su_2}}2 \,\, \underset{[Sp(2)]}{\overset{\mathfrak{g_2}}2} \,\, \overset{\mathfrak{su_2}}2 \,\, 2 \,\, 1 \,\, \overset{\mathfrak{e_8}}{(12)} \,\, ...[E_8]
$$

$Sp(2) \times U(1)$:

$$
[Sp(2)] \,\, \underset{[U(1)]}{\overset{\mathfrak{so_{10}}}3} \,\, \underset{[N_f=\frac{1}{2}]}{\overset{\mathfrak{sp_1}}1} \,\, \overset{\mathfrak{so_{9}}}4 \,\, 1 \,\, \overset{\mathfrak{g_2}}3 \,\,{\overset{\mathfrak{su_2}}2} \,\, 2 \,\, 1 \,\, \overset{\mathfrak{e_8}}{(12)} \,\, ...[E_8]
$$

$$
[Sp(2)] \,\, \overset{\mathfrak{so_{7}}}3  \,\, \underset{[U(1)]}1 \,\, {\overset{\mathfrak{g_2}}3} \,\, \overset{\mathfrak{su_2}}2 \,\, 2 \,\, 1 \,\, \overset{\mathfrak{e_8}}{(12)} \,\, ...[E_8]
$$

$$
[Sp(2)] \,\, \overset{\mathfrak{so_{10}}}4 \,\, \underset{[U(1)]}1 \,\, \overset{\mathfrak{su_3}}3 \,\, 1 \,\, \overset{\mathfrak{f_4}}5 \,\, 1 \,\, \overset{\mathfrak{g_2}}3 \,\,{\overset{\mathfrak{su_2}}2} \,\, 2 \,\, 1 \,\, \overset{\mathfrak{e_8}}{(12)} \,\, ...[E_8]
$$

$$
[Sp(2)] \,\, \overset{\mathfrak{so_{11}}}4 \,\, \underset{[U(1)]}{\overset{\mathfrak{sp_1}}1} \,\, \overset{\mathfrak{g_2}}3 \,\, 1 \,\, \overset{\mathfrak{f_4}}5 \,\, 1 \,\, \overset{\mathfrak{g_2}}3 \,\,{\overset{\mathfrak{su_2}}2} \,\, 2 \,\, 1 \,\, \overset{\mathfrak{e_8}}{(12)} \,\, ...[E_8]
$$

$G_2$:
$$
 {\overset{\mathfrak{su_3}}3} \,\, 1 \,\, \overset{[G_2]}{\overset{1}{\overset{\mathfrak{f_4}}5}} \,\, 1 \,\, \overset{\mathfrak{g_2}}3 \,\,{\overset{\mathfrak{su_2}}2} \,\, 2 \,\, 1 \,\, \overset{\mathfrak{e_8}}{(12)} \,\, ...[E_8]
$$

$$
[G_2]\,\, 1 \,\, \overset{\mathfrak{f_4}}5 \,\, 1 \,\, {\overset{\mathfrak{su_3}}3} \,\, 1 \,\, \overset{\mathfrak{f_4}}5 \,\, 1 \,\, \overset{\mathfrak{g_2}}3 \,\,{\overset{\mathfrak{su_2}}2} \,\, 2 \,\, 1 \,\, \overset{\mathfrak{e_8}}{(12)} \,\, ...[E_8]
$$

$$
[G_2] \,\, \overset{\mathfrak{su}_2}2 \,\, 2  \,\, 1 \,\, \overset{\mathfrak{e_8}}{(9)} \,\, ...[E_8]
$$

$$
2 \,\, \underset{[G_2]}{\overset{\mathfrak{su_2}}2} \,\, 1 \,\, \overset{\mathfrak{e_7}}8 \,\, 1  \,\, {\overset{\mathfrak{su_2}}2} \,\, \overset{\mathfrak{g_2}}3 \,\, 1 \,\, \overset{\mathfrak{f_4}}5 \,\, 1 \,\, \overset{\mathfrak{g_2}}3 \,\,{\overset{\mathfrak{su_2}}2} \,\, 2 \,\, 1 \,\, \overset{\mathfrak{e_8}}{(12)} \,\, ...[E_8]
$$

$SU(3) \times SU(3)$:

$$
[SU(3)] \,\, 1 \,\, \overset{[SU(3)]}{\overset{1}{\overset{\mathfrak{e_6}}6}} \,\, 1  \,\, \overset{\mathfrak{su_3}}3 \,\, 1 \,\, \overset{\mathfrak{f_4}}5 \,\, 1 \,\, \overset{\mathfrak{g_2}}3 \,\,{\overset{\mathfrak{su_2}}2} \,\, 2 \,\, 1 \,\, \overset{\mathfrak{e_8}}{(12)} \,\, ...[E_8]
$$

$Sp(2) \times SU(2)$:
$$
[Sp(2) \times SU(2)] \,\,  {\overset{\mathfrak{so_9}}3}  \,\, 1 \,\, {\overset{\mathfrak{g_2}}3} \,\, \overset{\mathfrak{su_2}}2 \,\, 2 \,\, 1 \,\, \overset{\mathfrak{e_8}}{(11)} \,\, ...[E_8]
$$

$$
[Sp(2)] \,\, \overset{\mathfrak{so_{7}}}3 \,\,1 \,\, \underset{[SU(2)]}{\overset{\mathfrak{so_{9}}}4} \,\, 1 \,\, {\overset{\mathfrak{g_2}}3} \,\, \overset{\mathfrak{su_2}}2 \,\, 2 \,\, 1 \,\, \overset{\mathfrak{e_8}}{(12)} \,\, ...[E_8]
$$

$G_2 \times SU(2)$:
$$
[G_2]\,\, 1 \,\, \overset{\mathfrak{f_4}}5 \,\, 1 \,\, \underset{[SU(2)]}{\overset{\mathfrak{g_2}}3} \,\, 1 \,\, \overset{\mathfrak{f_4}}5 \,\, 1 \,\, \overset{\mathfrak{g_2}}3 \,\,{\overset{\mathfrak{su_2}}2} \,\, 2 \,\, 1 \,\, \overset{\mathfrak{e_8}}{(12)} \,\, ...[E_8]
$$

$$
[SU(2)] \,\, 2 \,\, \underset{[G_2]}1 \,\, \overset{\mathfrak{f_4}}5 \,\, 1 \,\, \overset{\mathfrak{g_2}}3 \,\,{\overset{\mathfrak{su_2}}2} \,\, 2 \,\, 1 \,\, \overset{\mathfrak{e_8}}{(11)} \,\, ...[E_8]
$$

$$
[G_2]  \,\,1 \,\, \underset{[SU(2)]}{\overset{\mathfrak{f_{4}}}4} \,\, 1 \,\, {\overset{\mathfrak{g_2}}3} \,\, \overset{\mathfrak{su_2}}2 \,\, 2 \,\, 1 \,\, \overset{\mathfrak{e_8}}{(11)} \,\, ...[E_8]
$$

$$
 [G_2] \,\, {\overset{\mathfrak{su_2}}2} \,\, 2 \,\, 1 \,\,  \overset{[SU(2)]}{\overset{2}{\overset{1}{\overset{\mathfrak{e_8}}{(11)}}}} \,\, ...[E_8]
$$

$$
 [G_2] \,\, {\overset{\mathfrak{su_2}}2} \,\, 2 \,\, 1 \,\,  \overset{[SU(2)]}{\overset{2}{\overset{2}{\overset{1}{\overset{\mathfrak{e_8}}{(12)}}}}} \,\, ...[E_8]
$$

$$
[SU(2)]  \,\, {\overset{\mathfrak{g_2}}3} \,\, 1 \,\, \overset{[G_2]}{\overset{1}{\overset{\mathfrak{f_4}}5}} \,\, 1 \,\, \overset{\mathfrak{g_2}}3 \,\,{\overset{\mathfrak{su_2}}2} \,\, 2 \,\, 1 \,\, \overset{\mathfrak{e_8}}{(12)} \,\, ...[E_8]
$$

$$
[G_2]\,\, \overset{\mathfrak{su_2}}2 \,\, 2 \,\, \underset{[SU(2)]}1\,\, \overset{\mathfrak{e_7}}8\,\,1 \,\, \overset{\mathfrak{su_2}}2 \,\, \overset{\mathfrak{g_2}}3 \,\, 1 \,\, \overset{\mathfrak{f_4}}5 \,\, 1 \,\, \overset{\mathfrak{g_2}}3 \,\, {\overset{\mathfrak{su_2}}2} \,\, 2 \,\, 1 \,\, \overset{\mathfrak{e_8}}{(12)} \,\, ...[E_8]
$$

$Sp(3)$:

$$
[Sp(3)] \,\, {\overset{\mathfrak{g_2}}2} \,\, \overset{\mathfrak{su_2}}2 \,\, 2 \,\, 1 \,\, \overset{\mathfrak{e_8}}{(10)} \,\, ...[E_8]
$$

$$
[Sp(3)] \,\, \overset{\mathfrak{so_{12}}}4 \,\, \underset{[N_f=1/2]}{\overset{\mathfrak{sp_1}}1} \,\, \overset{\mathfrak{g_2}}3 \,\, 1 \,\, \overset{\mathfrak{f_4}}5 \,\, 1 \,\, \overset{\mathfrak{g_2}}3 \,\,{\overset{\mathfrak{su_2}}2} \,\, 2 \,\, 1 \,\, \overset{\mathfrak{e_8}}{(12)} \,\, ...[E_8]
$$

$$
[Sp(3)] \,\, \underset{[N_s=1/2]}{\overset{\mathfrak{so_{11}}}3} \,\,{\overset{\mathfrak{sp_1}}1} \,\, {\overset{\mathfrak{so_{9}}}4} \,\, 1 \,\, {\overset{\mathfrak{g_2}}3} \,\, \overset{\mathfrak{su_2}}2 \,\, 2 \,\, 1 \,\, \overset{\mathfrak{e_8}}{(12)} \,\, ...[E_8]
$$

$$
[Sp(3)]\,\, \overset{\mathfrak{g_2}}2 \,\, \overset{\mathfrak{su_2}}2 \,\, 1 \,\, \overset{\mathfrak{e_7}}8\,\,1 \,\, \overset{\mathfrak{su_2}}2 \,\, \overset{\mathfrak{g_2}}3 \,\, 1 \,\, \overset{\mathfrak{f_4}}5 \,\, 1 \,\, \overset{\mathfrak{g_2}}3 \,\, {\overset{\mathfrak{su_2}}2} \,\, 2 \,\, 1 \,\, \overset{\mathfrak{e_8}}{(12)} \,\, ...[E_8]
$$

$SO(7)$:

$$
[SO(7)]  \,\,  \overset{\mathfrak{su_2}}2  \,\, 1\,\, \underset{[N_f=1/2]}{\overset{\mathfrak{e_7}}7} \,\,1 \,\, {\overset{\mathfrak{su_2}}2} \,\, \overset{\mathfrak{g_2}}3 \,\, 1 \,\, \overset{\mathfrak{f_4}}5 \,\, 1 \,\, \overset{\mathfrak{g_2}}3 \,\,{\overset{\mathfrak{su_2}}2} \,\, 2 \,\, 1 \,\, \overset{\mathfrak{e_8}}{(12)} \,\, ...[E_8]
$$

$SO(7)\times U(1)$:

$$
[SO(7)]  \,\,1 \,\, {\overset{\mathfrak{so_{9}}}4} \,\, \underset{[U(1)]}{\overset{\mathfrak{sp_1}}1} \,\, {\overset{\mathfrak{so_{9}}}4} \,\, 1 \,\, {\overset{\mathfrak{g_2}}3} \,\, \overset{\mathfrak{su_2}}2 \,\, 2 \,\, 1 \,\, \overset{\mathfrak{e_8}}{(11)} \,\, ...[E_8]
$$

$$
[SO(7)]  \,\,  \overset{\mathfrak{su_2}}2  \,\, \underset{[U(1)]}1 \,\, {\overset{\mathfrak{e_6}}6} \,\, 1  \,\, \overset{\mathfrak{su_3}}3 \,\, 1 \,\, \overset{\mathfrak{f_4}}5 \,\, 1 \,\, \overset{\mathfrak{g_2}}3 \,\,{\overset{\mathfrak{su_2}}2} \,\, 2 \,\, 1 \,\, \overset{\mathfrak{e_8}}{(12)} \,\, ...[E_8]
$$

$SU(4)$:
$$
[SU(4)]\,\,   {\overset{\mathfrak{su_3}}2} \,\, {\overset{\mathfrak{su_2}}2} \,\, 2 \,\, 1 \,\, \overset{\mathfrak{e_8}}{(10)} \,\, ...[E_8]
$$

$SU(4) \times U(1)$:

$$
[SU(4)]\,\,   {\overset{\mathfrak{su_4}}2} \,\,  \underset{[U(1)]}{\overset{\mathfrak{su_4}}2} \,\,  {\overset{\mathfrak{su_3}}2} \,\, {\overset{\mathfrak{su_2}}2} \,\, 2 \,\, 1 \,\, \overset{\mathfrak{e_8}}{(12)} \,\, ...[E_8]
$$

$$
[SU(4)]\,\, \overset{\mathfrak{su_3}}2 \,\, \underset{[U(1)]}{\overset{\mathfrak{su_2}}2} \,\, 1 \,\, \overset{\mathfrak{e_7}}8\,\,1 \,\, \overset{\mathfrak{su_2}}2 \,\, \overset{\mathfrak{g_2}}3 \,\, 1 \,\, \overset{\mathfrak{f_4}}5 \,\, 1 \,\, \overset{\mathfrak{g_2}}3 \,\, {\overset{\mathfrak{su_2}}2} \,\, 2 \,\, 1 \,\, \overset{\mathfrak{e_8}}{(12)} \,\, ...[E_8]
$$

$SO(8)$:
$$
[SO(8)]\,\, 1 \,\, \overset{\mathfrak{so_8}}4  \,\, 1 \,\, \overset{\mathfrak{g_2}}3 \,\,{\overset{\mathfrak{su_2}}2} \,\, 2 \,\, 1 \,\, \overset{\mathfrak{e_8}}{(11)} \,\, ...[E_8]
$$

$$
[SO(8)]\,\, 1 \,\, \overset{\mathfrak{so_8}}4 \,\, 1 \,\, \overset{\mathfrak{so_8}}4  \,\, 1 \,\, \overset{\mathfrak{g_2}}3 \,\,{\overset{\mathfrak{su_2}}2} \,\, 2 \,\, 1 \,\, \overset{\mathfrak{e_8}}{(12)} \,\, ...[E_8]
$$

$Sp(2) \times Sp(2)$:
$$
[Sp(2)]\,\,  1 \,\, \underset{[Sp(2)]}{\overset{\mathfrak{so_{11}}}4} \,\, {\overset{\mathfrak{sp_1}}1} \,\, \overset{\mathfrak{so_{9}}}4 \,\, 1 \,\, \overset{\mathfrak{g_2}}3 \,\, \overset{\mathfrak{su_2}}2 \,\, 2 \,\, 1 \,\, \overset{\mathfrak{e_8}}{(12)} \,\, ...[E_8]
$$

$SU(4)\times SU(2)$:
$$
[SU(4)]\,\,   {\overset{\mathfrak{su_3}}2} \,\, {\overset{\mathfrak{su_2}}2} \,\, 2 \,\, 1 \,\, \overset{[SU(2)]}{\overset{2}{\overset{1}{\overset{\mathfrak{e_8}}{(12)}}}} \,\, ...[E_8]
$$

$$
[SU(4)]\,\, 1 \,\, \underset{[SU(2)]}{\overset{\mathfrak{so_{10}}}4} \,\, \underset{[N_f=1/2]}{\overset{\mathfrak{sp_1}}1} \,\, \overset{\mathfrak{so_9}}4  \,\, 1 \,\, \overset{\mathfrak{g_2}}3 \,\,{\overset{\mathfrak{su_2}}2} \,\, 2 \,\, 1 \,\, \overset{\mathfrak{e_8}}{(11)} \,\, ...[E_8]
$$

$SU(5)$:
$$
[SU(5)]\,\,  {\overset{\mathfrak{su_4}}2} \,\,  {\overset{\mathfrak{su_3}}2} \,\, {\overset{\mathfrak{su_2}}2} \,\, 2 \,\, 1 \,\, \overset{\mathfrak{e_8}}{(11)} \,\, ...[E_8]
$$

$F_4$:
$$
[F_4]\,\, 1 \,\, \overset{\mathfrak{g_2}}3 \,\,{\overset{\mathfrak{su_2}}2} \,\, 2 \,\, 1 \,\, \overset{\mathfrak{e_8}}{(10)} \,\, ...[E_8]
$$

$$
[F_4]\,\,1 \,\, \overset{\mathfrak{g_2}}3 \,\, \overset{\mathfrak{su_2}}2\,\, 1\,\, \overset{\mathfrak{e_7}}8\,\,1 \,\, \overset{\mathfrak{su_2}}2\,\, \overset{\mathfrak{g_2}}3 \,\, 1 \,\, \overset{\mathfrak{f_4}}5 \,\, 1 \,\, \overset{\mathfrak{g_2}}3 \,\,{\overset{\mathfrak{su_2}}2} \,\, 2 \,\, 1 \,\, \overset{\mathfrak{e_8}}{(12)} \,\, ...[E_8]
$$

$G_2 \times G_2$:
$$
[G_2] \,\,  1 \,\, \underset{[G_2]}{\underset{1}{\overset{\mathfrak{f_4}}5}} \,\, 1 \,\, \overset{\mathfrak{g_2}}3 \,\,{\overset{\mathfrak{su_2}}2} \,\, 2 \,\, 1 \,\, \overset{\mathfrak{e_8}}{(12)} \,\, ...[E_8]
$$

$$
[G_2] \,\, \overset{\mathfrak{su}_2}2 \,\, 2  \,\, 1 \,\, \overset{[G_2]}{\overset{\mathfrak{su_2}}{\overset{2}{\overset{2}{\overset{1}{\overset{\mathfrak{e_8}}{(12)}}}}}}  \,\, ...[E_8]
$$

$SO(7) \times SU(2)$:
$$
[SO(7)]\,\,   1\,\, \underset{[SU(2)]}{\overset{\mathfrak{so_9}}4}\,\,1 \,\, \overset{\mathfrak{g_2}}3 \,\,{\overset{\mathfrak{su_2}}2} \,\, 2 \,\, 1 \,\, \overset{\mathfrak{e_8}}{(11)} \,\, ...[E_8]
$$

$$
[SO(7)] \,\, {\overset{\mathfrak{su_2}}2} \,\, \underset{[SU(2)]}1 \,\, \overset{\mathfrak{f_4}}5 \,\, 1 \,\, \overset{\mathfrak{g_2}}3 \,\,{\overset{\mathfrak{su_2}}2} \,\, 2 \,\, 1 \,\, \overset{\mathfrak{e_8}}{(11)} \,\, ...[E_8]
$$

$$
[SO(7)] \,\, {\overset{\mathfrak{su_2}}2} \,\, 1 \,\, \overset{\mathfrak{e_7}}8 \,\,\underset{[SU(2)]}1 \,\,2 \,\, {\overset{\mathfrak{su_2}}2} \,\, \overset{\mathfrak{g_2}}3 \,\, 1 \,\, \overset{\mathfrak{f_4}}5 \,\, 1 \,\, \overset{\mathfrak{g_2}}3 \,\,{\overset{\mathfrak{su_2}}2} \,\, 2 \,\, 1 \,\, \overset{\mathfrak{e_8}}{(12)} \,\, ...[E_8]
$$

$$
[SO(7)]\,\, {\overset{\mathfrak{su_2}}2}  \,\, 1 \,\, \overset{[SU(2)]}{\overset{1}{\overset{\mathfrak{e_7}}8}} \,\,1 \,\, {\overset{\mathfrak{su_2}}2} \,\, \overset{\mathfrak{g_2}}3 \,\, 1 \,\, \overset{\mathfrak{f_4}}5 \,\, 1 \,\, \overset{\mathfrak{g_2}}3 \,\,{\overset{\mathfrak{su_2}}2} \,\, 2 \,\, 1 \,\, \overset{\mathfrak{e_8}}{(12)} \,\, ...[E_8]
$$

$Sp(3) \times SU(2)$:
$$
[Sp(3)]\,\,   {\overset{\mathfrak{g_2}}2} \,\, {\overset{\mathfrak{su_2}}2} \,\, 2 \,\, 1 \,\, \overset{[SU(2)]}{\overset{2}{\overset{1}{\overset{\mathfrak{e_8}}{(12)}}}} \,\, ...[E_8]
$$

$$
[Sp(3) \times SU(2)] \,\, \overset{\mathfrak{so_7}}2 \,\, {\overset{\mathfrak{su_2}}2} \,\, 1 \,\, \overset{\mathfrak{e_7}}8 \,\, 1  \,\, {\overset{\mathfrak{su_2}}2} \,\, \overset{\mathfrak{g_2}}3 \,\, 1 \,\, \overset{\mathfrak{f_4}}5 \,\, 1 \,\, \overset{\mathfrak{g_2}}3 \,\,{\overset{\mathfrak{su_2}}2} \,\, 2 \,\, 1 \,\, \overset{\mathfrak{e_8}}{(12)} \,\, ...[E_8]
$$

$Sp(4)$:
$$
[Sp(4)] \,\, \overset{\mathfrak{so_{13}}}4 \,\, {\overset{\mathfrak{sp_1}}1} \,\, \overset{\mathfrak{g_2}}3 \,\, 1 \,\, \overset{\mathfrak{f_4}}5 \,\, 1 \,\, \overset{\mathfrak{g_2}}3 \,\,{\overset{\mathfrak{su_2}}2} \,\, 2 \,\, 1 \,\, \overset{\mathfrak{e_8}}{(12)} \,\, ...[E_8]
$$

$$
[Sp(4)] \,\, {\overset{\mathfrak{g_2}}2} \,\, 1 \,\, \overset{\mathfrak{f_4}}5 \,\, 1 \,\, \overset{\mathfrak{g_2}}3 \,\,{\overset{\mathfrak{su_2}}2} \,\, 2 \,\, 1 \,\, \overset{\mathfrak{e_8}}{(11)} \,\, ...[E_8]
$$

$SO(10)$:
$$
[SO(10)]  \,\,{\overset{\mathfrak{sp_1}}1} \,\, {\overset{\mathfrak{so_{10}}}4} \,\, \underset{[N_f=1/2]}{\overset{\mathfrak{sp_1}}1} \,\, {\overset{\mathfrak{so_{9}}}4} \,\, 1 \,\, {\overset{\mathfrak{g_2}}3} \,\, \overset{\mathfrak{su_2}}2 \,\, 2 \,\, 1 \,\, \overset{\mathfrak{e_8}}{(12)} \,\, ...[E_8]
$$

$SO(9) \times SU(2)$:
$$
[SO(9)]  \,\,{\overset{\mathfrak{sp_1}}1} \,\, \underset{[SU(2)]}{\overset{\mathfrak{so_{11}}}4} \,\, {\overset{\mathfrak{sp_1}}1} \,\, {\overset{\mathfrak{so_{9}}}4} \,\, 1 \,\, {\overset{\mathfrak{g_2}}3} \,\, \overset{\mathfrak{su_2}}2 \,\, 2 \,\, 1 \,\, \overset{\mathfrak{e_8}}{(12)} \,\, ...[E_8]
$$

$$
[SO(9)] \,\, 1  \,\, \underset{[SU(2)]}{\overset{\mathfrak{so_7}}3} \,\, \overset{\mathfrak{su_2}}2\,\, 1\,\, \overset{\mathfrak{e_7}}8\,\,1 \,\, \overset{\mathfrak{su_2}}2\,\, \overset{\mathfrak{g_2}}3 \,\, 1 \,\, \overset{\mathfrak{f_4}}5 \,\, 1 \,\, \overset{\mathfrak{g_2}}3 \,\,{\overset{\mathfrak{su_2}}2} \,\, 2 \,\, 1 \,\, \overset{\mathfrak{e_8}}{(12)} \,\, ...[E_8]
$$

$F_4 \times SU(2)$:
$$
[F_4]\,\, 1 \,\, \overset{\mathfrak{g_2}}3 \,\,{\overset{\mathfrak{su_2}}2} \,\, 2 \,\, 1 \,\,  \overset{[SU(2)]}{\overset{2}{\overset{1}{\overset{\mathfrak{e_8}}{(12)}}}}  \,\, ...[E_8]
$$

$$
[F_4]\,\, 1 \,\, \underset{[SU(2)]}{\overset{\mathfrak{g_2}}3} \,\, 1 \,\, \overset{\mathfrak{f_4}}5 \,\, 1 \,\, \overset{\mathfrak{g_2}}3 \,\,{\overset{\mathfrak{su_2}}2} \,\, 2 \,\, 1 \,\, \overset{\mathfrak{e_8}}{(11)} \,\, ...[E_8]
$$

$SO(11)$:
$$
[SO(11)]\,\,   \overset{\mathfrak{sp_1}}1\,\, \overset{\mathfrak{so_9}}4\,\,1 \,\, \overset{\mathfrak{g_2}}3 \,\,{\overset{\mathfrak{su_2}}2} \,\, 2 \,\, 1 \,\, \overset{\mathfrak{e_8}}{(11)} \,\, ...[E_8]
$$

$SU(6)$:
$$
[SU(6)]\,\,   {\overset{\mathfrak{su_5}}2} \,\,  {\overset{\mathfrak{su_4}}2} \,\,  {\overset{\mathfrak{su_3}}2} \,\, {\overset{\mathfrak{su_2}}2} \,\, 2 \,\, 1 \,\, \overset{\mathfrak{e_8}}{(12)} \,\, ...[E_8]
$$

$$
[SU(6)] \,\, {\overset{\mathfrak{su_3}}2} \,\, 1 \,\, \overset{\mathfrak{f_4}}5 \,\, 1 \,\, \overset{\mathfrak{g_2}}3 \,\,{\overset{\mathfrak{su_2}}2} \,\, 2 \,\, 1 \,\, \overset{\mathfrak{e_8}}{(11)} \,\, ...[E_8]
$$

$$
[SU(6)]\,\, \overset{\mathfrak{su_4}}2 \,\, \overset{\mathfrak{su_2}}2\,\, 1\,\, \overset{\mathfrak{e_7}}8\,\,1 \,\, \overset{\mathfrak{su_2}}2\,\, \overset{\mathfrak{g_2}}3 \,\, 1 \,\, \overset{\mathfrak{f_4}}5 \,\, 1 \,\, \overset{\mathfrak{g_2}}3 \,\,{\overset{\mathfrak{su_2}}2} \,\, 2 \,\, 1 \,\, \overset{\mathfrak{e_8}}{(12)} \,\, ...[E_8]
$$

$$
[SU(6)]\,\, \overset{\mathfrak{su_3}}2\,\, 1\,\, \overset{\mathfrak{e_6}}6\,\,1 \,\, \overset{\mathfrak{su_3}}3 \,\, 1 \,\, \overset{\mathfrak{f_4}}5 \,\, 1 \,\, \overset{\mathfrak{g_2}}3 \,\,{\overset{\mathfrak{su_2}}2} \,\, 2 \,\, 1 \,\, \overset{\mathfrak{e_8}}{(12)} \,\, ...[E_8]
$$

$SO(12)$:
$$
[SO(12)] \,\, {\overset{\mathfrak{sp_1}}1}  \,\, \overset{\mathfrak{so_7}}3 \,\, \overset{\mathfrak{su_2}}2\,\, 1\,\, \overset{\mathfrak{e_7}}8\,\,1 \,\, \overset{\mathfrak{su_2}}2\,\, \overset{\mathfrak{g_2}}3 \,\, 1 \,\, \overset{\mathfrak{f_4}}5 \,\, 1 \,\, \overset{\mathfrak{g_2}}3 \,\,{\overset{\mathfrak{su_2}}2} \,\, 2 \,\, 1 \,\, \overset{\mathfrak{e_8}}{(12)} \,\, ...[E_8]
$$

$SO(13)$:
$$
[SO(13)]  \,\,{\overset{\mathfrak{sp_2}}1} \,\, {\overset{\mathfrak{so_{11}}}4} \,\, {\overset{\mathfrak{sp_1}}1} \,\, {\overset{\mathfrak{so_{9}}}4} \,\, 1 \,\, {\overset{\mathfrak{g_2}}3} \,\, \overset{\mathfrak{su_2}}2 \,\, 2 \,\, 1 \,\, \overset{\mathfrak{e_8}}{(11)} \,\, ...[E_8]
$$

$$
[SO(13)]\,\,  {\overset{\mathfrak{sp_1}}1} \,\, {\overset{\mathfrak{g_2}}3} \,\, 1 \,\, \overset{\mathfrak{f_4}}5 \,\, 1 \,\, \overset{\mathfrak{g_2}}3 \,\,{\overset{\mathfrak{su_2}}2} \,\, 2 \,\, 1 \,\, \overset{\mathfrak{e_8}}{(11)} \,\, ...[E_8]
$$

$E_6$:
$$
[E_6]\,\,  1 \,\, {\overset{\mathfrak{su_3}}3} \,\, 1 \,\, \overset{\mathfrak{f_4}}5 \,\, 1 \,\, \overset{\mathfrak{g_2}}3 \,\,{\overset{\mathfrak{su_2}}2} \,\, 2 \,\, 1 \,\, \overset{\mathfrak{e_8}}{(11)} \,\, ...[E_8]
$$

$$
[E_6]\,\, 1 \,\, \overset{\mathfrak{su_3}}3\,\, 1 \,\, \overset{\mathfrak{e_6}}6\,\,1 \,\, \overset{\mathfrak{su_3}}3 \,\, 1 \,\, \overset{\mathfrak{f_4}}5 \,\, 1 \,\, \overset{\mathfrak{g_2}}3 \,\, {\overset{\mathfrak{su_2}}2} \,\, 2 \,\, 1 \,\, \overset{\mathfrak{e_8}}{(12)} \,\, ...[E_8]
$$

$E_7$:
$$
[E_7]\,\, 1 \,\,{\overset{\mathfrak{su_2}}2} \,\, \overset{\mathfrak{g_2}}3 \,\, 1 \,\, \overset{\mathfrak{f_4}}5 \,\, 1 \,\, \overset{\mathfrak{g_2}}3 \,\,{\overset{\mathfrak{su_2}}2} \,\, 2 \,\, 1 \,\, \overset{\mathfrak{e_8}}{(11)} \,\, ...[E_8]
$$

$$
[E_7]\,\,1 \,\,  \overset{\mathfrak{su_2}}2 \,\,  \overset{\mathfrak{so_7}}3 \,\, \overset{\mathfrak{su_2}}2 \,\, 1\,\, \overset{\mathfrak{e_7}}8\,\,1 \,\, \overset{\mathfrak{su_2}}2\,\, \overset{\mathfrak{g_2}}3 \,\, 1 \,\, \overset{\mathfrak{f_4}}5 \,\, 1 \,\, \overset{\mathfrak{g_2}}3 \,\,{\overset{\mathfrak{su_2}}2} \,\, 2 \,\, 1 \,\, \overset{\mathfrak{e_8}}{(12)} \,\, ...[E_8]
$$

$E_8$:
$$
[E_8]\,\, 1\,\, 2 \,\,{\overset{\mathfrak{su_2}}2} \,\, \overset{\mathfrak{g_2}}3 \,\, 1 \,\, \overset{\mathfrak{f_4}}5 \,\, 1 \,\, \overset{\mathfrak{g_2}}3 \,\,{\overset{\mathfrak{su_2}}2} \,\, 2 \,\, 1 \,\, \overset{\mathfrak{e_8}}{(11)} \,\, ...[E_8]
$$

%%%%%%%%%%%%%%%%%%%%END HERE%%%%%%%%%%%%%%%%%%%%%%%%

\newpage

\bibliographystyle{utphys}
\bibliography{ref}

\end{document}